\documentclass[twocolumn]{aastex631}

\usepackage{graphicx}
\usepackage{txfonts}
\usepackage{hyperref}
\hypersetup{
    colorlinks = true,
    linkbordercolor = white,
    citecolor = blue,
    linkcolor=red
}

\usepackage{multirow}
\usepackage{longtable}  
\usepackage{xtab,booktabs}   %
\usepackage{threeparttable}  

\definecolor{mviola}{rgb}{0.54, 0.16, 0.8}
\definecolor{molive}{rgb}{0.33, 0.42, 0.18}
\definecolor{mgrey}{rgb}{0.1, 0.1, 0.3}

\shorttitle{The short GRB 200826A}
\shortauthors{Rossi et al.}

\graphicspath{{./}{figures/}}

\begin{document} 

\title{The peculiar short-duration GRB 200826A and its supernova\footnote{Based on data obtained with the LBT programs: LBT-2019B-18 (PI A. Rossi), DDT-2019B-9 (PI P. D'Avanzo), and LD-2020B-0100 (PI. B. Rothberg); and the TNG programme A41\_TAC15 (PI V. D'Elia).}
}

\author[0000-0002-8860-6538]{A. Rossi}
\affiliation{INAF - Osservatorio di Astrofisica e Scienza dello Spazio, via Piero Gobetti 93/3, 40129 Bologna, Italy}

\author[0000-0003-2283-2185]{B. Rothberg} 
\affiliation{LBT Observatory, University of Arizona, 933 N.Cherry Ave,Tucson AZ 85721, USA} 
\affiliation{George Mason University, Department of Physics \& Astronomy, MS 3F3, 4400 University Drive, Fairfax, VA 22030, USA}  
\author[0000-0002-8691-7666]{E. Palazzi}
\affiliation{INAF - Osservatorio di Astrofisica e Scienza dello Spazio, via Piero Gobetti 93/3, 40129 Bologna, Italy}

\author[0000-0003-2902-3583]{D. A. Kann}
\affiliation{Instituto  de  Astrof\'isica  de  Andaluc\'ia  (IAA-CSIC),  Glorieta  de  la Astronom\'ia, S/N, 18008 Granada, Spain}

\author{P. D'Avanzo}
\affiliation{INAF - Osservatorio Astronomico di Brera, Via E. Bianchi 46, I-23807, Merate (LC), Italy}

\author{L. Amati}
\affiliation{INAF - Osservatorio di Astrofisica e Scienza dello Spazio, via Piero Gobetti 93/3, 40129 Bologna, Italy}

\author[0000-0001-8413-7917]{S. Klose}
\affiliation{Th\"uringer Landessternwarte Tautenburg, Sternwarte 5, 07778 Tautenburg, Germany}

\author[0000-0002-0936-8237]{A. Perego}
\affiliation{Dipartimento di Fisica, Universit\'a di Trento, Via Sommarive 14, 38123 Trento, Italy}
\affiliation{INFN-TIFPA, Trento Institute for Fundamental Physics and Applications, via Sommarive 14, I-38123 Trento, Italy}

\author{E. Pian}
\affiliation{INAF - Osservatorio di Astrofisica e Scienza dello Spazio, via Piero Gobetti 93/3, 40129 Bologna, Italy}

\author{C. Guidorzi}
\affiliation{Department of Physics and Earth Science, University of Ferrara, via Saragat 1, I-44122, Ferrara, Italy}
\affiliation{INFN – Sezione di Ferrara, Via Saragat 1, 44122 Ferrara, Italy}
\affiliation{INAF - Osservatorio di Astrofisica e Scienza dello Spazio, via Piero Gobetti 93/3, 40129 Bologna, Italy}

\author{A. S. Pozanenko}
\affiliation{Space Research Institute of the Russian Academy of Sciences (IKI), 84/32 Profsoyuznaya Str, Moscow 117997, Russia}
\affiliation{Moscow Institute of Physics and Technology (MIPT), 9 Institutskiy per., Dolgoprudny, Moscow Oblast 141701, Russia}
\affiliation{National Research University Higher School of Economics, Moscow 101000, Russia}

\author{S. Savaglio}
\affiliation{Physics  Department,  University  of  Calabria,  87036  Arcavacata  diRende, Italy}

\author{G. Stratta}
\affiliation{INAF - Osservatorio di Astrofisica e Scienza dello Spazio, via Piero Gobetti 93/3, 40129 Bologna, Italy}
\affiliation{INAF - Istituto di Astrofisica e Planetologia Spaziali, via Fosso del Cavaliere 100, I-00133 Roma, Italy}

\author[0000-0002-6382-2613]{G. Agapito}
\affiliation{INAF - Osservatorio Astrofisico di Arcetri, Largo E. Fermi 5, 50125-Firenze, Italy}

\author{S. Covino}
\affiliation{INAF - Osservatorio Astronomico di Brera, Via E. Bianchi 46, I-23807, Merate (LC), Italy}

\author{F. Cusano}
\affiliation{INAF - Osservatorio di Astrofisica e Scienza dello Spazio, via Piero Gobetti 93/3, 40129 Bologna, Italy}

\author{V. D'Elia}
\affiliation{Space Science Data Center (SSDC) - Agenzia Spaziale Italiana (ASI), I-00133 Roma, Italy}
\affiliation{INAF - Osservatorio Astronomico di Roma, via Frascati 33, 00040 Monte Porzio Catone, Italy}

\author{M. De Pasquale}
\affiliation{University of Messina, MIFT Department, Polo Papardo, Viale F.S. D'Alcontres 31, 98166 Messina, Italy}
\affiliation{Department of Astronomy and Space Sciences, Faculty of Science, Istanbul University, 34119 Istanbul, Turkey}

\author{M. Della Valle}
\affiliation{INAF, Osservatorio Astronomico di Capodimonte, Salita Moiariello 16, 80131 Naples, Italy}

\author{O. Kuhn}
\affiliation{LBT Observatory, University of Arizona, 933 N.Cherry Ave,Tucson AZ 85721, USA}

\author{L. Izzo}
\affiliation{DARK, Niels Bohr Institute, University of Copenhagen, Jagtvej 128, 2200 Copenhagen, Denmark}

\author{E. Loffredo}
\affiliation{Gran Sasso Science Institute, Viale F. Crispi 7, I-67100, L'Aquila (AQ), Italy}
\affiliation{INFN - Laboratori Nazionali del Gran Sasso, I-67100, L'Aquila (AQ), Italy}

\author[0000-0001-9487-7740]{N. Masetti}
\affiliation{INAF - Osservatorio di Astrofisica e Scienza dello Spazio, via Piero Gobetti 93/3, 40129 Bologna, Italy}
\affiliation{Departamento de Ciencias F\'isicas, Universidad Andr\'es Bello, Fern\'andez Concha 700, Las Condes, Santiago, Chile}

\author{A. Melandri}
\affiliation{INAF - Osservatorio Astronomico di Brera, Via E. Bianchi 46, I-23807, Merate (LC), Italy}

\author{P. Y. Minaev}
\affiliation{Space Research Institute of the Russian Academy of Sciences (IKI), 84/32 Profsoyuznaya Str, Moscow 117997, Russia}
\affiliation{Moscow Institute of Physics and Technology (MIPT), 9 Institutskiy per., Dolgoprudny, Moscow Oblast 141701, Russia}
\affiliation{P. N. Lebedev Physical Institute of the Russian Academy of Sciences, Leninskii pr. 53, Moscow, 119991, Russia}

\author[0000-0002-6856-9813]{A. Nicuesa Guelbenzu}
\affiliation{Th\"uringer Landessternwarte Tautenburg, Sternwarte 5, 07778 Tautenburg, Germany}


\author{D. Paris}
\affiliation{INAF - Osservatorio Astronomico di Roma, via Frascati 33, 00040 Monte Porzio Catone, Italy}


\author{S. Paiano}
\affiliation{INAF - Osservatorio Astronomico di Roma, via Frascati 33, 00040 Monte Porzio Catone, Italy}
\affiliation{INAF - Istituto di Astrofisica Spaziale e Fisica Cosmica di Palermo, via Ugo La Malfa, 153, 90146 Palermo, Italy}
\affiliation{INAF - Istituto di Astrofisica Spaziale e Fisica cosmica di Milano, via Alfonso Corti 12, 20133 Milano, Italy}

\author[0000-0002-3898-4004]{C. Plantet}
\affiliation{INAF - Osservatorio Astrofisico di Arcetri, Largo E. Fermi 5, 50125-Firenze, Italy}

\author[0000-0002-3544-1629]{F. Rossi}
\affiliation{INAF - Osservatorio Astrofisico di Arcetri, Largo E. Fermi 5, 50125-Firenze, Italy}

\author{R. Salvaterra}
\affiliation{INAF - Istituto di Astrofisica Spaziale e Fisica cosmica di Milano, via Alfonso Corti 12, 20133 Milano, Italy}

\author{S. Schulze}
\affiliation{The Oskar Klein Centre, Physics Department of Physics, Stockholm University, Albanova University Center, SE 106 91 Stockholm, Sweden}

\author{C. Veillet}
\affiliation{LBT Observatory, University of Arizona, 933 N.Cherry Ave,Tucson AZ 85721, USA}

\author{A. A. Volnova}
\affiliation{Space Research Institute of the Russian Academy of Sciences (IKI), 84/32 Profsoyuznaya Str, Moscow 117997, Russia}

\email{andrea.rossi@inaf.it}

 \begin{abstract}
   Gamma-ray bursts (GRBs) are classified as long and short events.
   Long GRBs (LGRBs) are associated with the end states of very massive stars, while short GRBs (SGRBs) are linked to the merger of compact objects.  
GRB 200826A was a peculiar event, because by definition it was a SGRB, with a rest-frame duration of $\sim0.5$ s.
   However, this event was energetic and soft, which is consistent with LGRBs. 
   The relatively low redshift ($z=0.7486$) motivated a comprehensive, multi-wavelength follow-up campaign  to characterize its host, search for a possible associated supernova (SN),
 and thus understand the origin of this burst.
   To this aim we obtained a combination of deep near-infrared (NIR) and optical imaging together with spectroscopy. 
   Our analysis reveals an  optical and NIR bump in the light curve
   whose luminosity and evolution is in agreement with several LGRB-SNe. 
   Analysis of the prompt GRB shows that this event follows the $E_{\rm p,i}-E_{\rm iso}$ relation found for LGRBs. 
The host galaxy is a low-mass star-forming galaxy, typical for LGRBs, but
with one of the  highest star-formation rates (SFR), especially with respect to its mass
($\log M_\ast/M_\odot = 8.6$, SFR $\sim 4.0 \,M_\odot$/yr).
 We conclude that GRB 200826A is a typical collapsar event 
 in the low tail of the duration distribution
 of LGRBs.
  
 These findings support theoretical predictions that events produced by collapsars can be as short as 0.5 s in the host frame and further confirm that duration alone is not an efficient discriminator for the progenitor class of a GRB.
 \end{abstract}


%

\section{Introduction \label{sec:int}}

Traditionally, gamma-ray bursts (GRBs) are divided in two classes based on their duration
  and their spectral hardness in the gamma energy range. The majority of observed GRBs have durations
  longer than $2$ s, a soft high-energy spectrum and are termed ``long GRBs'' (LGRBs), as opposed to ``short GRBs'' (SGRBs) with a usually sub-second duration and a harder high-energy spectrum \citep{Mazets1981a,Kouveliotou1993a}.
  However, there is a substantial overlap between the distributions of the durations of the two GRB classes (\citealt{Kouveliotou1993a}, \citealt{Bromberg2013a}, \citealt{Horvath2016a}).

Long GRBs are typically associated with highly energetic broad-lined type Ic 
supernovae (SNe) and have therefore been associated confidently with the deaths of very 
massive stars \citep[e.g.,][]{WoosleyBloom2006a,Cano2017a}. On the other hand, short GRBs 
are thought to be connected with the merger of compact objects, as was firmly 
demonstrated in spectacular fashion by the SGRB 170817A which was accompanied by the binary 
neutron star (NS) merger gravitational-wave source GW170817 detected by the LIGO and Virgo interferometers as well as by the 
kilonova (KN) AT2017gfo \citep[e.g.,][]{Abbott2017c,Abbott2017b,Abbott2017a,Pian2017a,Smartt2017a}. 
 
Improving our understanding of the connection between LGRBs and SNe has been one of the most 
intense areas of research in the field of GRBs for over 20 years. More than 1400 bursts have been 
discovered by the \textit{Neil Gehrels \textit{Swift} Observatory} (\textit{Swift} hereafter) in the 
last 16 years,
among which only $40-50$ GRBs have an associated SN identified by the late bumps in their optical  
afterglow light curves, and to date, just 28 have been spectroscopically confirmed 
(Rossi et al. 2022, in prep., \citealt{Izzo2019a,Melandri2022AA},
\citealt{Cano2017a} and references therein, \citealt{Klose2019a,Cano2017AA,Ashall2019a,Melandri2019a,Hu2021a}). The burst duration for 
all these events lasted for more than 2 s, therefore they are considered LGRBs. While these 
 results show a simple and clear correlation between SNe and LGRBs, 
the reality may be more murky.

 A first break in the LGRB-SN association was already 
  found in 2006 with the discovery of the LGRBs 060505 and 060614\footnote{GRBs 060505 and 060614 have durations (T$_{90}$) of $4\pm1$s \citep{HullingerGCN2006a} and $\sim102\pm5$ s \citep{BarthelmyGCN2006a}, respectively.}. Their relatively low redshifts ($z=0.089$ and $z=0.125$, respectively) allowed for the search of the possibly associated SNe, but despite achieving
  deep photometric limits no SN component as luminous as those observed in the case of other long GRBs was detected \citep{DellaValle2006a,Gal-Yam2006a,Fynbo2006a,Ofek2007a,McBreen2008a,Xu2009a}\footnote{ More recently, KNe have been claimed to have been detected in the afterglows of GRB 060614 \citep{Yang2015a,Jin2015a} and GRB 060505 \citep{Jin2021a}. This would imply they are actually merger events (see \citealt{Kann2011a} for more discussion).}.
  Later, LGRB 111005A at $z=0.0133$ \citep{Tanga2018a,Michalowski2018a} also joined this group.
 On the other hand, in the case of GRB 040924, classified as short for its rest-frame duration ($\sim1$ s), an associated SN was detected \citep{Wiersema2008a,Soderberg2006ApJ} thus identifying its origin as being the collapse of a massive star. Another peculiar case is GRB 090426, which has a duration shorter than 2 s, but with a high-energy spectrum and afterglow brightness typical of long GRBs (\citealt{Antonelli2009a,Thone2011MNRAS,Nicuesa2011a,Nicuesa2012a}; Kann et al. 2022, in prep.).
Indeed, although classifying a burst by its duration as either long or short is still the most simple and adopted approach, the above results show that this is not enough to reveal the nature of the GRB progenitor.
Even considering the spectral hardness often may not help in the classification \citep[e.g.,][]{MinaevPozanenko2020a,Agui2021a}, 
and thus understanding the origin of the burst. 
Indeed, the most secure way 
is to determine whether the event is associated with a supernova (SN) and is a LGRB, or with a KN and is a SGRB.

We here present follow-up observations and analysis of GRB 200826A, including the search for an 
associated SN  and characterization of the host galaxy. This very intense, but short-duration burst was detected by several 
space-based GRB detectors \citep[see][]{Hurley2020GCNipn}, including Konus-\emph{Wind} 
\citep{Ridnaia2020GCNkw,Svinkin2020GCNenergy}, \emph{Fermi}/GBM \citep{Mangan2020GCNgbm}, 
\emph{AGILE}/MCAL \citep{Pittori2020GCNagile} and \emph{AstroSat}/CZTI 
\citep{Gupta2020GCNastrosat}. Stemming from its observed short duration (GBM $T_{90}=1.14$ s, 
\citealt{Mangan2020GCNgbm}; Konus-\emph{Wind} $T_{90}=0.772$ s, 
\citealt{Svinkin2020GCNenergy}), it was initially thought to be a genuine SGRB. 
 The ZTF survey project \citep{Bellm2019a} launched an observing campaign to search for the optical counterpart \citep{Carracedo2020GCN} within the large error circle of \emph{Fermi}/GBM. Among 
the several candidates detected, ZTF20abwysqy was confirmed to be the afterglow of GRB 200826A \citep{Ahumada2020GCNdisco,Belkin2020GCNkitab,Dichiara2020GCNlowell}. 
An X-ray afterglow was also detected by \emph{Swift}/XRT \citep{Dai2020GCNswiftxrt}.
The GRB spectroscopic redshift ($z=0.748$) was obtained thanks to LBT/MODS observations 
\citep{Rothberg2020GCNlbt}. The provided distance was used to calculate the real 
luminosities and energy values, which showed that GRB 200826A was a very energetic event 
with an isotropic energy release and a peak energy more consistent with classical LGRBs than 
with typical SGRBs \citep{Svinkin2020GCNenergy}. 
 A SN bump in the afterglow light curve was finally reported by \cite{Ahumada2020GCNsnbump1}.
Additional optical observations of the afterglow and the SN have been reported in \cite{Ahumada2021a,Zhang2021a}. 

Throughout this work, the flux density of  the afterglow is described as $F_\nu (t) \propto 
t^{-\alpha} \nu^{-\beta}$, and a $\Lambda$CDM world model with $\Omega_M = 0.308$, 
$\Omega_{\Lambda} = 0.692$, and $H_0 = 67.8$ km s$^{-1}$ Mpc$^{-1}$ \citep{Planck2016a} has 
been assumed. 
All data are in observer frame, unless differently specified.

\begin{figure*}
\begin{center}
\includegraphics[width=0.37\textwidth,angle=0]{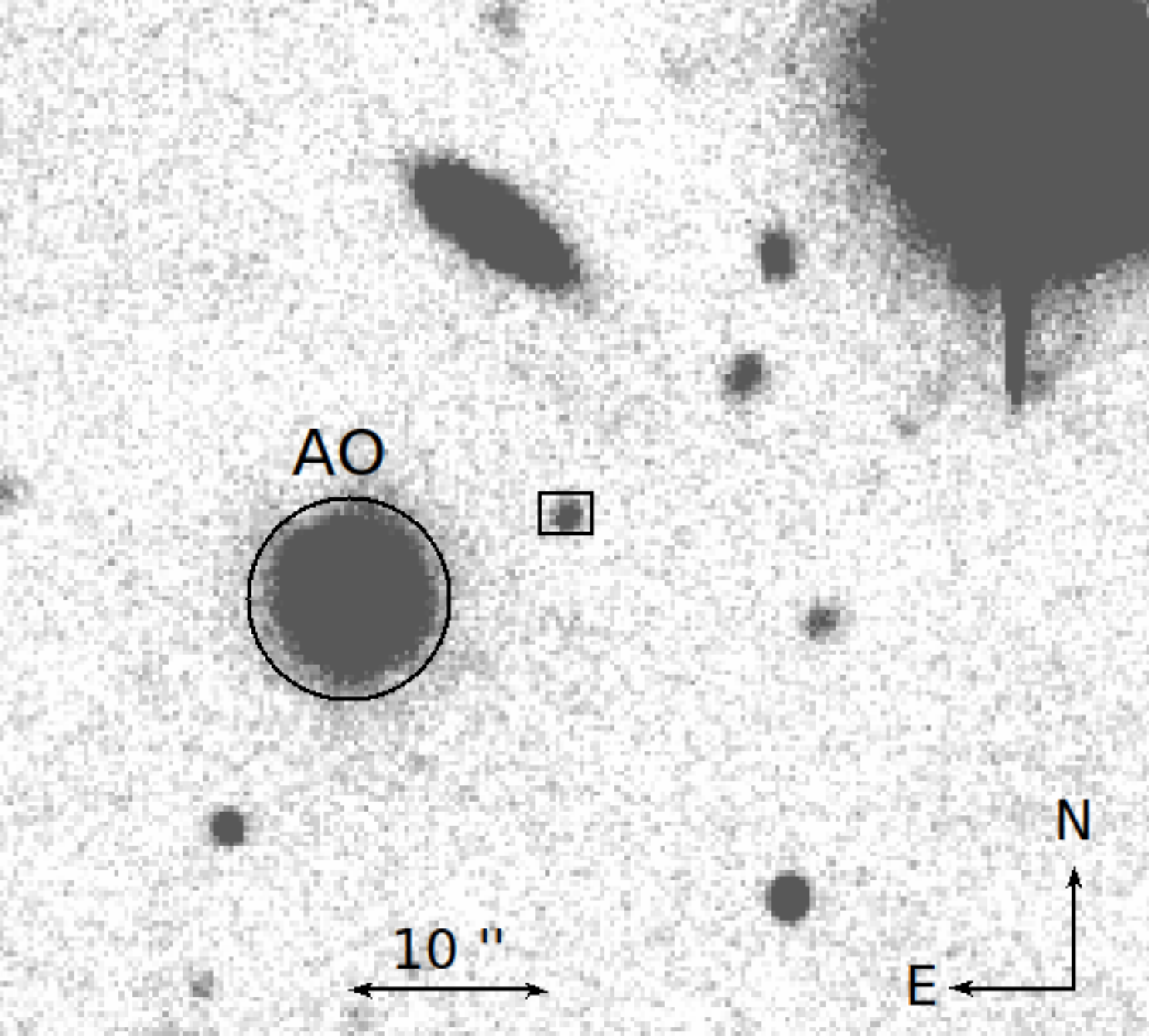}
\includegraphics[width=0.605\textwidth,angle=0]{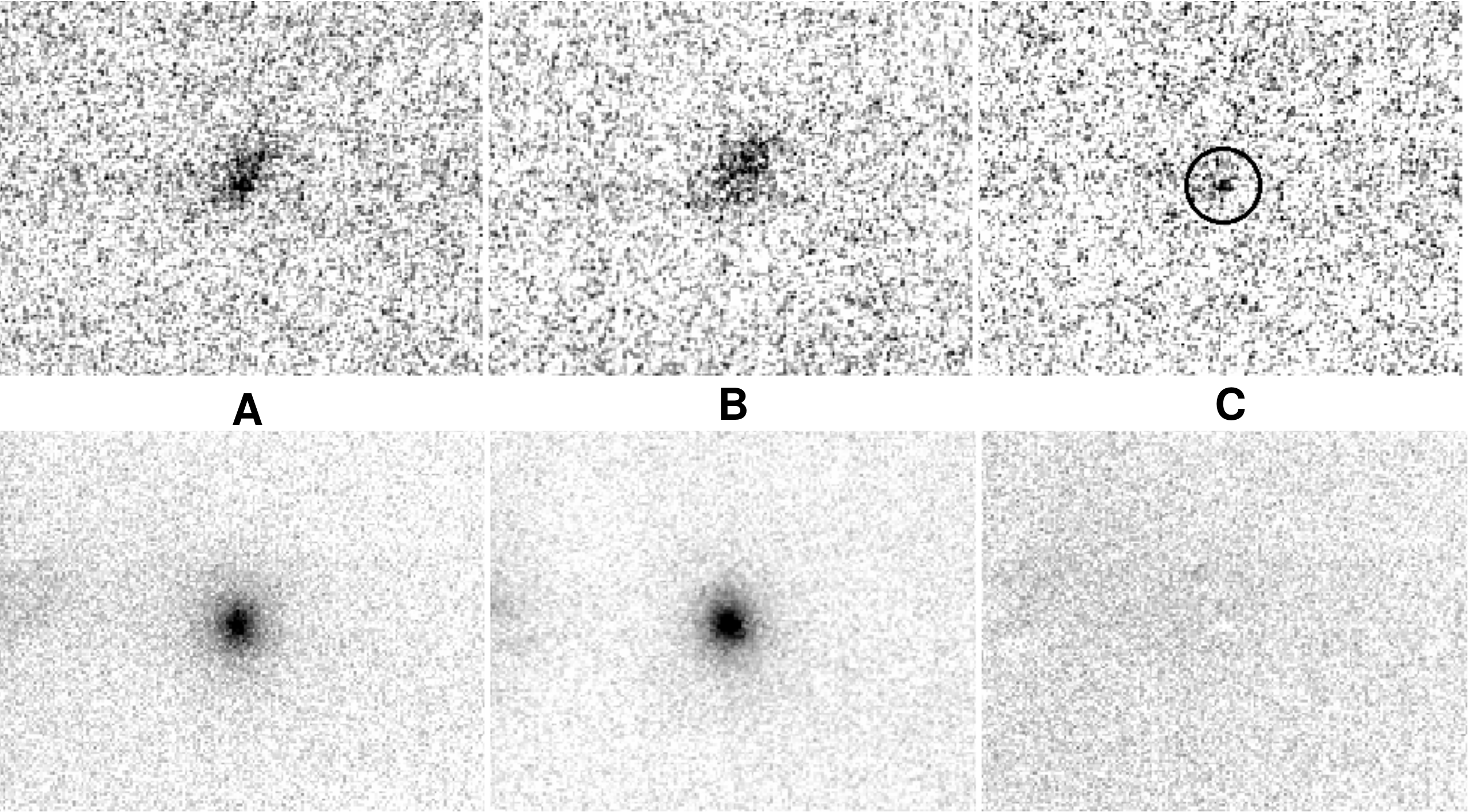}
\caption{{\bf Left: } $1\arcmin\times1\arcmin$ field-of-view (FoV) finding chart derived from the TNG $r^\prime$-band imaging obtained on 2020-09-27 UT. The star used for AO observations with SOUL and the LBT/LUCI camera is highlighted. The rectangular region is centered on the location of the afterglow, and its angular size is the same as the panels on the right. {\bf Right: } Results of the $H$-band follow-up using the second generation SOUL Adaptive Optics with LUCI-1 at the Large Binocular Telescope. The top row shows the location of the host and the GRB (see the region in the left panel). For comparison, the bottom row shows the location of a star in the field. From left to right: A) the LUCI-1 AO $H$-band imaging obtained on UT 02-10-2020 (i.e., close to the peak of the SN); B) LUCI-1 AO $H$-band image obtained four months later when only the host is visible; and C) the subtraction between these two epochs, where  the transient is highlighted with a circle (top) and  
 the field star is cleanly subtracted (bottom).
The FoV of each panel is $2\farcs6\times2\farcs0$.  The LUCI-1 AO FoV is $30\arcsec\times30\arcsec$. In each panel North is up, East is to the left. The transient lies at a projected distance of 0.75 kpc from the center of its host galaxy as discussed in \S\ref{sec:host}.
}
\label{fig:ao}%
\end{center}
\end{figure*}

\section{Observations and data analysis \label{sec:data}}

\subsection{Optical Spectroscopy} \label{sec:spec}
Optical spectroscopy of GRB 200826A was obtained with the Large Binocular Telescope (LBT) on Mt. Graham, Arizona, USA, on August 28, 2020 at UT 09:06:45 (MJD 59089.379692) using both optical Multi-Object Double Spectrographs \citep[MODS,][]{Pogge2010a}. Each MODS was configured using a 1\arcsec  wide slit (R $\sim1500$). MODS-1 was configured with the red grating ($0.58\,\mu\textnormal{m}-1.0\,\mu$m coverage) due to issues at the time with the blue channel, and MODS-2 was configured with the dual grating ($0.32\,\mu\textnormal{m}-1.05\,\mu$m coverage). Observations consisted of $4\times900$ second exposures for the three channels, which yielded an effective exposure time of 7200 seconds for $0.58\,\mu\textnormal{m}-1.0\,\mu$m (the red channels).  A position angle $\textnormal{(PA)}=17.0^{\circ}$ was used in order to cover both the reported location of the GRB 200826A afterglow and a nearby galaxy $15\arcsec$ NE.  Observations were obtained at or near transit with an effective airmass of 1.0005. The seeing for the spectroscopic observations ranged from $0\farcs7-1\farcs0$ and conditions were photometric.

The MODS data were reduced first with modsCCDRed version 2.04 to remove the bias and flat-field the data.  Next, custom IRAF scripts were used to: extract along the central slit using a stellar trace; wavelength calibrate and rectify the tilt in both X and Y directions using arc-lamp lines; flux calibrate the data using the spectro-photometric standard star G191-B2B; and remove telluric features from the red channels using the spectro-photometric standard.  The data from the three channels were combined into a single spectrum and re-binned to a common scale of $\Delta\lambda=0.85$ \AA{} pixel$^{-1}$.  
An afterglow spectral energy distribution (SED; \S \ref{sec:res}) was then subtracted from the spectrum and the absolute flux calibration was fine-tuned using the late-time $g^\prime$, $r^\prime$, $\it i^\prime$, $\it z^\prime$ MODS  and the Large Binocular Cameras
\citep[LBCs;][]{Giallongo2008a} photometry (Table \ref{tab:photall}), which is free from SN and afterglow emission (see \S \ref{sec:ag}).

\subsection{Optical and Near-Infrared Imaging } \label{sec:opt}

The optical follow-up was performed with
LBT in the Sloan filters $g^\prime$ and $r^\prime$ bands using the two twin MODS instruments, and in the Sloan $r^\prime$ filter with the Device Optimized for the Low Resolution (DOLoRes) optical imager and spectrograph mounted on the Telescopio Nazionale Galileo (TNG) on La Palma (Canary Islands, Spain).
 We have also observed GRB 200826A twice with AZT-22 1.5m telescope equipped with the SNUCAM CCD camera \citep{Im2010} at the Maidanak Astronomical Observatory (MAO, Uzbekistan).

All MODS observations were taken under good seeing conditions ($0\farcs5-1\farcs0$). Due to the binocular capabilities of LBT, the observations in $g^\prime$ and $r^\prime$ were obtained simultaneously with {\it both} MODS instruments. 
 During summer of 2021, with the aim of expanding our host-galaxy photometry data set, we obtained further $U r^\prime i^\prime z^\prime$ and $R_C$ imaging with
the LBCs under variable seeing conditions ($0\farcs9-1\farcs5$), followed by further MAO observations in the $R_C$ band.
LBT/MODS  and LBC imaging data were reduced using the data reduction pipeline developed at INAF - Osservatorio
Astronomico di Roma \citep{Fontana2014a} which includes bias subtraction and flat-fielding, bad pixel and cosmic ray masking, astrometric calibration, and coaddition. 
We have also downloaded the Gemini/GMOS data at 28 and 46 d reported in \cite{Ahumada2021a}\footnote{ PI L. Singer, program GN-2020B-DD-104.}, and afterwards reduced them using the dedicated Gemini pipeline DRAGONS \citep{dragons2019}.
 Finally, we have downloaded the $r^\prime$-band  Gran Telescopio Canarias (GTC) data at $3.99$ d reported in \cite{Zhang2021a}\footnote{ PI A. J. Castro-Tirado, program GTCMULTIPLE4B-20A}.
TNG/DOLoRes, GTC and MAO images were reduced in a standard manner using PyRAF/IRAF tasks \citep{Tody1993}.
 
The first $H$-band observations were obtained using the LBT Utility Camera in the Infrared \citep[LUCI,][]{Seifert2003a} imager and spectrograph in conjunction with the Single conjugate adaptive Optics Upgrade for LBT \citep[SOUL;][]{Pinna2021a}, the 2nd generation adaptive optics (AO) system at LBT. Data were obtained 37.1 d after the GRB, close to the expected maximum light of the possible associated SN. Compared to $J$ and $K$ bands, the $H$ band was preferred for three reasons: 1) lower sky and thermal background  with respect to the $K$ band; 2)  the AO performance is better than in $J$, allowing the system to achieve a full-width at half-maximum (FWHM) $<0\farcs1$; 3) there are no strong emission lines from non-stellar sources or from star-formation regions  (like H$\alpha$ in the $J$-band at the redshift of our target), and the host continuum is dominated by light from old, late-type stars. 
A relatively bright foreground star ($R=15.06$ mag) $11\farcs4$ away from the GRB location (and within the field-of-view) was used as the ``AO Reference star'' for the AO system. 
The AO system achieved an average FWHM of the point-spread function (PSF) $\sim0\farcs13$. Subsequent observations were able to achieve a FWHM $\sim0\farcs08-0\farcs13$ (see \S \ref{sec:sub} for more details).
 The observations were repeated on December 1st (2020) and February 7th (2021), 97 and 158 d respectively, after the GRB event.

 In September 2021 we also obtained $J$-band imaging of the host with LBT/LUCI.
All NIR data have been reduced using the same data reduction pipeline as for LBT MODS data (\S \ref{sec:opt}), which in case of LUCI includes dark subtraction, flat-fielding, bad pixel and cosmic ray masking, and sky subtraction. Finally, the images were aligned to match the position of the AO reference star and co-added using IRAF tasks. 
The astrometry was calibrated against field stars in the GAIA DR2 catalogue \citep{Gaia2018a} and has an astrometric precision of 0\farcs15. 

All data were analyzed by performing aperture photometry using DAOPHOT and APPHOT under PyRAF/IRAF.
For the optical data, the photometric calibration was performed using a set of field stars observed during the fifth MODS epoch (which had photometric conditions) and calibrated against the SDSS DR12 catalog magnitudes of brighter nearby stars \citep{Alam2015a}. NIR photometry was calibrated against the AO reference star and a pair of 2MASS stars observed immediately after the GRB field during the first epoch.
 The photometric calibration has a standard deviation of $0.02$ mags. 
To err on the side of caution, $2\times$ this value was used as the photometric uncertainty.


\begin{table}
\caption{Photometry of the Transient and Host Galaxy.}
\begin{threeparttable}
\setlength{\tabcolsep}{0.4em}
\begin{tabular}{lccccc} 
\toprule
$\Delta$t\tnote{a} & Magnitude    & Filter   & Telescope & Ref.\tnote{c}  \\ 
(d)    & AB\tnote{b}           &          &             & \\  
\midrule  
0.2112    &   $20.86\pm0.04$   &  $g^\prime$    &  ZTF 	   & [1]    \\
0.2304    &   $20.70\pm0.05$   &  $r^\prime$    &  ZTF 	   & ''           \\
0.2779    &   $20.96\pm0.16$   &  $g^\prime$    &  ZTF 	   & ''           \\
0.7663    &   $>20.2$           &  $R_C$          &  Kitab       & [2]    \\
1.1458    &   $22.75\pm0.26$    &  $g^\prime$    &  ZTF 	   & [1]           \\
1.2125    &   $>21.3$           &  $r^\prime$    &  ZTF 	   & ''           \\
1.2875    &   $>21.2$           &  $g^\prime$    &  ZTF         & ''          \\
1.7351    &   $>20.4$           &  $R_C$          &  Kitab       & [2]    \\
2.183\tnote{d}&$23.39\pm0.16$   &  $r^\prime$    & LBT/MODS & [3]      \\
2.183\tnote{d}&$24.08\pm0.19$   &  $g^\prime$    & LBT/MODS & ''  \\ 
2.27      &   $>23.3$           &  $r^\prime$    & FTN+LCO     & [1] \\
2.28      &   $>23.4$           &  $g^\prime$    & FTN+LCO     & '' \\
3.23      &   $24.46\pm0.12$    &  $r^\prime$    & Lowell      & ''    \\
3.99      &   $24.51\pm0.14$   & $r^\prime$     & GTC          & [3,4]    \\
7.21      &   $>23.5$           & $r^\prime$     & LBT/MODS         & [3]   \\
7.21      &   $>23.6$           & $g^\prime$     & LBT/MODS         & ''    \\  
15.7      &   $>24.3$           & $R_C$          & MAO        & ''    \\  
28.3      &   $24.53\pm0.21$    & $i^\prime$     & Gemini      & [1,3]   \\
28.3      &   $>25.6$           & $r^\prime$     & Gemini      & [1,3]   \\
31.9      &   $>25.3$           & $r^\prime$     &  TNG        & [3]    \\
33.3      &   $>25.7$           & $r^\prime$     &  LBT/MODS        & ''    \\
33.3      &   $>25.8$           & $g^\prime$     &  LBT/MODS        & ''    \\  
37.1      &   $24.06\pm0.20$    & $H$              & LBT/LUCI         & ''    \\
46.1      &   $25.36\pm0.26$    & $i^\prime$     & Gemini      &  [1,3]   \\
46.1      &   $>25.5$           & $r^\prime$     & Gemini      &  [1,3]   \\
50.88     &   $>25.6$           & $r^\prime$     &  LBT/MODS        & [3]      \\
50.88     &   $>26.2$           & $g^\prime$     &  LBT/MODS        & ''      \\
97.0      &   $>24.7$           & $H$              & LBT/LUCI         & ''      \\
\midrule
\midrule
74.8      &   $22.85\pm0.11$    & $r^\prime$    & TNG         & ''      \\
78.9      &   $22.83\pm0.04$    & $r^\prime$     & LBT/MODS         & ''     \\
78.9      &   $23.12\pm0.04$    & $g^\prime$     & LBT/MODS         & ''      \\
158.5 & $22.84\pm0.20$  &  $H$          &LBT/LUCI    & ''  \\
287      &   $23.36\pm0.10$    & U\tnote{e}  & LBT/LBC        & '' \\  
287      &   $22.92\pm0.08$    & $r^\prime$      & LBT/LBC        & '' \\  
287      &   $22.84\pm0.07$    & $R_C$           & LBT/LBC        & '' \\  
287      &   $22.58\pm0.05$    & $i^\prime$      & LBT/LBC        & '' \\  
287      &   $22.25\pm0.11$    & $z^\prime$      & LBT/LBC        & '' \\  
340      &   $22.85\pm0.07$    & $R_C$           & MAO            & '' \\
382      &   $23.29\pm0.08$    & $B$           & MAO            & '' \\
391       &  $22.65\pm0.14$    & $J$             &LBT/LUCI        & '' \\
\bottomrule                                 
\end{tabular}   
\begin{tablenotes}\footnotesize 
\item[a] Mid-time after the \emph{Fermi}/GBM trigger in the observer frame.
\item[b] The photometry is not corrected for Galactic extinction.  The first part (above the double line) is the result of image subtraction and represents the pure transient. The second part has been used to study the host and as references for the subtraction.  
\item[c] [1] \cite{Ahumada2021a}; [2] GCN 28306 \citep{Belkin2020GCNkitab}; [3] this work; [4] \cite{Zhang2021a}. 
\item[d] These are LBT/MODS acquisition images obtained using a $\sim2\arcmin\times2\arcmin$ FOV instead of 
the full $\sim6\arcmin\times6\arcmin$. MODS1 was equipped with the $g^\prime$ filter and MODS2 with the $r^\prime$ filter. 
 \item[e] We have used the $Uspec$ for its better sensitivity compared to the $U_{Bessel}$ filter.
\end{tablenotes}    
\label{tab:photall}
\end{threeparttable}         
\end{table}  

\subsection{Image subtraction and photometry of the transient}\label{sec:sub}

 To remove the contribution of the host to the afterglow/SN photometry, a reference frame free of any transient light is required.
To better investigate any possible contamination from SN light in these earlier observations, 
the GRB 200826A data sets were compared with GRB 980425/SN 1998bw \citep{Galama1998a},
which is the standard template for GRB-SNe.
Following \cite{Zeh2004ApJ} and \cite{Klose2019a}, the observed SN 1998bw light curves were first corrected for Galactic extinction, moved to rest-frame, interpolated with a polynomial curve at steps of 0.5 d and finally shifted to the GRB 200826A observed bands, $r^\prime$, $i^\prime$, and $H$, 
accounting for filter band differences and cosmological k-corrections.

At the time of the last MODS $r^\prime$ observation (78.9 d after the trigger) any SN 1998bw-like GRB-SN would have had $r^\prime \sim 27$ mag, 
more than 2 mag fainter than its peak magnitude ($r^\prime \sim24.7$ mag) at $20-40$ d, when the first observations had been obtained. 
Therefore, it is acceptable to consider the late $r^\prime$ MODS images free from any contamination by SN light and they can be used as templates for image subtraction. They can also be used to measure the flux of the host galaxy ($\sim4$ mag brighter in $r^\prime$ band) instead of the later LBC images, which have a lower signal-to-noise ratio. These LBC images were instead used as reference for the image subtraction of the $r^\prime$ imaging at $\sim50$~d.
Similarly, in the $H$ band\footnote{To derive the $H$-band flux of SN 1998bw at the redshift of GRB 200826A  
we had to extrapolate the rest-frame $I_C$ band data of SN 1998bw \citep[e.g.,][]{Zeh2004ApJ,Klose2019a}.}
any SN1998bw-like GRB-SN contribution at 160 d ($H\sim26$ mag at the time of our last LBT/LUCI observation) 
would have been $\sim2$ mag fainter than its peak magnitude ($H\sim24.3$ mag).
Differences of $\sim2$ mag translate to a contribution from the SN to the late-time image of $\sim0.2$ mag.
This contribution
does not impact the analysis after the image-subtraction, but is included in the photometric errors.
On the contrary, any SN1998bw-like GRB-SN would fade less than 1 mag between 97 and 160 d, thus the last $H-$band observation (97 d) may also be affected by over-subtraction, and hence a photometric value for the transient that is too faint at 97 d, though the SN was likely too faint to contribute substantially at this point. For completeness we report the resulting upper limit.

For the $i^\prime$ band 
 we note that
at the time of the final Gemini observation (at 75 d, see \citealt{Ahumada2020GCNcorrection}) any SN1998bw-like GRB-SN would have had $i^\prime\sim25.5$ mag, just $\sim1.5-2$ mag fainter than the peak magnitude at $\sim30$ d ($\sim24$ mag) which could be acceptable. Instead, this difference is certainly less in the case of the Gemini observation at 46 d (see Table \ref{tab:photall} and Fig.~\ref{fig:snlc}).
Therefore, we caution that using the value at 75 d as pure host can produce an over-subtraction, in particular at 46 d. Therefore, in the image subtraction of the $i^\prime$band we use the LBC imaging obtained at 287 d.

 Before applying image subtraction,
input and reference images were aligned using the WCS\-REMAP package\footnote{http://tdc-www.harvard.edu/wcstools}. Image subtraction was then performed using a routine based on \texttt{HOTPANTS}\footnote{https://github.com/acbecker/hotpants} \citep{Becker2015a}. This algorithm matches the PSF and count flux of both input images. The PSF is modeled via Gaussian functions in sub-regions of the original image. The software outputs a noise map of the resulting difference image, which is used to derive the uncertainties in the measured fluxes. The routine creates a grid of input Gaussian widths and sub-region sizes and searches for the solution with the lowest noise.

In the case of the $H$-band imaging, the average FWHM during the first epoch was 0\farcs13 as measured from the AO reference star in the LUCI-1 FOV\footnote{The PSF is best modeled with a Moffat profile.} and was constant for nearly the entire observation.  The small amount of data with larger FWHMs were not used for analysis. A total of 1180 s of exposure time with FWHM $\sim$ 0\farcs13 was used for analysis. For the other epochs the FWHMs ranged from 0\farcs08 to $<0\farcs15$. Therefore, only the exposures with a FWHM of the AO reference star $<0\farcs13$ (to match the first epoch) were selected.
This yielded a combined image with a total integration time of 2400 s and 2700 s for the second and last epochs, respectively. In particular, the AO reference star in the last epoch has a FWHM $\sim0\farcs10$, and this was used as the template for image subtraction. 
In contrast to the optical images, only two stars (the AO reference star and another star close to it) were useful to map the PSF and flux. Unfortunately, this impacts the optimal use of HOTPANTS to scale the template image to the flux of the science image. Instead, a better result was obtained using the {\tt PSF-MATCH} task under IRAF and scaling the images to have the same flux for the two stars. The image subtraction of the two $H$-band images shows a transient located at RA (J2000) = 00$^h$27$^m$ 08$\fs$55, Dec. (J2000) = +34$^{\circ}$01\arcmin38\farcs2 with an uncertainty of 0\farcs15 and calibrated against the MODS images. This is consistent with the position reported by \cite{Ahumada2020GCNdisco}.

 All image-subtracted data have been analyzed using aperture photometry
as detailed in \S \ref{sec:opt}.
 Table~\ref{tab:photall} provides a summary of all the photometry of the transient and the host galaxy including values collected from the literature.

Apparent magnitudes were corrected for Galactic extinction using the \citet{Cardelli1989} interstellar extinction curve, a total-to-selective extinction of $R_{\rm V}$ = 3.1, and a reddening along the line of sight of $E(B-V)$ = 0.058 mag \citep{SchlaflyFinkbeiner2011a}.

\begin{figure}
\centering
\includegraphics[width=0.5\textwidth,angle=0]{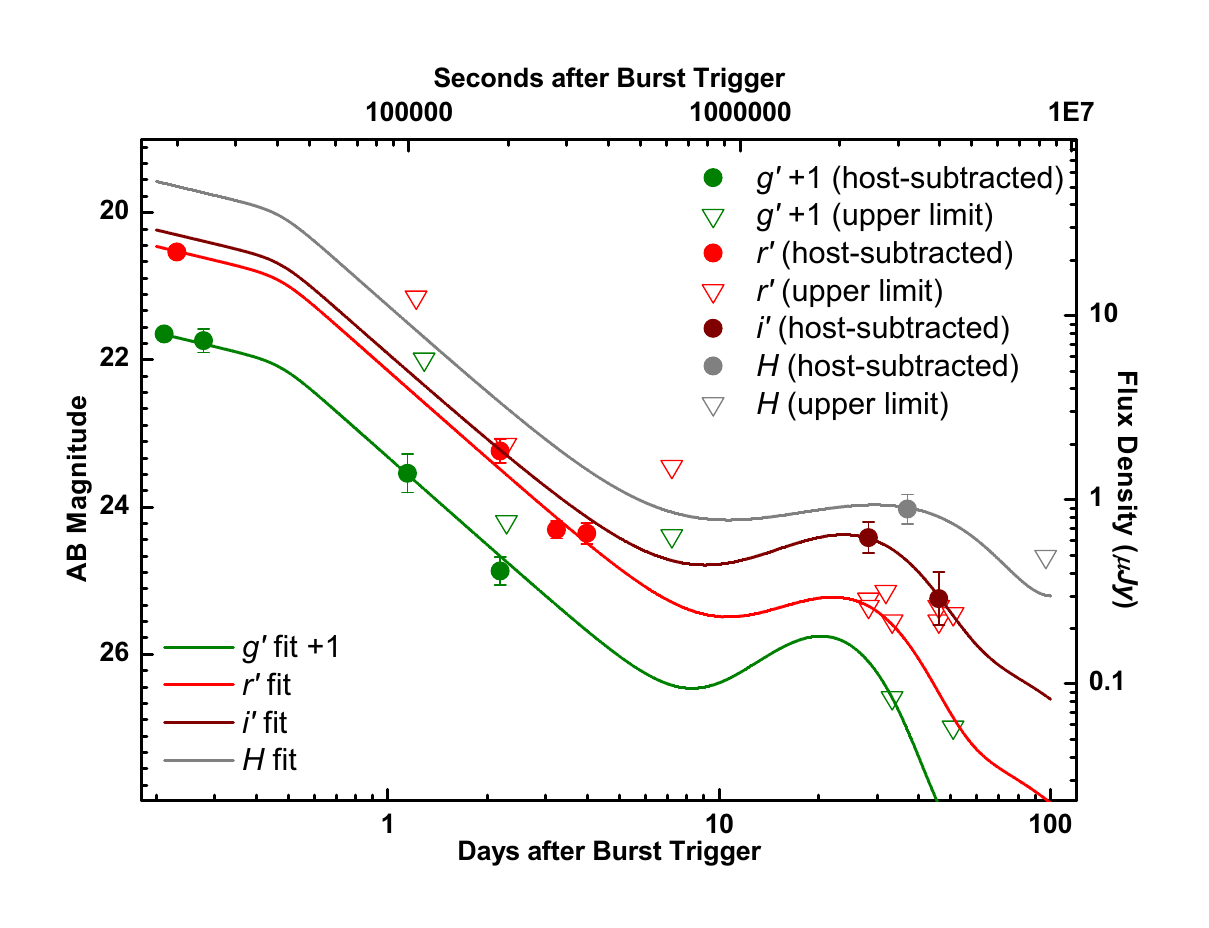}  
\caption{Light curves of the afterglow and SN component of GRB 200826A in the $g^\prime r^\prime i^\prime H$ bands, after subtraction of the host-galaxy contribution. The data have been modeled with a smoothly broken power-law for the afterglow, and with a SN 1998bw template redshifted to $z=0.7486$ for the SN phase. 
In the $g^\prime$ and $r^\prime$ bands, we forced the individual SN light curve components to go through the most significant deep upper limit in each band to obtain upper limits  on the potential SN luminosity. Note the $g^\prime$ band is offset by 1 mag for reasons of clarity. 
See Tab.~\ref{tab:photall}, \S \ref{sec:ag} and \S \ref{sec:sn} for more information.
}
\label{fig:aglc}%
\end{figure}

\subsection{Modelling of the light curve and SED}\label{sec:ag}

Although the $g^\prime\,r^\prime$ light curves are quite sparse, they can be modeled with a single power-law with $\alpha = 0.94\pm0.05$. However, this model implies that the late afterglow is brighter than the deep $r^\prime$ upper limits obtained at $\sim30$ d after subtraction of the host-galaxy contribution. Therefore, the data were modeled with a smoothly broken power-law \citep{Beuermann1999a} (Fig. \ref{fig:aglc}).
The last detections in the $g^\prime$ and $r^\prime$ bands are at $\approx2.2$ and $\approx4$ d (observer-frame).
Using SN 1998bw as a SN template (see below 
for more details), the SN itself would have been $g^\prime=29.8$ mag and $r^\prime=26.9$ mag  at $2.2$ d and $4$ d, respectively.  Therefore, it can safely be assumed to not contribute significantly to the optical transient and can be ignored in the afterglow fitting.
This yielded decay slopes $\alpha_1 = 0.43\pm0.55$, $\alpha_2=1.59\pm0.20$, with a break time $t_b = 0.47\pm0.24$ d, 
break smoothness $n=10$ fixed.
The post-break decay slope easily accommodates the magnitude upper limits at later times, with values beyond $r^\prime\sim27.8$ mag  and $g^\prime\sim28.3$ mag at $\sim30$ d. Therefore, any late emission detected via subtraction is not from the afterglow. 

To investigate whether the optical data are affected by local dust extinction along the line of sight, a spectral energy distribution
(SED) was modeled spanning the optical to the X-ray bands. All \textit{Swift} XRT data available were used, which have a mean arrival time of 65840 seconds (0.762 d), along with the simultaneous $g^\prime$ and $r^\prime$ band obtained from the broken power-law fit (see above) which is free from host contamination.
The redshift was fixed to match that of the host galaxy and the Galactic hydrogen column density was set to $N_H=6.02\times10^{20}\, \textnormal{cm}^{-2}$ \citep{Willingale2013a}. 
The optical-to-X-ray SED is well-fit  ($\chi^2/d.o.f.=33/47$) by a power-law with spectral index $\beta=0.79\pm0.03$ without need for additional dust 
extinction and gas absorption to improve the fit. The negligible dust extinction is well-supported by the optical-to-X-ray spectral index $\beta_{OX}>0.5$ \citep{Jakobsson2004a,Rossi2012a}. 
The optical-to-X-ray spectral slope was then used to extrapolate the late-time brightness in the $i^\prime$ and $H$ bands.
For the first $i^\prime$ detection at 28 d, the brightness of the projected afterglow was 27.6 mag, 
3.2 mag fainter than the detection. 
The $H$-band afterglow is  $H\sim27.4$ mag at 37 d,  
3.4 mag fainter than the detection.

To consider also the contribution of an emerging SN component, we have followed the work of  \cite{Zeh2004ApJ} and \cite{Klose2019a}, and modeled the afterglow and SN component in a joint
$g^\prime r^\prime i^\prime H$ fit 
(rest-frame far-$UV$ and $U B z^\prime$ bands). The analytical light curve of SN 1998bw and the $k,s$ parametrization were used (see \S\ref{sec:ag}). 
The parameter $k$ is the luminosity factor and describes the observed luminosity ratio between the GRB-SN at peak time and the SN template in the considered band. The parameter $s$ is a stretch factor with respect to the used template\footnote{As in \S\ref{sec:sub}, the cosmological effects such as k-correction were considered when moving the SN template to the redshift of our GRB. For further details on the method see \cite{Zeh2004ApJ}, \cite{Klose2019a}, and references therein.}.
The observer-frame $g^\prime$ band (rest-frame far-UV) is not covered by observations in the case of SN 1998bw (in contrast to the observer-frame $r^\prime$ band, which is the rest-frame $U$ band), so we assume that the flux density of the
SN falls as $\nu^{-3}$
to extrapolate and derive the corresponding far-UV SN 1998bw template \citep[e.g.,][]{Klose2019a}.  
Note that the $g^\prime$ and $r^\prime$ data points in Figure \ref{fig:aglc} after $2.2$ and $4$ d (observer-frame) are upper limits only.
In order to constrain the flux density of a possible SN contribution in these bands, we considered these limits as proxy detections in the modelling.
We also remind the reader that the $i^\prime$- and $H$-band afterglow light curves have been spectrally extrapolated (see above) because there are no observations in these bands during the afterglow phase.
The sparse data {in $i^\prime H$} preclude determining the precise temporal evolution of the late emission, but do not preclude determining more information on the luminosity. The stretch factor was fixed at $s=1$ for all bands and the luminosity factor $k$ was allowed to vary freely and individually in each band. 
The results of this joint fit are presented and discussed in  \S\ref{sec:sn}.

\section{Results  \label{sec:res}}

A summary of the photometry of the transient resulting from the host-subtraction analyses is presented in Table \ref{tab:photall}. NIR emission (rest-frame $\sim0.93\,\mu$m) is clearly detected 37 d (21.22 d rest frame) after the burst (see also Fig. \ref{fig:ao}).
Thanks to our late reference images obtained with the LBCs, which are free
from SN light, we also detect the bump in the $i^\prime$-band at 28.3 and 46.1 d (14.9 and 26.3 d rest frame). 
\cite{Ahumada2021a} report a bump that is $\sim1$ magnitude fainter and they do not find the transient in the second epoch. This last detection allows us to  better  constrain the late evolution of the transient.
 In the $r^\prime$-band, the presence of the bump could only be constrained beyond the $3\sigma$ magnitude limits of $r^\prime>25.5$ mag, $r^\prime>25.3$ mag, and $r^\prime>25.7$ mag in the first Gemini, TNG and LBT epochs, respectively (the results presented here supersede the preliminary results of \citealt{Rossi2020GCN}).

\begin{figure}
\centering
\includegraphics[width=\columnwidth,angle=0]{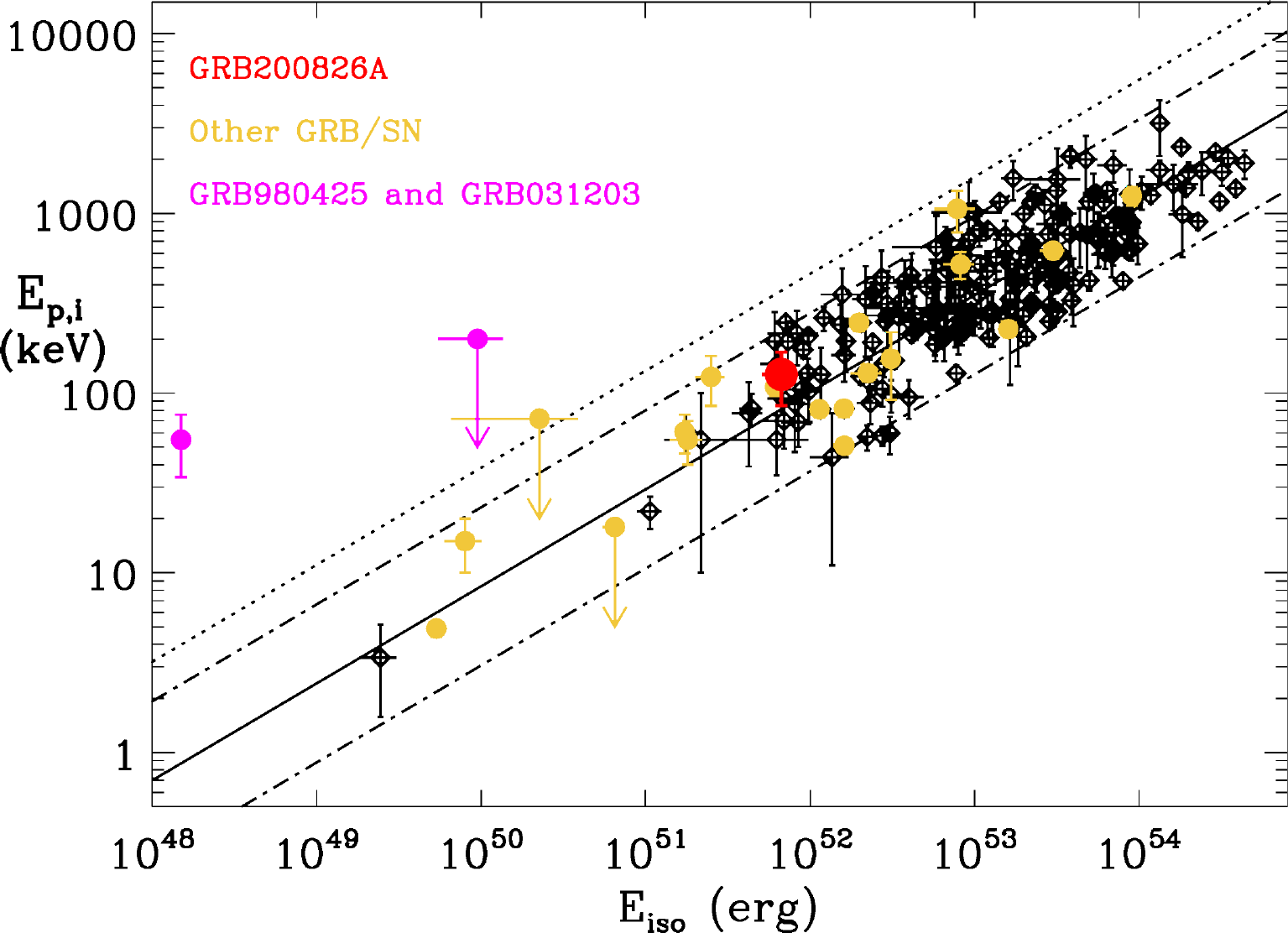}
\caption{GRB 200826A (red) in the $E_{\rm p,i}-E_{\rm iso}$ plane. GRBs with an associated SN are highlighted in green, outliers in blue (see \S\ref{sec:grb}). GRB data is from \cite{Amati2019a}. 
}
\label{fig:epieiso}%
\end{figure}

\begin{figure}
\centering
\includegraphics[width=\columnwidth,angle=0]{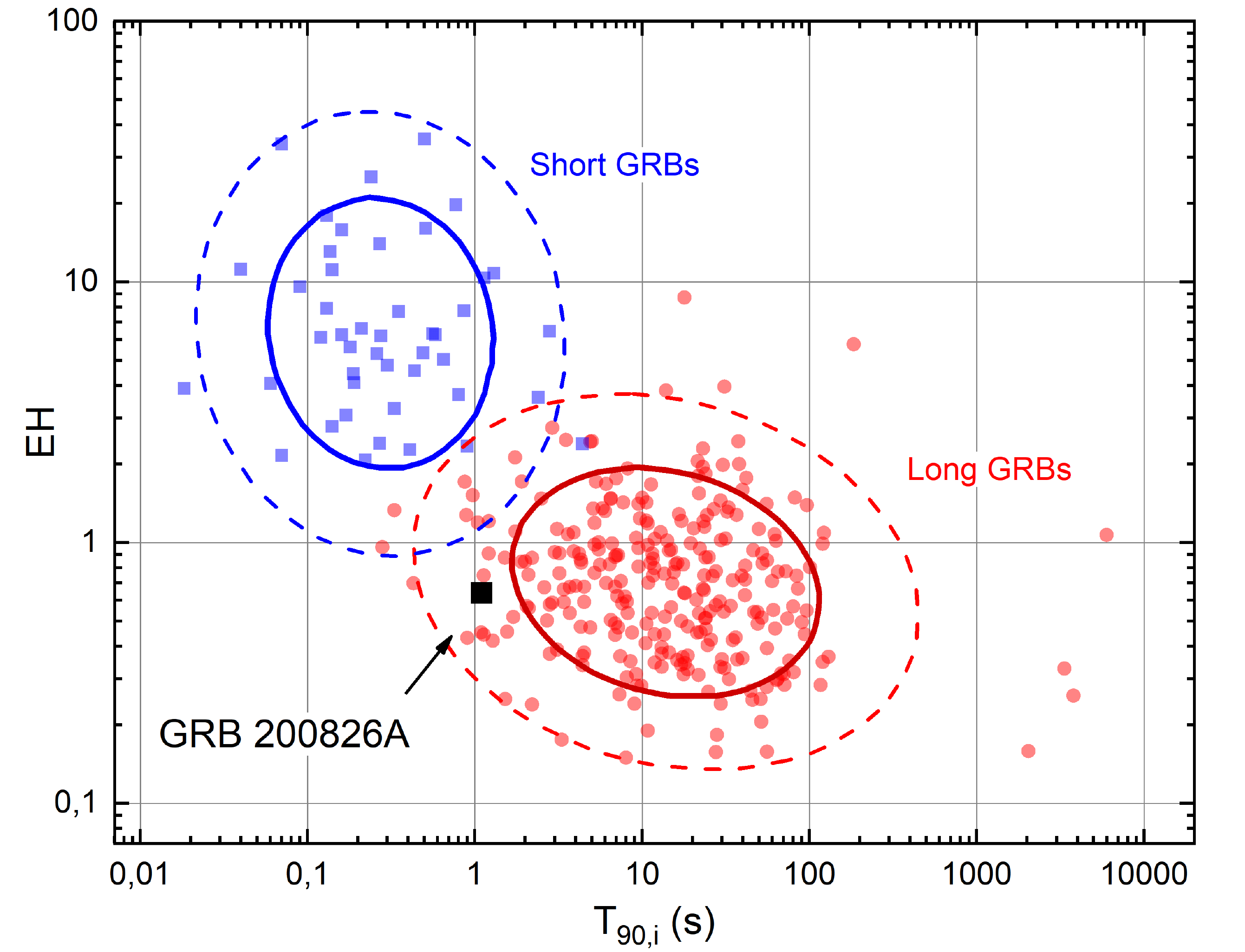}
\caption{ The $T_{\rm 90,i}$ -- $EH$ diagram for SGRBs (blue squares) and LGRBs (red circles) with corresponding cluster analysis
results. 68\% and 95\% confidence regions are shown by bold solid and thin dashed curves. Data from \citet{MinaevPozanenko2020a,MinaevPozanenko2020b}.}
\label{fig:ehd}
\end{figure}

\subsection{GRB 200826A} \label{sec:grb}

The duration of GRB 200826A, as reported by the different detectors and in different energy bands, is $1-2$ s, placing it right at the divide between SGRBs and LGRBs in the duration distribution commonly used to distinguish between these two classes \citep[but see][]{deUgartePostigo2011a,Bromberg2012a}. 

There are other events that lie in between the two populations, but none has an observed $T_{90}$ shorter than 2~s and an associated SN which clearly points to a collapsar origin. For example, GRB 090426 ($z=2.609$) has a duration shorter than 2~s but also has
characteristics typical of LGRBs including softer spectra, a bright afterglow and a blue and low-mass host  (\citealt{Antonelli2009a,Thone2011MNRAS,Nicuesa2011a,Nicuesa2012a}, Kann et al. 2022, in prep.). 
It could be argued that one should consider the intrinsic duration in the rest frame ($T_{\rm 90,i}$). Doing so, the duration of GRB 200826A would be between 0.39 (Konus-\textit{Wind}) and 0.57 (GBM) s, i.e., well below 2~s.  The only other GRB with an associated SN that could be classified as short because of its intrinsic duration ($ T_{\rm 90,i} =2.4$ s, $z=0.86$) is GRB 040924 \citep{Wiersema2008a,Soderberg2006ApJ,Huang2005a}.
Many other cases exist at high redshift like GRB 080913 \citep{Greiner2009a,Zhang2009ApJ}, but the redshift is too high to search for the associated SN (at least with the current facilities). 

\cite{Bromberg2012a} showed that the 2~s duration commonly used to separate collapsars and non-collapsars is inconsistent with the duration distributions of \emph{Swift} and \emph{Fermi} GRBs and only holds for old \emph{CGRO}/BATSE GRBs. However, the analysis of the Konus-\emph{Wind} data \citep{Svinkin2020GCNenergy} showed that GRB 200826A has a duration typical of SGRBs. Thus, the short duration of GRB 200826A is instrument-independent.  In this case T$_{90}$ alone is insufficient to determine which population GRB 200826A is best associated with. 
Another feature typical of LGRBs is a non-negligible spectral lag \citep{Norris1996a}.  
   This parameter has been calculated and the details are provided in Appendix~\ref{sec:lag}. We obtain a spectral lag of $96\pm38$~ms, which lies in between the values found by \citet{Ahumada2021a} and \citet{Zhang2021a}. In agreement with their analysis, we conclude that the measured spectral lag is more typical of LGRBs.

It is only when both the spectral properties of GRB 200826A and its location in the hardness-duration plane are considered that it appears to be more similar to ``long'' GRBs.  However, if this is indeed the shortest ``long'' GRB associated with a SN, then additional evidence is needed to demonstrate that it does not belong to the SGRB class.  Such a result would challenge the standard paradigm for SGRB and LGRB progenitors.

In this respect, a very useful tool is the location of a GRB in the 
$E_{\rm p,i}-E_{\rm iso}$ plane\footnote{$E_{\rm p,i}=E_{\rm p,obs}(1+z)$ is the  rest-frame photon  energy at which the $\nu$F$_\nu$ spectrum peaks, and $E_{\rm iso}$ is the isotropic-equivalent radiated energy as measured in a ``bolometric'' band, usually 1 keV -- 10 MeV in the rest frame.}, where
LGRBs follow a strong correlation known as the ``Amati Relation'' \citep{Amati2002a,Amati2006a}.  In comparison, SGRBs populate a different region in this plane \citep[e.g.,][]{Davanzo2014a, MinaevPozanenko2020a}. Therefore, the fluence and spectral parameters reported by the Konus-\emph{Wind} and \emph{Fermi}/GBM teams were used
 \citep{Ridnaia2020GCNkw,Svinkin2020GCNenergy,Mangan2020GCNgbm} to compute the values of $E_{\rm p,i}$ and $E_{\rm iso}$, and their uncertainties, for GRB 200826A. As can be seen in Fig. \ref{fig:epieiso}, this event is fully consistent with the $E_{\rm p,i}-E_{\rm iso}$ relation followed by LGRBs.  This further supports the hypothesis that it is actually a ``long'' GRB with a duration in the short tail of the distribution. 

Recently, the energy-hardness ($EH$) parameter – a combination of Ep,i and Eiso ($EH$ = $ E_{\rm p,i} E_{\rm iso}^{-0.4} $) –
was introduced in \cite{MinaevPozanenko2020a}.
 Although GRB 200826A is placed inside the 95\% cluster region of long bursts in the $ T_{\rm 90,i} $ -- $ EH $ diagram (Fig.~\ref{fig:ehd}), it is evident that this does not permit an unambiguous classification of a burst being close to the cluster region intersection
(see also \citealt{Zhang2021a} for a similar analysis and conclusion). 
Also the position in the $E_{\rm p,i}-E_{\rm iso}$ plane may not be sufficient for classification. 
Indeed, interesting examples such as GRB 201221D\footnote{ With a rest-frame \textit{Fermi}/GBM $\rm T_{90}=0.068$ s \citep{Hamburg2020a}, and \textit{Swift}/BAT $\rm T_{90}=0.078\pm0.0196$ s \citep{Krimm2020a}.} \citep{Agui2021a} exist which follow the
$E_{\rm p,i}-E_{\rm iso}$ relation for LGRBs but are in the SGRB region in the  $ T_{\rm 90,i}$ -- $EH$ diagram.

\begin{figure}
\centering
\includegraphics[width=0.46\textwidth,angle=0]{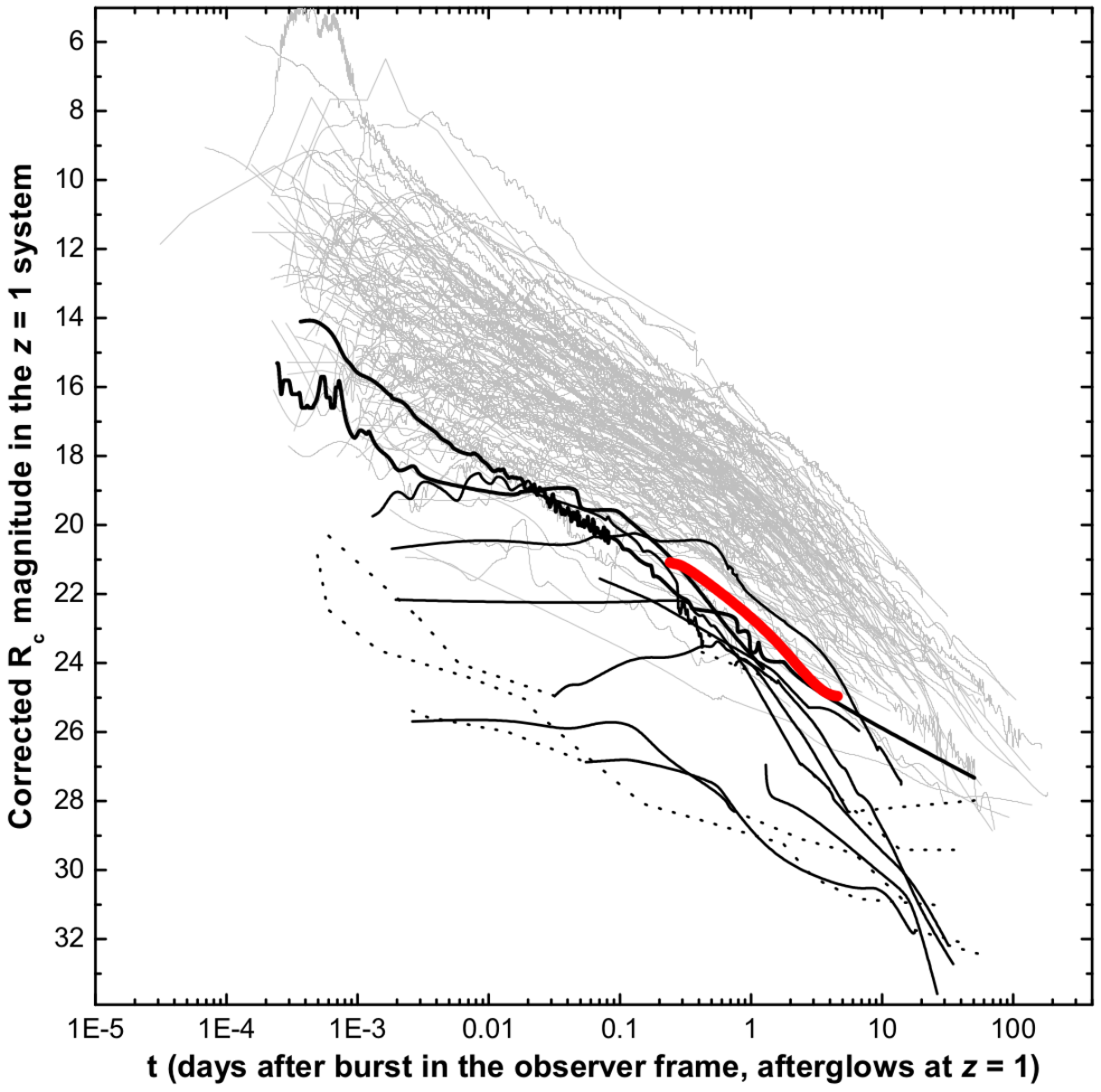}
\caption{The optical afterglow of GRB 200826A (red line) compared with the sample of extinction-corrected afterglows shifted to $z=1$ from \citet[][2022, in prep.]{Kann2010a,Kann2011a}. Hereby, time and magnitudes are given in the observer frame, but assuming all GRBs are at $z=1$ in a perfectly transparent universe.  
Light gray are LGRBs, thicker black lines SGRBs with secure redshifts (solid lines connect detections, dashed lines connect upper limits). All magnitudes are Vega magnitudes. The afterglow of GRB 200826A lies intermediately between those of secure LGRBs and secure SGRBs.
}
\label{fig:kann}%
\end{figure}

\begin{figure}
\centering
\includegraphics[width=0.46\textwidth,angle=0]{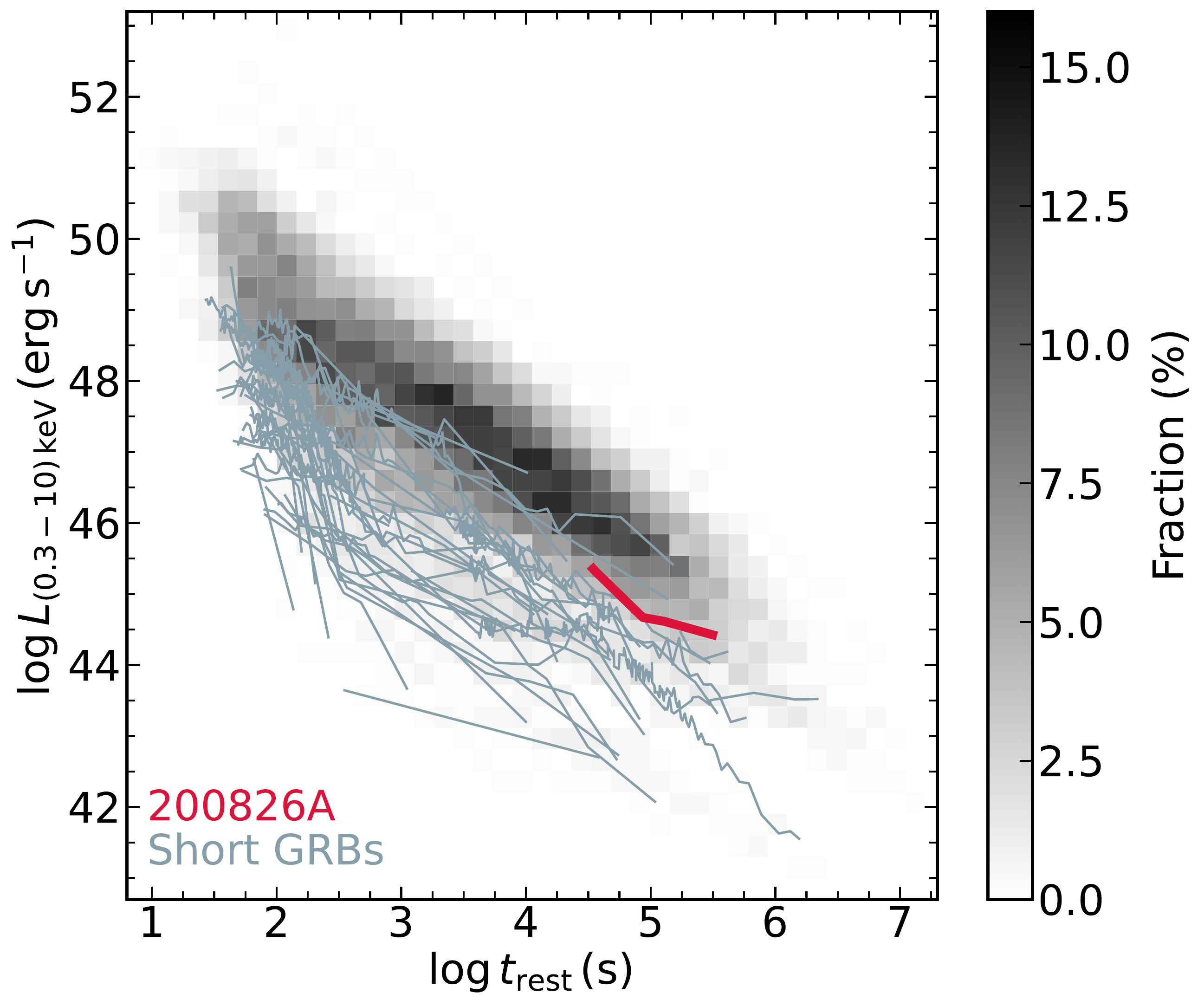}
\caption{ The X-ray afterglow of GRB 200826A (red line) in the context of the X-ray afterglows of 449 \textit{Swift} GRBs with known redshifts. This event could be followed-up by Swift/XRT only after 10ks, but the few data points obtained show that the luminosity of GRB 200826A is in between the long and short populations. 
The grayscale on the right is used to convert a given luminosity and time into a fraction of bursts.
}
\label{fig:schulze}%
\end{figure}


\subsection{The afterglow of GRB 200826A}

A well known-result is that most SGRB afterglows are much fainter than LGRB afterglows \citep{Gehrels2008ApJ,Nysewander2009ApJ,Kann2010a,Kann2011a}. Following the method of \cite{Kann2006ApJ}, using a spectral slope of $\beta=0.79$ and no line-of-sight extinction, in Fig.~\ref{fig:kann}
the optical afterglow of GRB 200826A was compared with a large sample of SGRB and LGRB optical afterglows (taken from \citealt{Kann2006ApJ,Kann2010a,Kann2011a,Agui2021a}, Kann et al. 2022, in prep.). 
 Similarly, in Fig.~\ref{fig:schulze} the X-ray emission of GRB 200826A was compared with that of $\sim500$ GRB afterglows (of which 38 are SGRBs) detected in at least two epochs and with known redshift. The X-ray light curves where obtained from the \textit{Swift} Burst analyzer\footnote{
\href{https://www.swift.ac.uk/burst_analyzer/}{https://www.swift.ac.uk/burst\_analyzer/}} \citep{Evans2010a} and put in the same frame using the method described in \cite{Schulze2014a}.
The afterglow of GRB 200826A is among the faintest afterglows of LGRBs but also among the brightest ones of SGRBs , both in the X-ray and optical bands. Thus, the afterglow luminosity is not a good method for discriminating between, nor understanding the nature of this burst.

\begin{figure*}
\centering
\includegraphics[width=0.86\textwidth,angle=0]{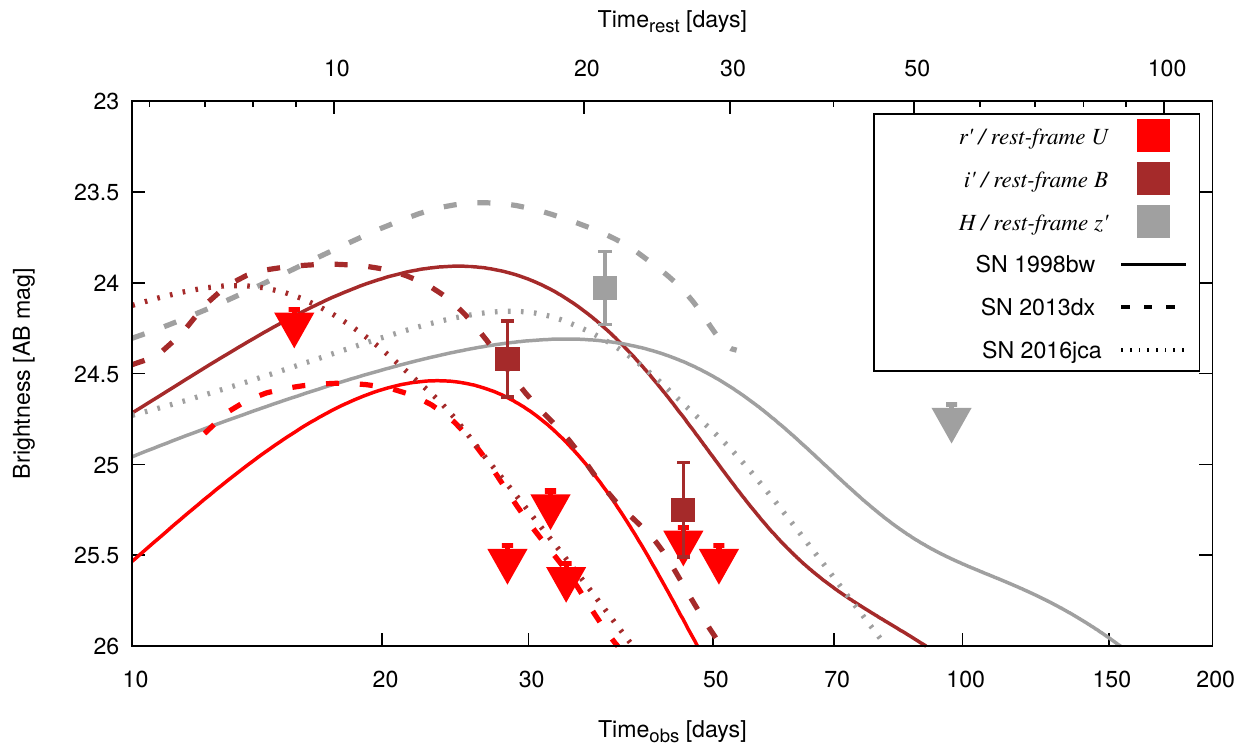}
\caption{Light curve in $i^\prime$ and $H$ bands after removing the host component via image subtraction (see Table \ref{tab:photall}). Only data after 10 d (observer frame) are shown. The data are corrected for Galactic extinction (see \S~\ref{sec:sub}). The SN light curve templates obtained from SN 1998bw, 2013dx, and SN 2016jca are shown for comparison and to stress the large range of variability and color of GRB-SNe (See \S\ref{sec:sn}). Downward-pointing triangles are upper limits.
}
\label{fig:snlc}%
\end{figure*}


\subsection{The late bump as supernova emission \label{sec:sn}}

Although the sparse data precludes determining the evolution of the late  bump seen in the transient light curve, a comparison can be made with other GRB-SNe to look for similarities. 
This comparison includes:  
GRB 980425/SN 1998bw \citep{Galama1998a}  
GRB 060218/SN2006aj \citep{Ferrero2006AA};
GRB 100316D/SN1010bh \citep{Olivares2012a}; 
GRB 130702A/SN 2013dx  \citep{Delia2015AA,Toy2016ApJ,Volnova2017MNRAS,Mazzali2021a}; and GRB 161219B/SN 2016jca  \citep{Cano2017AA,Ashall2019MNRAS}. 
In Fig.~\ref{fig:snlc} we show only those GRB-SNe with comparable peak luminosities and colors, including SN 1998bw which 
has been used to also fit extreme events such as SN 2011kl \citep{Greiner2015Nat,Kann2019AA}  and the SN bump of GRB 140506A \citep{Kann2021a}. 
Due to the non-negligible redshift of GRB 200826A, the observed $r^\prime$-, $i^\prime$- and $H$-band data sets correspond approximately to the rest-frame $U$, $B$, and $z^\prime$-bands, respectively. 
In the following comparison, we will refer to these rest-frame bands.

 The SN associated with GRB 200826A appears fainter than SN 1998bw in the rest-frame $B$-band than SN 1998bw,   but is of comparable luminosity or slightly brighter ($\lesssim0.5$ mag) than SN 1998bw in rest-frame $z$-band. The most striking feature is the suppression in the rest-frame $U$-band, at least $1$~mag compared to SN 1998bw at 28 d.
 This prompted a further investigation into GRB-SNe with redder colors, and produced a possible agreement with GRB 130702A/SN 2013dx.
 This GRB-SN (like GRB 161219B/SN 2016jca and possibly GRB 050525A/SN 2005nc, \citealt{DellaValle2006a}) is characterized by a rise-to-maximum time $\Delta t <\sim15$ d in the respective rest frame \citep[e.g.,][]{Belkin2020a}, which is a parameter related to properties of the progenitor stars \citep[e.g.,][]{Gonzalez-Gaitan2015a}. 

In section \ref{sec:ag} we have modelled the late light curve considering the analytical light curve of SN 1998bw and the $k,s$ parametrization following the work of  \cite{Zeh2004ApJ} and \cite{Klose2019a}, which allows us to place quantitative constraints on the above analysis.
We remind the reader that there are no detections in $g^\prime r^\prime$ after $2.2$ and $4$ d (observer-frame), thus with our modelling we can only place deep upper limits on the flux density of the SN contribution in these bands (see Fig.~\ref{fig:aglc}).
In terms of luminosity, a potential SN contribution in the rest frame far-UV (observed $g^\prime$ band) is found as $k_{g^\prime}<1.5$, an unremarkable limit to its luminosity. Therefore, although the spectral slope of the extrapolation mentioned above is insecure, this does not have any significant impact on our modeling.
However, there is clear UV suppression in the rest frame $U$-band,
we find $k_{r^\prime} <0.43$, implying that the SN associated with GRB 200826A is less than half as luminous as SN 1998bw in the same rest-frame band.
 In the rest frame $B$-band the suppression is also strong compared to SN 1998bw with $k_{i^\prime}=0.56\pm0.10$.
The rest-frame $z^\prime$-band detection (observed $H$ band)
is, however, marginally brighter than SN 1998bw  ($k_H=1.18\pm0.22$). 
Moreover, it lies close to the SN peak for $s=1$, and thus a significantly faster or slower SN would be even brighter. 
When compared with the list of $k,s$ values of \cite{Klose2019a} and \cite{Cano2017a} the $k$ factor ranges between 0.5 and 2. This provides plausible support for a SN.

\begin{figure}
\includegraphics[width=\columnwidth,angle=0]{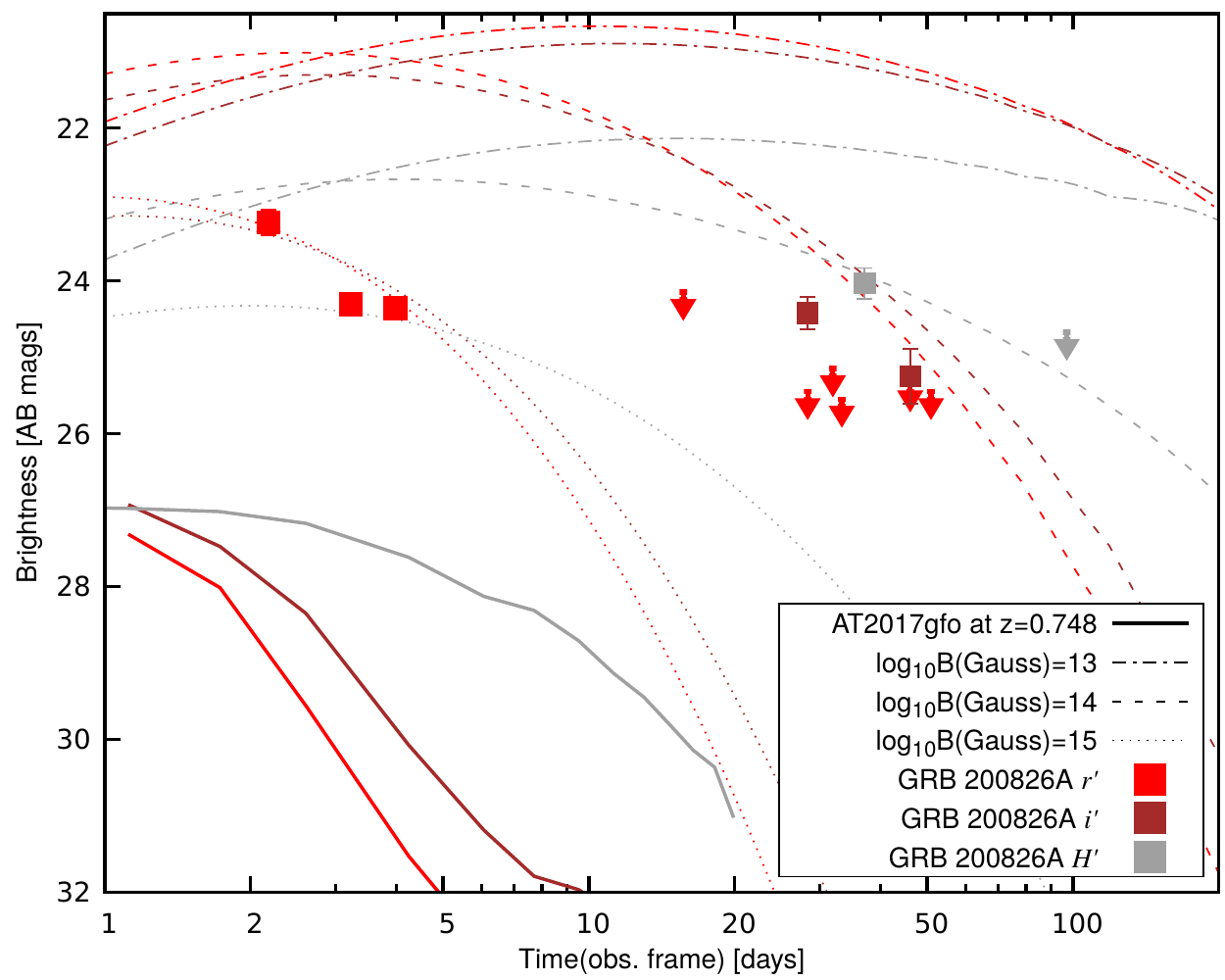}
\caption{Light curves of GRB 200826A after host subtraction compared with the light curves of AT2017gfo (solid lines) from \cite{Rossi2020a} and with theoretical KN light curves boosted by a magnetar from \cite{Perego2017a} with three different magnetic field strengths (dotted, dashed, dot-dashed lines; \S\ref{sec:kn}). Downward-pointing triangles are upper limits.
}
\label{fig:kn}
\end{figure}

\subsection{Possible Kilonova emission \label{sec:kn}}

In Fig.~\ref{fig:kn}, the late-time photometry of the transient is compared with the light curve of AT2017gfo, the KN associated with the binary NS merger GW 170817 and the SGRB 170817A \citep{Abbott2017a}.
The KN light curves are obtained from observed spectra shifted and scaled to the same redshift as GRB 200826A, following the method described in \citet{Rossi2020a}.
 The comparison shows that the luminosity of the transient was at all times at least 3 mag brighter than the peak luminosity of AT2017gfo. This is not extraordinary, because KNe $\gtrsim10$ times more luminous in the optical have been claimed \citep[e.g.,][]{Gao2017a,Fong2021a}.
However, what makes this event very different from all KNe observed so far is its slower temporal evolution: 
assuming that the rest-frame $B$-band detection ($i^\prime$ band in observer-frame) detection is close to maximum light, i.e. $\sim16$ d in the rest frame, then the peak would be more than $\sim15$ d later than that of AT2017gfo. 
The analysis below investigates whether such behavior is theoretically possible.

Modeling of KNe powered by radioactive decay of $r$-process nuclei showed that KNe produced by nearly equal-mass binary NS mergers are expected to have peak times and magnitudes similar (within a few d and one or two magnitudes, respectively) to the ones observed in AT2017gfo, especially in the case of face-on mergers characterized by lanthanide-free high-latitude ejecta \citep[e.g.,][]{Radice2018a,Kawaguchi2020}. Viewing angle and composition effects possibly decrease the KN luminosity compared with face-on configurations \citep{Korobkin:2021}. 
In the case of very unequal-mass binary NS or black-hole NS mergers, the peak evolution can be slower, especially in the rest-frame NIR bands, but with similar peak magnitudes \citep[e.g.,][]{Barbieri2019,Bernuzzi2020,Kawaguchi2020,Zhu2020a}. 
These theoretical results seem to also disfavour the interpretation of the observed transient as a KN powered by radioactive decay.

However, it is known 
from several studies 
\citep[e.g.,][]{Gompertz2018a,Ascenzi2018a,Rossi2020a,Rastinejad2021a}, 
that the distribution of luminosities covers a large range, 
and some KNe can be ten times more luminous than AT2017gfo,
such as in the cases of those associated with GRBs 060614 and 050724 \citep{Gao2017a}.  
 Even more extreme cases that are $\sim100$ times more luminous have been claimed in association with GRB 070714B \citep{Gao2017a} and more recently GRB 200522A \citep{Fong2021a,Oconnor2021a}, although the evidence is not very strong.
To achieve this large blue luminosity, the KN can be boosted by energy deposition from a magnetar remnant. Indeed,
long-lived merger remnants could have a significant impact on KN light curves. While larger amounts of ejecta, expelled due to angular momentum excess, can increase the peak magnitudes, times, and widths only marginally \citep[e.g.,][]{Radice2018b}, the presence of a fast-rotating magnetar (in addition to being a possible GRB central engine) can boost the KN luminosity via 
spin-down luminosity \citep{Yu2013a,MetzgerPiro2014a,Siegel2016a}. 

To explore this scenario, the KN model presented in \citet{Perego2017a} was used and extended to consider spin-down energy injection, as well as sources at cosmological distances in which the fluxes, the (thermal) spectra, and the observer time are corrected for the cosmological redshift. The computed KN light curves are shown in Fig. \ref{fig:kn}.
For simplicity, a spherically symmetric ejecta model is considered for which $0.02\,M_\odot$ of matter expands homologously at $0.2c$. Given that the adopted model foresees the presence of a long-lived magnetar, an effective gray opacity $\kappa=1$ cm$^2$ g$^{-1}$, typical of lanthanide-poor ejecta, was assumed.
$P_0=0.7$ ms was fixed as a typical initial rotation period \citep{Radice2018b}, while the strength of the magnetic field $B$ and the energy deposition efficiency for the spin-down luminosity $\epsilon_{\rm th}$ were varied to obtain a plausible agreement with the observations. 
A reasonable match for $B \sim 10^{14}$ G
and $\epsilon_{\rm th} = 0.03$ are shown in Fig.~\ref{fig:kn}. For the considered bands, larger magnetic field strengths ($B = 10^{15}$ G) result in lower, more rapidly declining peaks, and a plausible match with the data would require $\epsilon_{\rm th} \gtrsim 1$. Weaker field strengths ($B = 10^{13}$ G), on the other hand, produce more luminous, slower evolving peaks, which plausibly match with the data for $\epsilon \lesssim 0.001$. The ejecta properties also influence the light curves, but less significantly. 

The calculations suggest that the observed transient could be compatible with the late emission of a spin-down powered KN for plausible values of the magnetic field and thermalization efficiency.
Unfortunately, all of the magnetically-boosted KN models fail to explain the red color of the transient, and in particular the strong UV suppression observed in the observer-frame $r^\prime$ band, although modeling greatly simplifies the ejecta, the emission properties, and in particular its opacity. But most importantly, during the first 10 d the theoretical KN is comparable or brighter than the observed data without accounting for the afterglow. Therefore, the magnetically-boosted KN models are excluded.


\begin{figure}[t]
\centering
\includegraphics[width=\columnwidth,angle=0]{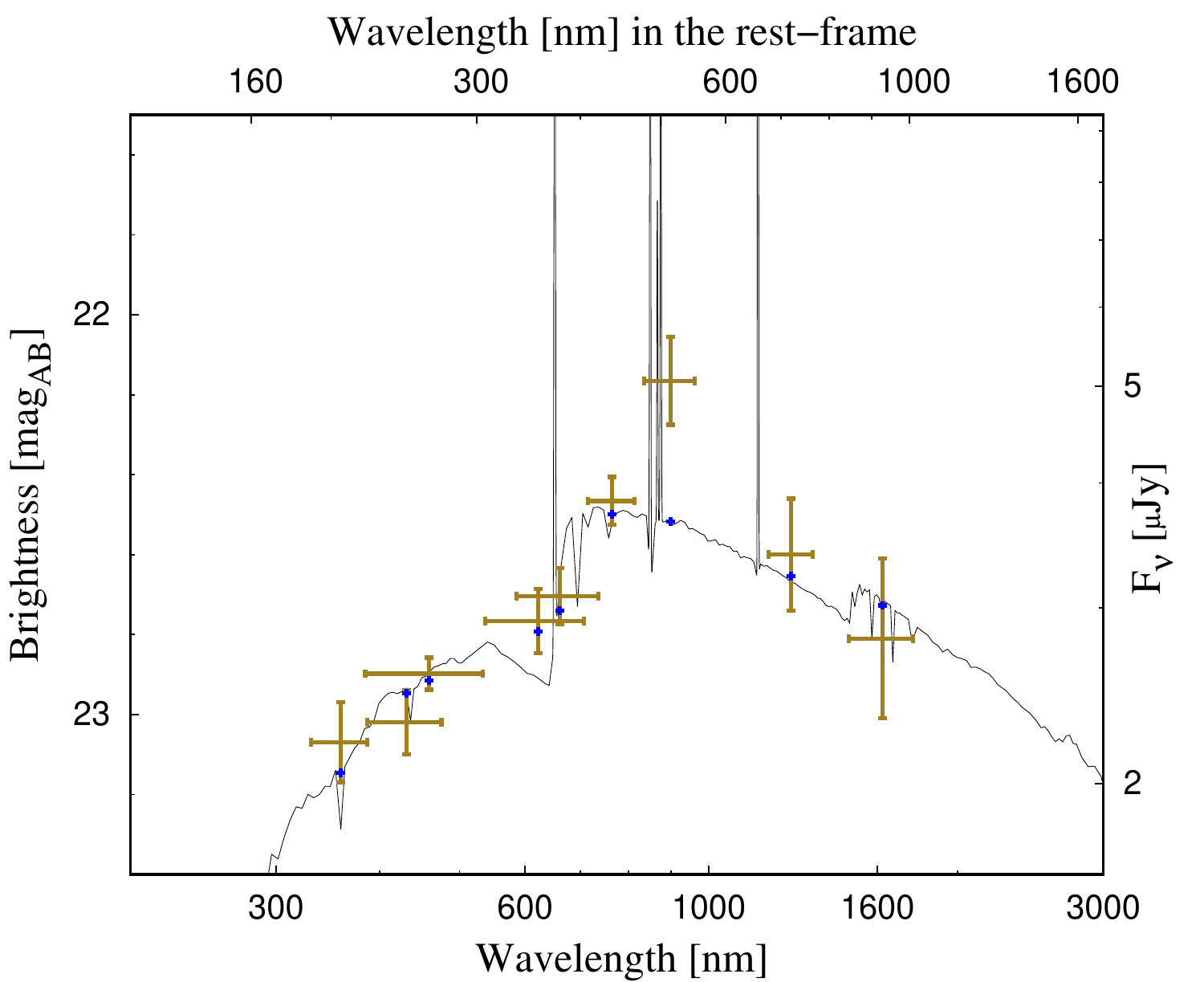}
\caption{\texttt{Le PHARE} modeling discussed in \S\ref{sec:host} of the SED obtained from the final LBT/LBC $U R_{C} i^\prime z^\prime$, MAO $B$, LBT/MODS $g^\prime r^\prime$,  LBT/LUCI $J$, and LBT/LUCI+SOUL $H$ band imaging (see Table \ref{tab:photall}). 
The blue points indicate where the photometry would be placed according to the best-fit model and without contribution from emission lines.
}
\label{fig:host}
\end{figure}


\begin{figure*}[h!t]
\centering
\includegraphics[width=0.55\textwidth,angle=90]{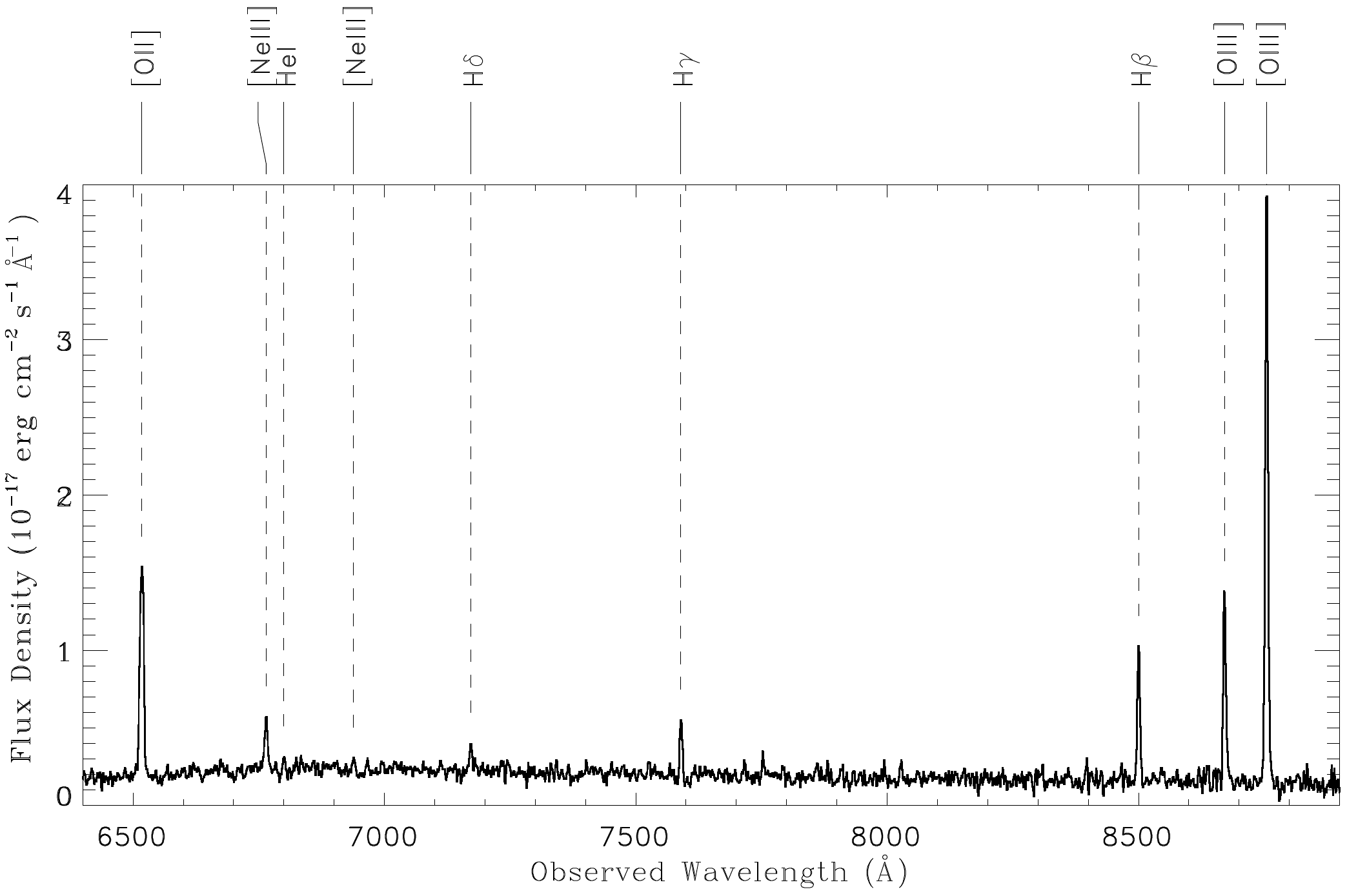}
\caption{LBT/MODS observed spectrum of the host galaxy of GRB~200826A
binned by 2.5~{\AA} for presentation purposes. The detected emission lines are marked.}
\label{fig:spectrum}
\end{figure*}

\subsection{The host galaxy} \label{sec:host}

\begin{table}
\centering
\caption{Emission lines and their measured fluxes corrected for Galactic extinction.}
\begin{tabular}{lc} 
\toprule  
Lines         &   flux       \\
              &[10$^{-17}$ erg/cm$^2$/s] \\
\midrule
$[${O}{II}] $\lambda$3727 & $12.6\pm0.4$    \\
$[$NeIII]           $\lambda$3869 & $2.1\pm0.4$  \\
HeI           $\lambda$3889 & $1.2\pm0.4$  \\
$[$NeIII]         $\lambda$3968 & $1.2\pm0.4$  \\
H$\delta$    $\lambda$4102 & $1.8\pm0.4$  \\
H$\gamma$    $\lambda$4340 & $2.8\pm0.4$  \\
H$\beta $    $\lambda$4861 & $6.4\pm0.6$  \\
$[$OIII] $\lambda$4959 & $8.0\pm0.3$  \\
$[$OIII] $\lambda$5007 & $26.5\pm0.9$  \\
\bottomrule
\end{tabular}
\begin{tablenotes}\footnotesize 
\item The FWHM of the lines is $\sim6.1$ \AA. The mean redshift of the emission lines is $z=0.748577\pm0.000065$. 
\end{tablenotes}
\label{tab:lines}
\end{table}

As first reported in \cite{Rothberg2020GCNlbt, 2020SPIE11447E..06R}, the continuum
of the host galaxy was clearly detected between
$0.38\,\mu\textnormal{m}-0.98\,\mu$m. A redshift of
$z=0.748577\pm0.000065$ was determined from the simultaneous detection of the following emission
lines: [\ion{O}{2}], [\ion{O}{3}] and [\ion{Ne}{3}] doublets, \ion{He}{1} $\lambda$3889,
and Balmer H$\delta$ $\lambda$4101, H$\gamma$ $\lambda$4340, and H$\beta$ $\lambda$4861 (Fig.~\ref{fig:spectrum}). Fluxes of the lines were measured using the
\texttt{slinefit} code \citep{Schreiber2018a} which also allows correction of the Balmer-line fluxes for the underlying stellar absorption. All the
measured fluxes are reported in Table \ref{tab:lines}. 
The absorption lines of the \ion{Mg}{2}$ \lambda\lambda$2796,2803 doublet were also detected at  a lower redshift ($z=0.7462\pm0.0002$) which may possibly imply that they are due to an intervening \ion{Mg}{2} absorber along the line of sight \citep{Vergani2009a,Christensen2017a}.
The rest-frame equivalent width for both lines in the doublet is $W_{rest} = 2.4\pm0.6$ \AA.

Using the derived redshift as a fixed input, the optical/NIR SED of the host galaxy was modeled (Table  \ref{tab:photall}) with the code \texttt{Le PHARE}
\citep{Arnouts1999a,Ilbert2006a}. 
 The best fit ($\chi^2/d.o.f.=2.2/9$) points to a galaxy dominated by a young population ($0.04_{-0.02}^{+0.08}$~Gyr) with global dust reddening E(B-V) $\sim0.2$ mag using the Small Magellanic Cloud (SMC) curves from \cite{Pei1992a} and low stellar mass $\log M_\ast/M_\odot = 8.6\pm 0.2$ (Fig.~\ref{fig:host}).
The $B$-band dust-corrected absolute magnitude is M$_B=-20.01$ mag 
and the star-formation rate is $\textnormal{SFR}= 13.0_{-6.0}^{+10.9}\,M_\odot/$yr. With this large uncertainty, this SFR is consistent with what is measured from the emission lines (see below).
The dust extinction is in agreement with the broad range of values found for LGRB hosts, also in comparison with the SFR \citep{Hunt2014a,Japeli2016a}. It is different from the afterglow-derived negligible extinction, indicating that the GRB sight-line might be crossing a relatively dust-free region (e.g., by being placed at the front side of the host galaxy).
 Respective to \cite{Ahumada2021a}, we find a mass that is a factor ten lower. 
The computed lower mass is most likely because the NIR observations were obtained sufficiently late to be free of any contributions from the transient.

Using the Balmer decrement, and assuming the theoretical values for the H$\gamma$/H$\beta$ and
H$\delta$/H$\beta$ ratios, in the absence of dust, for case B recombination at a gas temperature of 10$^4$ K \citep{Osterbrock1989}, and a Milky Way (MW) extinction law\footnote{Note that the difference between
the MW and the SMC or LMC extinction laws is small in the wavelength range of H$\delta$ to H$\beta$.} 
\citep{Pei1992a} with $R_{\rm V}=3.1$, we derive an $\textnormal{E(B-V)}$ of $\sim0.16$ mag and $A_V\sim0.5$ mag.  
Using extinction corrected H$\beta$ and [\ion{O}{2}]  fluxes and the prescriptions provided by 
\cite{Kennicutt1998a} and  \cite{Savaglio2009a}, the resulting star-formation rate (SFR) of the host galaxy is  SFR $\sim 4.0 \,M_\odot$/yr.
 The SFR as determined from the line fluxes is within $2\sigma$ of that derived from SED fitting. The high SFR obtained from two separate methods implies a specific star-formation rate (sSFR, SFR per unit stellar mass) of 1--3$\times 10^{-8}$ yr$^{-1}$, 
among the highest measured so far in the LGRB host population 
\citep[e.g.,][]{Savaglio2009a,Hunt2014a,Perley2013a,Vergani2015a,Japeli2016a,Schulze2018a,Modjaz2020a}. For comparison, this sSFR is about two orders of magnitude higher than for the Large Magellanic Cloud. Moreover, if the derived SFR were constant, this would imply that the time required for the host to build the measured stellar mass would be about 0.03--0.1 Gyr, which is consistent with the upper limit obtained from the SED best-fit.
 Using several diagnostic tools in \cite{Lamareille2010a},
\cite{Trouille2011a}, and \cite{Marocco2011a} the presence of an AGN, which may mimic a high SFR, can be confidently excluded.

The diagnostic ratios of [\ion{O}{2}], [\ion{O}{3}], and H$\beta$ emission lines were used to derive the gas metallicity in the host galaxy. Many calibrators have been proposed in the past \citep[for a review, see][]{MaiolinoMannucci2019a}; the method proposed by 
\citet{Curti2017a} 
was adopted which computes the best-fit metallicity by minimizing $\chi^2$ in the space defined by the different possible diagnostics given the available emission lines. This yielded $12+\log(\textnormal{O/H})=8.31\pm0.02$. This is in agreement with the value of $12+\log(\textnormal{O/H})\simeq8.3$  obtained from using the
[\ion{Ne}{3}]$\lambda$3869/[\ion{O}{2}] line ratio \citep[e.g.,][]{Bian2018a}, a method which has more uncertainties. 
The resulting sub-solar metallicity ($Z/Z_\odot \simeq 0.4$)  is consistent with the value derived for the LGRB hosts at similar redshift \citep{Japeli2016a}.
However, we note that the redshift evolution of the mass-metallicity relation in normal star-forming galaxies derived by \cite{Bellstedt+2021} shows that galaxies at $z = 0.62$ or 0.8 and stellar mass similar to the GRB 200826A host have a typical metallicity of about $12+\log(\textnormal{O/H})=7.85$ or 7.80, respectively, significantly below our measured value.
Taken together, these results indicate that the host galaxy of GRB 200826A is characterized  by a typical metallicity but larger than usual SFR and sSFR for a LGRB host.

For the adopted cosmological parameters 
and using the last deep $H$-band observation, the half-light radius of the host galaxy is $R_{h}=0\farcs16$ or 1.2 kpc at the redshift of the GRB. This size is consistent with LGRB hosts of similar masses \citep{Kelly2014a}. The high angular resolution from the LUCI-AO observations also makes it possible to measure the GRB-SN position within the host. The transient has an offset of $R=0\farcs1$ from the center of
the host, which corresponds to 0.75 kpc. This corresponds to a normalized offset
$R/R_h=0.62$ which is also consistent with the vast majority of LGRBs \citep[e.g.,][]{Bloom2002a,Blanchard2016a}.


\section{Discussion \label{sec:dis}}

Additional support for the collapsar scenario was recently reported by \citet{Rhodes2021a}.  By analyzing the radio and X-ray light curves, they found evidence of a sharp rise peaking at $4-5$ d after the trigger in the radio bands. 
 Within the  LGRB/collapsar scenario, 
they propose that the peak is produced by the synchrotron self-absorption frequency moving through the radio band, resulting from the forward shock propagating into a wind medium.
Is our modeling of the afterglow and optical-to-Xray SED in agreement with their claim? The spectral slope of $\beta=0.79$ we have found in \S \ref{sec:ag} favors the emission in both optical and X-rays to be below the cooling frequency, i.e., $\beta=(p-1)/2$, with $p=2.58\pm0.06$ following \cite{Sari1998a}, a typical value for synchrotron emission in GRB afterglows. If the afterglow is in the post-break phase when we measure the SED, the spectral and decay slopes are difficult to justify within the standard afterglow theory. 
 Therefore, assuming that the afterglow is still in the pre-break phase (spherical expansion) the temporal decay index for emission below the cooling break is $\alpha = 3(p-1)/4$ and $\alpha = (3p-1)/4$ for  an  ISM  and  wind  environment,  respectively. Using $\alpha=1.59\pm0.20$, we obtain $p=3.12\pm0.27$ and $p=2.45\pm0.27$, respectively \citep{GranotSari2002a}, and thus we find that propagation into a wind environment could well explain both spectral index and light-curve decay, in support of the analysis of \cite{Rhodes2021a}.
{While steep, a pre-break decay of $\alpha=1.59$ is not unheard of, the sample of \cite{Zeh2006a} contains two out of 16 well-defined light curves with $\alpha_1\approx1.75$ followed by clear breaks and much steeper post-break decay slopes $\alpha_2\approx3$. In such a case, the early shallower decay of the afterglow of GRB 200826A could represent an optical plateau \citep[e.g.,][Ronchini et al., in prep.]{Dainotti2020a}.}
In this scenario the late shallow decay observed in the X-rays can be explained as energy injection from a new-born magnetar \citep[see also][]{Zhang2021a}, but also by the receding cooling frequency after the passage of the wind termination-shock radius \citep[e.g.,][]{Schulze2011a}.

 All arguments presented in this work point to a massive-star origin
consistent with a collapsar event despite the very short duration. Several theoretical routes have been presented to explain this.
In one of the scenarios discussed by \cite{Zhang2021a} and by \cite{Ahumada2021a} to explain the short duration of the burst, if the progenitor was a massive star it is possible that the total duration of the engine was longer than the rest-frame duration of 1~s, but somehow the jet could not emit gamma-rays for a longer time. For example the jet could spend a longer time penetrating the envelope of the star. Alternatively, 
the jet might not be highly relativistic for longer than 1~s, so that no bright $\gamma$-rays could be produced after this time like, e.g., in the case of a newborn magnetar that initially injects a baryon-loaded, neutrino-driven wind \citep[see also][]{Ghisellini2020a}. 
\cite{Zhang2021a} also note that the energy injection from the newborn magnetar could explain the late shallow decay observed in the X-rays.
Other alternatives exist like the mechanism known as
the inhomogeneous jet model, proposed by \cite{Yamazaki2004a}, in which the jet consists of multiple sub-jets which we observe as short GRBs. Therefore, a fraction of SGRBs may come from collapsars.
\cite{Peng2021a} have recently proposed that the burst could have originated from the collapse of a Thorne-{\.Z}ytkow-like Object (T{\.Z}lO), which consists of a central NS and envelope formed after a coalescence with a white dwarf. They found the collapse of such a T{\.Z}lO can naturally explain the short duration of GRB 200826A and the interaction between the disk wind and the ejected material can explain the ``supernova bump". However, this model needs to be proven with the updated photometry presented here, and in particular with our result that the SN is similar to other GRB-SNe.
We have also demonstrated the importance of Adaptive Optics (AO) to studying GRBs.  The NIR follow-up observations presented here are the {\it first} to detect and pinpoint the location of a SN associated with a GRB, track the SN evolution, as well as determine the properties of the host galaxy.
The last published attempts were more than 11 years ago and served only to provide information on the host galaxy of GRB 060418 \citep{Pollack2009a}, and in case of GRB 070610 to determine that a NIR transient was actually the counterpart of a galactic source \citep{Castro-Tirado2008a}, making that GRB association uncertain \citep{Kasliwal2008a}.  The only other approach similar to that outlined in this paper was the case of GRB 071003 \citep{Perley2008a}. In that case, AO observations made possible to detect the afterglow which was
localized between two bright stars, but 
those observations were too early and the redshift too high to detect the associated SN ($z=1.60435$, and $\sim$16 d, respectively). The results presented here demonstrate clearly that AO can provide a powerful tool for studying GRBs.


\section{Summary and conclusions \label{sec:con}}

GRB 200826A is a temporally short GRB at $z=0.748577\pm0.000065$ with a rest-frame duration of $\sim0.5$ s, below the threshold of 2 s commonly used to separate SGRBs and LGRBs.
To better understand the nature of this event, we initiated a follow-up campaign spanning a period of 1 year
which involved the LBT telescope in Arizona (USA), the TNG telescope on La Palma (Canary Islands, Spain) and the Maidanak Astronomical Observatory (Uzbekistan). 
Taking advantage of the adaptive optics capabilities of LBT, we were able to obtain deep $H$-band observations between 37 and 159 d (observer frame), corresponding to the rest-frame $z^\prime$ band at 21 to 91 d after the burst trigger. Image subtraction shows a faint transient within its host galaxy.
Moreover, image subtraction of archival Gemini $i^\prime$-band images with late reference observations obtained with the LBT revealed an optical transient 
which is less affected by over-subtraction than had been when reference images obtained at an earlier time after the GRB had been used.
Finally, we were able to put strong upper limits on the UV rest-frame luminosity thanks to our LBT and TNG $r^\prime$-band observations.
Our results show that:
\begin{itemize}
    
    \item Despite its short duration, this event is consistent with the $E_{\rm p,i}-E_{\rm iso}$ ``Amati'' relation followed by LGRBs. The spectral lag is also more typical of LGRBs.
    
    \item It was followed by a relatively faint optical and X-ray afterglow with a luminosity that lies in between those of LGRB and SGRB afterglows.
    
    \item The evolution and color of the late bump is in good agreement with other GRB-SNe, and especially with the fast rising GRB 130702A/SN 2013dx. GRB 200826A is one of the cosmologically most remote GRB-SNe detected to date, close to the sensitivity limit of the present generation of 8 to 10 m class optical telescopes.
    
    \item The possible alternative scenario of a genuine SGRB followed by a KN like AT2017gfo is not supported by the different evolution and luminosity of the light curve of the observed transient.
    
    \item 
    The host galaxy of GRB 200826A is remarkable because it is typical of an LGRB host galaxy, but with higher SFR and sSFR rates than expected.
    
    \item The GRB lies at a projected distance of 0.75 kpc from the center of its host galaxy, which is consistent with the majority of LGRBs. 
    
\end{itemize}

In summary, with the obtained results 
we are faced with a GRB of short duration which exploded in a star-forming galaxy, with a moderately faint afterglow, emitted by a jet most likely propagating into a wind environment, and followed by a bump in the light curve whose color and luminosity are typical for a GRB-SN. Thus we firmly classify this burst as a collapsar event.
This evidence, together with the analysis of the energetics of this  burst, further weakens the effectiveness of simple duration as an indicator of the source of a GRB. In addition, strong support is provided to theoretical predictions that collapsar-produced events may have an observable duration well short of the classical short/long divide (about 2 s), and down to 0.5 s or less \citep[e.g.,][]{Bromberg2013a}.

In the next years future missions, like the Space Variable Objects Monitor \citep[SVOM;][]{svom2011a} and the Gamow Explorer \citep{White2021SPIE} will offer a combination of extended sensitivity and energy bands that will increase the number of known GRBs. This, when coupled with the improved capabilities of the new generation of extremely large telescopes will allow us to observe both the GRB and SN components resulting from collapsar explosions in increasing numbers and at higher redshifts \citep[e.g.,][]{Maiorano2018a,Rossi2018a}.
In this scenario, the first detection of a SN with AO observations  represents what ground-based telescopes can achieve: they will not just offer a sharper view of the GRB-SN location within its host, but have the necessary depth to discover GRB-SNe at larger redshift than what has been possible up to now from the ground, and yet at similar wavelengths in rest-frame.
Eventually, these future facilities will assess whether peculiar events like GRB 200826A are actually the result of the rich variety of the collapsar phenomenon \citep{Amati2021a}.

\begin{acknowledgements}
We thank the anonymous referee for providing  thoughtful comments.
We thank E. Pinna, D. Miller, J. Power and G. Taylor for their support during AO observations.
We thank M. Curti for providing us with his code.  
A. Rossi acknowledges support from the INAF project Premiale Supporto Arizona \& Italia.
B. Rothberg would like to acknowledge the assistance and support of R. T. Gatto. D.A.K. acknowledges support from Spanish National Research Project RTI2018-098104-J-I00 (GRBPhot). 
P.D.A. acknowledges funding from the Italian Space Agency, contract ASI/INAF n. I/004/11/4.
L.I. was supported by two grants from VILLUM FONDEN (project number 16599 and 25501).
A.N.G. and S.K. acknowledge financial support by grants DFG Kl 766/16-3, DFG Kl 766/18-1, and DFG RA 2484/1-3. A.N.G. acknowledges support by the Th\"uringer Landessternwarte Tautenburg.
A.S.P., P.Y.M. and A.A.V. acknowledge a support of the RSF grant 18-12-00378.
S.S. acknowledges support from the G.R.E.A.T research environment, funded by {\em Vetenskapsr\aa det},  the Swedish Research Council, project number 2016-06012.
A.R., E.Pal., P.D.A., L.A., A.P., E.Pi., S.S., G.S., S.C., V.D.E., M.D.V., and A.M. acknowledge support from PRIN-MIUR 2017 (grant 20179ZF5KS).
The LBT is an international collaboration among institutions in the United States, Italy and Germany. LBT Corporation partners are: The University of Arizona on behalf of the Arizona Board of Regents; Istituto Nazionale di Astrofisica, Italy; LBT Beteiligungsgesellschaft, Germany, representing the Max-Planck Society, The Leibniz Institute for Astrophysics Potsdam, and Heidelberg University; The Ohio State University, representing OSU, University of Notre Dame, University of Minnesota and University of Virginia.  
This work made use of data supplied by the UK \textit{Swift} Science Data center at the University of Leicester.
\end{acknowledgements}

\facilities{Swift(XRT), LBT, TNG, MAO}

\software{IRAF \citep{Tody1993}, 
PyRAF \citep{pyraf2012}, 
DRAGONS \citep{dragons2019}, 
HOTPANTS \citep{Becker2015a}, 
Le PHARE \citep{Arnouts1999a,Ilbert2006a},
WCSTools \citep{wcstools2019}
}

\appendix

\section{Spectral lag} \label{sec:lag}

We first  extracted the Fermi/GBM light curves with 1~ms resolution for the best exposed NaI detectors 6, 7, 8, 9, and 11 and calculated the corresponding total light curve in the same energy bands as \cite{Zhang2021a}: 10-20~keV, and 250-300~keV, respectively. We linearly interpolated the background counts of the 16-ms binned curves by fitting the time windows $[-10:-2]\cup [7:10]$~s.

\begin{figure}
\centering
\includegraphics[width=0.48\textwidth,angle=0]{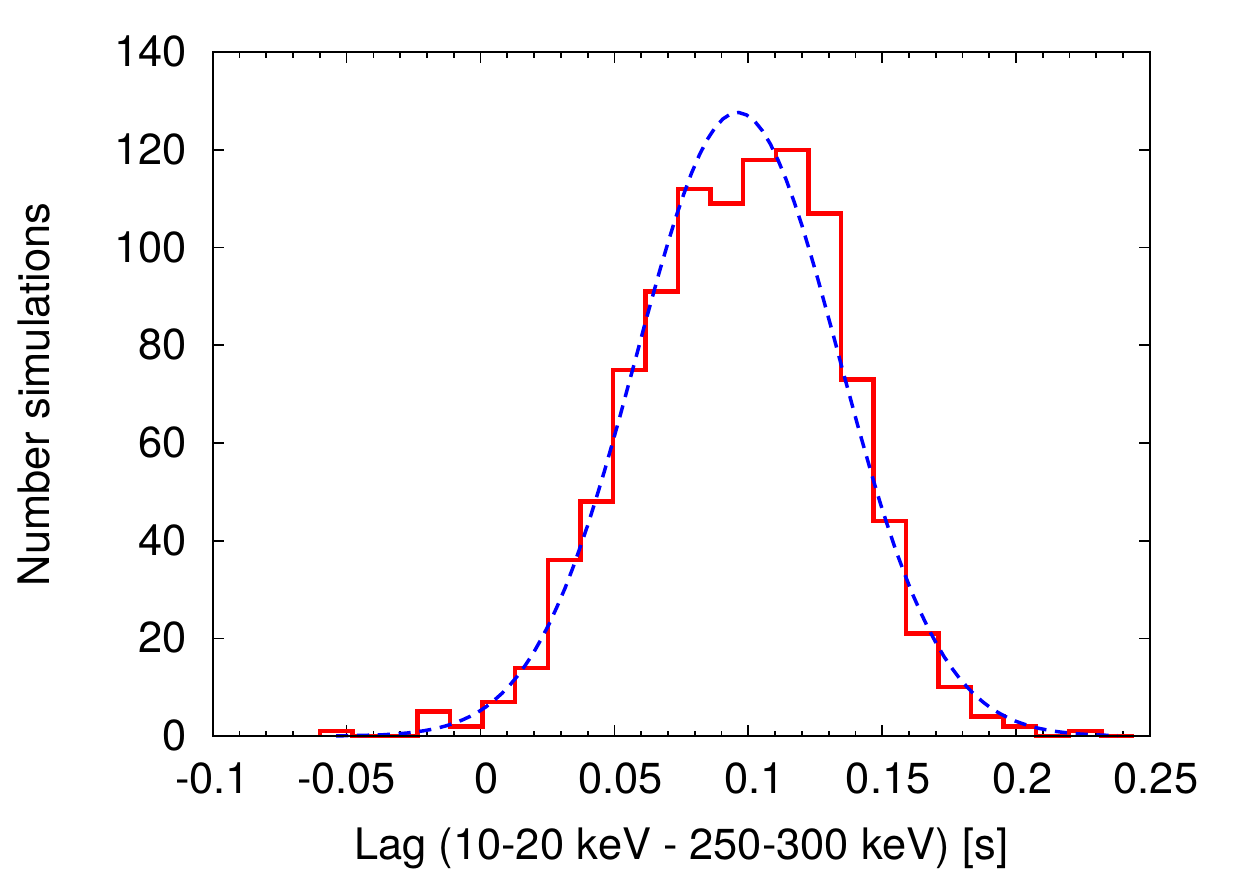}
\caption{Distribution of the spectral lag analysis.}
\label{fig:lag}%
\end{figure}

For the spectral lag calculation, we then restricted the following analysis to the time interval $[-0.5:4]$~s and calculated the cross-correlation function (CCF) between the background-subtracted profiles of the two energy bands, both using the original binning of 1~ms as well as that of 16~ms. A positive lag corresponds to the softer band lagging behind the harder one. The peak of the CCF was found to be about 100~ms. To provide a more accurate estimate, we fitted the CCF around the peak with a third-degree polynomial between $-0.4$ and $0.6$~s and it gave 95~ms. To estimate the uncertainty, we carried out the following simulations: starting from the 16-ms profiles, we obtained smooth versions of these curves by adopting the L1 trend filtering by \citet{Politsch2020a} properly adapted to model GRB light curves (e.g., see \citealt{Ghirlanda21_WPTHESEUS}). We then obtained 1000 random realisations of both profiles by assuming uncorrelated Poisson noise, assuming the total expected counts for each temporal bin (i.e., interpolated background plus smoothed GRB signal). For each pair of simulated profiles the corresponding CCF was fitted around the peak following the same procedure as for the real CCF. We finally collected the resulting distribution of 1000 lags shown in Figure~\ref{fig:lag}: this is approximately normal with mean and standard deviation $(96\pm38)$~ms, which therefore represents our estimate with uncertainty for the lag between the chosen energy bands.
Our value looks slightly shorter than the analogous result by \citet{Zhang2021a}, who obtained $(157\pm51)$~ms, even though statistically compatible and well in accord with values obtained by those authors for adjacent energy bands (see their Extended Data Figure 4).

We finally repeated the same simulations using the 1-ms light curves, still adopting the same smoothed versions properly calibrated to 1-ms bins, and obtained a very similar result.

\bibliographystyle{aasjournal}
\bibliography{biblio}{} 

\begin{thebibliography}{}
\expandafter\ifx\csname natexlab\endcsname\relax\def\natexlab#1{#1}\fi
\providecommand{\url}[1]{\href{#1}{#1}}
\providecommand{\dodoi}[1]{doi:~\href{http://doi.org/#1}{\nolinkurl{#1}}}
\providecommand{\doeprint}[1]{\href{http://ascl.net/#1}{\nolinkurl{http://ascl.net/#1}}}
\providecommand{\doarXiv}[1]{\href{https://arxiv.org/abs/#1}{\nolinkurl{https://arxiv.org/abs/#1}}}

\bibitem[{{Abbott} {et~al.}(2017{\natexlab{a}}){Abbott}, {Abbott}, {Abbott},
  {Acernese}, {Ackley}, {Adams}, {Adams}, {Addesso}, {Adhikari}, {Adya},
  {Affeldt}, {Afrough}, {Agarwal}, {Agathos}, {Agatsuma}, {Aggarwal}, {Aguiar},
  {Aiello}, {Ain}, \& et~al.}]{Abbott2017c}
{Abbott}, B.~P., {Abbott}, R., {Abbott}, T.~D., {et~al.} 2017{\natexlab{a}},
  \apjl, 848, L13, \dodoi{10.3847/2041-8213/aa920c}

\bibitem[{{Abbott} {et~al.}(2017{\natexlab{b}}){Abbott}, {Abbott}, {Abbott},
  {Acernese}, {Ackley}, {Adams}, {Adams}, {Addesso}, {Adhikari}, {Adya},
  {Affeldt}, {Afrough}, {Agarwal}, {Agathos}, {Agatsuma}, {Aggarwal}, {Aguiar},
  {Aiello}, {Ain}, \& et~al.}]{Abbott2017b}
---. 2017{\natexlab{b}}, \apjl, 848, L12, \dodoi{10.3847/2041-8213/aa91c9}

\bibitem[{{Abbott} {et~al.}(2017{\natexlab{c}}){Abbott}, {Abbott}, {Abbott},
  {Acernese}, {Ackley}, {Adams}, {Adams}, {Addesso}, {Adhikari}, {Adya}, \&
  et~al.}]{Abbott2017a}
---. 2017{\natexlab{c}}, Physical Review Letters, 119, 161101,
  \dodoi{10.1103/PhysRevLett.119.161101}

\bibitem[{{Ag{\"u}{\'\i} Fern{\'a}ndez} {et~al.}(2021){Ag{\"u}{\'\i}
  Fern{\'a}ndez}, {Th{\"o}ne}, {Kann}, {de Ugarte Postigo}, {Selsing},
  {Schady}, {Yates}, {Greiner}, {Oates}, {Malesani}, {Xu}, {Klotz}, {Campana},
  {Rossi}, {Perley}, {Blazek}, {D'Avanzo}, {Giunta}, {Hartmann}, {Heintz},
  {Jakobsson}, {Kirkpatrick}, {Kouveliotou}, {Melandri}, {Pugliese},
  {Salvaterra}, {Starling}, {Tanvir}, {Vergani}, \& {Wiersema}}]{Agui2021a}
{Ag{\"u}{\'\i} Fern{\'a}ndez}, J.~F., {Th{\"o}ne}, C.~C., {Kann}, D.~A.,
  {et~al.} 2021, arXiv e-prints, arXiv:2109.13838.
\newblock \doarXiv{2109.13838}

\bibitem[{{Ahumada} {et~al.}(2020{\natexlab{a}}){Ahumada}, {Anand}, {Sten},
  {Kumar}, {Sagues Carracedo}, {Ztf Collaboration}, \& {Growth
  Collaboration}}]{Ahumada2020GCNdisco}
{Ahumada}, T., {Anand}, S., {Sten}, R., {et~al.} 2020{\natexlab{a}}, GRB
  Coordinates Network, 28295, 1

\bibitem[{{Ahumada} {et~al.}(2020{\natexlab{b}}){Ahumada}, {Kumar}, {Fremling},
  {Singer}, {Ztf Collaboration}, \& {Growth
  Collaboration}}]{Ahumada2020GCNcorrection}
{Ahumada}, T., {Kumar}, H., {Fremling}, C., {et~al.} 2020{\natexlab{b}}, GRB
  Coordinates Network, 29029, 1

\bibitem[{{Ahumada} {et~al.}(2020{\natexlab{c}}){Ahumada}, {Singer}, {Kumar},
  {Reusch}, {Ztf}, \& {Growth Collaborations}}]{Ahumada2020GCNsnbump1}
{Ahumada}, T., {Singer}, L., {Kumar}, H., {et~al.} 2020{\natexlab{c}}, GRB
  Coordinates Network, 28727, 1

\bibitem[{{Ahumada} {et~al.}(2021){Ahumada}, {Singer}, {Anand}, {Coughlin},
  {Kasliwal}, {Ryan}, {Andreoni}, {Cenko}, {Fremling}, {Kumar}, {Pang},
  {Burns}, {Cunningham}, {Dichiara}, {Dietrich}, {Svinkin}, {Almualla},
  {Castro-Tirado}, {De}, {Dunwoody}, {Gatkine}, {Hammerstein}, {Iyyani},
  {Mangan}, {Perley}, {Purkayastha}, {Bellm}, {Bhalerao}, {Bolin}, {Bulla},
  {Cannella}, {Chandra}, {Duev}, {Frederiks}, {Gal-Yam}, {Graham}, {Ho},
  {Hurley}, {Karambelkar}, {Kool}, {Kulkarni}, {Mahabal}, {Masci}, {McBreen},
  {Pandey}, {Reusch}, {Ridnaia}, {Rosnet}, {Rusholme}, {Carracedo}, {Smith},
  {Soumagnac}, {Stein}, {Troja}, {Tsvetkova}, {Walters}, \&
  {Valeev}}]{Ahumada2021a}
{Ahumada}, T., {Singer}, L.~P., {Anand}, S., {et~al.} 2021, Nature Astronomy,
  5, 917, \dodoi{10.1038/s41550-021-01428-7}

\bibitem[{{Alam} {et~al.}(2015){Alam}, {Albareti}, {Allende Prieto}, {Anders},
  {Anderson}, {Anderton}, {Andrews}, {Armengaud}, {Aubourg}, {Bailey}, \&
  et~al.}]{Alam2015a}
{Alam}, S., {Albareti}, F.~D., {Allende Prieto}, C., {et~al.} 2015, \apjs, 219,
  12, \dodoi{10.1088/0067-0049/219/1/12}

\bibitem[{{Amati}(2006)}]{Amati2006a}
{Amati}, L. 2006, \mnras, 372, 233, \dodoi{10.1111/j.1365-2966.2006.10840.x}

\bibitem[{{Amati}(2021)}]{Amati2021a}
---. 2021, Nature Astronomy, \dodoi{10.1038/s41550-021-01401-4}

\bibitem[{{Amati} {et~al.}(2019){Amati}, {D'Agostino}, {Luongo}, {Muccino}, \&
  {Tantalo}}]{Amati2019a}
{Amati}, L., {D'Agostino}, R., {Luongo}, O., {Muccino}, M., \& {Tantalo}, M.
  2019, \mnras, 486, L46, \dodoi{10.1093/mnrasl/slz056}

\bibitem[{{Amati} {et~al.}(2002){Amati}, {Frontera}, {Tavani}, {in't Zand},
  {Antonelli}, {Costa}, {Feroci}, {Guidorzi}, {Heise}, {Masetti}, {Montanari},
  {Nicastro}, {Palazzi}, {Pian}, {Piro}, \& {Soffitta}}]{Amati2002a}
{Amati}, L., {Frontera}, F., {Tavani}, M., {et~al.} 2002, \aap, 390, 81,
  \dodoi{10.1051/0004-6361:20020722}

\bibitem[{{Antonelli} {et~al.}(2009){Antonelli}, {D'Avanzo}, {Perna}, {Amati},
  {Covino}, {Cutini}, {D'Elia}, {Gallozzi}, {Grazian}, {Palazzi},
  {Piranomonte}, {Rossi}, {Spiro}, {Stella}, {Testa}, {Chincarini}, {di Paola},
  {Fiore}, {Fugazza}, {Giallongo}, {Maiorano}, {Masetti}, {Pedichini},
  {Salvaterra}, {Tagliaferri}, \& {Vergani}}]{Antonelli2009a}
{Antonelli}, L.~A., {D'Avanzo}, P., {Perna}, R., {et~al.} 2009, \aap, 507, L45,
  \dodoi{10.1051/0004-6361/200913062}

\bibitem[{{Arnouts} {et~al.}(1999){Arnouts}, {Cristiani}, {Moscardini},
  {Matarrese}, {Lucchin}, {Fontana}, \& {Giallongo}}]{Arnouts1999a}
{Arnouts}, S., {Cristiani}, S., {Moscardini}, L., {et~al.} 1999, \mnras, 310,
  540, \dodoi{10.1046/j.1365-8711.1999.02978.x}

\bibitem[{{Ascenzi} {et~al.}(2019){Ascenzi}, {Coughlin}, {Dietrich}, {Foley},
  {Ramirez-Ruiz}, {Piranomonte}, {Mockler}, {Murguia-Berthier}, {Fryer},
  {Lloyd-Ronning}, \& {Rosswog}}]{Ascenzi2018a}
{Ascenzi}, S., {Coughlin}, M.~W., {Dietrich}, T., {et~al.} 2019, \mnras, 486,
  672, \dodoi{10.1093/mnras/stz891}

\bibitem[{{Ashall} {et~al.}(2019{\natexlab{a}}){Ashall}, {Mazzali}, {Pian},
  {Woosley}, {Palazzi}, {Prentice}, {Kobayashi}, {Holmbo}, {Levan}, {Perley},
  {Stritzinger}, {Bufano}, {Filippenko}, {Melandri}, {Oates}, {Rossi},
  {Selsing}, {Zheng}, {Castro-Tirado}, {Chincarini}, {D'Avanzo}, {De Pasquale},
  {Emery}, {Fruchter}, {Hurley}, {Moller}, {Nomoto}, {Tanaka}, \&
  {Valeev}}]{Ashall2019a}
{Ashall}, C., {Mazzali}, P.~A., {Pian}, E., {et~al.} 2019{\natexlab{a}},
  \mnras, 487, 5824, \dodoi{10.1093/mnras/stz1588}

\bibitem[{{Ashall} {et~al.}(2019{\natexlab{b}}){Ashall}, {Mazzali}, {Pian},
  {Woosley}, {Palazzi}, {Prentice}, {Kobayashi}, {Holmbo}, {Levan}, {Perley},
  {Stritzinger}, {Bufano}, {Filippenko}, {Melandri}, {Oates}, {Rossi},
  {Selsing}, {Zheng}, {Castro-Tirado}, {Chincarini}, {D'Avanzo}, {De Pasquale},
  {Emery}, {Fruchter}, {Hurley}, {Moller}, {Nomoto}, {Tanaka}, \&
  {Valeev}}]{Ashall2019MNRAS}
---. 2019{\natexlab{b}}, \mnras, 487, 5824, \dodoi{10.1093/mnras/stz1588}

\bibitem[{{Barbieri} {et~al.}(2019){Barbieri}, {Salafia}, {Perego}, {Colpi}, \&
  {Ghirlanda}}]{Barbieri2019}
{Barbieri}, C., {Salafia}, O.~S., {Perego}, A., {Colpi}, M., \& {Ghirlanda}, G.
  2019, \aap, 625, A152, \dodoi{10.1051/0004-6361/201935443}

\bibitem[{{Barthelmy} {et~al.}(2006){Barthelmy}, {Barbier}, {Cummings},
  {Fenimore}, {Gehrels}, {Hullinger}, {Krimm}, {Koss}, {Markwardt}, {Palmer},
  {Parsons}, {Sakamoto}, {Sato}, {Stamatikos}, \&
  {Tueller}}]{BarthelmyGCN2006a}
{Barthelmy}, S., {Barbier}, L., {Cummings}, J., {et~al.} 2006, GRB Coordinates
  Network, 5256, 1

\bibitem[{{Becker}(2015)}]{Becker2015a}
{Becker}, A. 2015, {HOTPANTS: High Order Transform of PSF ANd Template
  Subtraction}.
\newblock \doeprint{1504.004}

\bibitem[{{Belkin} {et~al.}(2020{\natexlab{a}}){Belkin}, {Zhornichenko},
  {Pozanenko}, {Pankov}, {Mazaeva}, {Volnova}, {Ehgamberdiev}, \&
  {IKI-GRB-FuN}}]{Belkin2020GCNkitab}
{Belkin}, S., {Zhornichenko}, A., {Pozanenko}, A., {et~al.} 2020{\natexlab{a}},
  GRB Coordinates Network, 28306, 1

\bibitem[{{Belkin} {et~al.}(2020{\natexlab{b}}){Belkin}, {Pozanenko},
  {Mazaeva}, {Volnova}, {Minaev}, {Tominaga}, {Grebenev}, {Chelovekov},
  {Buckley}, {Blinnikov}, {Volvach}, {Volvach}, {Inasaridze}, {Klunko},
  {Molotov}, {Reva}, {Rumyantsev}, \& {Chestnov}}]{Belkin2020a}
{Belkin}, S.~O., {Pozanenko}, A.~S., {Mazaeva}, E.~D., {et~al.}
  2020{\natexlab{b}}, Astronomy Letters, 46, 783,
  \dodoi{10.1134/S1063773720120014}

\bibitem[{{Bellm} {et~al.}(2019){Bellm}, {Kulkarni}, {Barlow}, {Feindt},
  {Graham}, {Goobar}, {Kupfer}, {Ngeow}, {Nugent}, {Ofek}, {Prince}, {Riddle},
  {Walters}, \& {Ye}}]{Bellm2019a}
{Bellm}, E.~C., {Kulkarni}, S.~R., {Barlow}, T., {et~al.} 2019, \pasp, 131,
  068003, \dodoi{10.1088/1538-3873/ab0c2a}

\bibitem[{{Bellstedt} {et~al.}(2021){Bellstedt}, {Robotham}, {Driver},
  {Thorne}, {Davies}, {Holwerda}, {Hopkins}, {Lara-Lopez},
  {L{\'o}pez-S{\'a}nchez}, \& {Phillipps}}]{Bellstedt+2021}
{Bellstedt}, S., {Robotham}, A. S.~G., {Driver}, S.~P., {et~al.} 2021, \mnras,
  503, 3309, \dodoi{10.1093/mnras/stab550}

\bibitem[{{Bernuzzi} {et~al.}(2020){Bernuzzi}, {Breschi}, {Daszuta},
  {Endrizzi}, {Logoteta}, {Nedora}, {Perego}, {Radice}, {Schianchi}, {Zappa},
  {Bombaci}, \& {Ortiz}}]{Bernuzzi2020}
{Bernuzzi}, S., {Breschi}, M., {Daszuta}, B., {et~al.} 2020, \mnras, 497, 1488,
  \dodoi{10.1093/mnras/staa1860}

\bibitem[{{Beuermann} {et~al.}(1999){Beuermann}, {Hessman}, {Reinsch},
  {Nicklas}, {Vreeswijk}, {Galama}, {Rol}, {van Paradijs}, {Kouveliotou},
  {Frontera}, {Masetti}, {Palazzi}, \& {Pian}}]{Beuermann1999a}
{Beuermann}, K., {Hessman}, F.~V., {Reinsch}, K., {et~al.} 1999, \aap, 352,
  L26.
\newblock \doarXiv{astro-ph/9909043}

\bibitem[{{Bian} {et~al.}(2018){Bian}, {Kewley}, \& {Dopita}}]{Bian2018a}
{Bian}, F., {Kewley}, L.~J., \& {Dopita}, M.~A. 2018, \apj, 859, 175,
  \dodoi{10.3847/1538-4357/aabd74}

\bibitem[{{Blanchard} {et~al.}(2016){Blanchard}, {Berger}, \&
  {Fong}}]{Blanchard2016a}
{Blanchard}, P.~K., {Berger}, E., \& {Fong}, W.-f. 2016, \apj, 817, 144,
  \dodoi{10.3847/0004-637X/817/2/144}

\bibitem[{{Bloom} {et~al.}(2002){Bloom}, {Kulkarni}, \&
  {Djorgovski}}]{Bloom2002a}
{Bloom}, J.~S., {Kulkarni}, S.~R., \& {Djorgovski}, S.~G. 2002, \aj, 123, 1111,
  \dodoi{10.1086/338893}

\bibitem[{{Bromberg} {et~al.}(2012){Bromberg}, {Nakar}, {Piran}, \&
  {Sari}}]{Bromberg2012a}
{Bromberg}, O., {Nakar}, E., {Piran}, T., \& {Sari}, R. 2012, \apj, 749, 110,
  \dodoi{10.1088/0004-637X/749/2/110}

\bibitem[{{Bromberg} {et~al.}(2013){Bromberg}, {Nakar}, {Piran}, \&
  {Sari}}]{Bromberg2013a}
---. 2013, \apj, 764, 179, \dodoi{10.1088/0004-637X/764/2/179}

\bibitem[{{Cano} {et~al.}(2017{\natexlab{a}}){Cano}, {Wang}, {Dai}, \&
  {Wu}}]{Cano2017a}
{Cano}, Z., {Wang}, S.-Q., {Dai}, Z.-G., \& {Wu}, X.-F. 2017{\natexlab{a}},
  Advances in Astronomy, 2017, 8929054, \dodoi{10.1155/2017/8929054}

\bibitem[{{Cano} {et~al.}(2017{\natexlab{b}}){Cano}, {Izzo}, {de Ugarte
  Postigo}, {Th{\"o}ne}, {Kr{\"u}hler}, {Heintz}, {Malesani}, {Geier},
  {Fuentes}, {Chen}, {Covino}, {D'Elia}, {Fynbo}, {Goldoni}, {Gomboc},
  {Hjorth}, {Jakobsson}, {Kann}, {Milvang-Jensen}, {Pugliese},
  {S{\'a}nchez-Ram{\'\i}rez}, {Schulze}, {Sollerman}, {Tanvir}, \&
  {Wiersema}}]{Cano2017AA}
{Cano}, Z., {Izzo}, L., {de Ugarte Postigo}, A., {et~al.} 2017{\natexlab{b}},
  \aap, 605, A107, \dodoi{10.1051/0004-6361/201731005}

\bibitem[{{Cardelli} {et~al.}(1989){Cardelli}, {Clayton}, \&
  {Mathis}}]{Cardelli1989}
{Cardelli}, J.~A., {Clayton}, G.~C., \& {Mathis}, J.~S. 1989, \apj, 345, 245,
  \dodoi{10.1086/167900}

\bibitem[{{Castro-Tirado} {et~al.}(2008){Castro-Tirado}, {de Ugarte Postigo},
  {Gorosabel}, {Jel{\'\i}nek}, {Fatkhullin}, {Sokolov}, {Ferrero}, {Kann},
  {Klose}, {Sluse}, {Bremer}, {Winters}, {Nuernberger},
  {P{\'e}rez-Ram{\'\i}rez}, {Guerrero}, {French}, {Melady}, {Hanlon},
  {McBreen}, {Leventis}, {Markoff}, {Leon}, {Kraus}, {Aceituno}, {Cunniffe},
  {Kub{\'a}nek}, {V{\'\i}tek}, {Schulze}, {Wilson}, {Hudec}, {Durant},
  {Gonz{\'a}lez-P{\'e}rez}, {Shahbaz}, {Guziy}, {Pandey}, {Pavlenko}, {Sonbas},
  {Trushkin}, {Bursov}, {Nizhelskij}, {S{\'a}nchez-Fern{\'a}ndez}, \&
  {Sabau-Graziati}}]{Castro-Tirado2008a}
{Castro-Tirado}, A.~J., {de Ugarte Postigo}, A., {Gorosabel}, J., {et~al.}
  2008, \nat, 455, 506, \dodoi{10.1038/nature07328}

\bibitem[{{Christensen} {et~al.}(2017){Christensen}, {Vergani}, {Schulze},
  {Annau}, {Selsing}, {Fynbo}, {de Ugarte Postigo}, {Ca{\~n}ameras}, {Lopez},
  {Passi}, {Cort{\'e}s-Zuleta}, {Ellison}, {D'Odorico}, {Becker}, {Berg},
  {Cano}, {Covino}, {Cupani}, {D'Elia}, {Goldoni}, {Gomboc}, {Hammer},
  {Heintz}, {Jakobsson}, {Japelj}, {Kaper}, {Malesani}, {M{\o}ller},
  {Petitjean}, {Pugliese}, {S{\'a}nchez-Ram{\'\i}rez}, {Tanvir}, {Th{\"o}ne},
  {Vestergaard}, {Wiersema}, \& {Worseck}}]{Christensen2017a}
{Christensen}, L., {Vergani}, S.~D., {Schulze}, S., {et~al.} 2017, \aap, 608,
  A84, \dodoi{10.1051/0004-6361/201731382}

\bibitem[{{Curti} {et~al.}(2017){Curti}, {Cresci}, {Mannucci}, {Marconi},
  {Maiolino}, \& {Esposito}}]{Curti2017a}
{Curti}, M., {Cresci}, G., {Mannucci}, F., {et~al.} 2017, \mnras, 465, 1384,
  \dodoi{10.1093/mnras/stw2766}

\bibitem[{{D'Ai} {et~al.}(2020){D'Ai}, {Sbarufatti}, {Oates}, {Burrows},
  {Gropp}, {Osborne}, {Page}, {Beardmore}, {Melandri}, {Sbarrato}, {Evans}, \&
  {Swift Team}}]{Dai2020GCNswiftxrt}
{D'Ai}, A., {Sbarufatti}, B., {Oates}, S.~R., {et~al.} 2020, GRB Coordinates
  Network, 28300, 1

\bibitem[{{Dainotti} {et~al.}(2020){Dainotti}, {Livermore}, {Kann}, {Li},
  {Oates}, {Yi}, {Zhang}, {Gendre}, {Cenko}, \& {Fraija}}]{Dainotti2020a}
{Dainotti}, M.~G., {Livermore}, S., {Kann}, D.~A., {et~al.} 2020, \apjl, 905,
  L26, \dodoi{10.3847/2041-8213/abcda9}

\bibitem[{{D'Avanzo} {et~al.}(2014){D'Avanzo}, {Salvaterra}, {Bernardini},
  {Nava}, {Campana}, {Covino}, {D'Elia}, {Ghirlanda}, {Ghisellini}, {Melandri},
  {Sbarufatti}, {Vergani}, \& {Tagliaferri}}]{Davanzo2014a}
{D'Avanzo}, P., {Salvaterra}, R., {Bernardini}, M.~G., {et~al.} 2014, \mnras,
  442, 2342, \dodoi{10.1093/mnras/stu994}

\bibitem[{{de Ugarte Postigo} {et~al.}(2011){de Ugarte Postigo}, {Horv{\'a}th},
  {Veres}, {Bagoly}, {Kann}, {Th{\"o}ne}, {Balazs}, {D'Avanzo}, {Aloy},
  {Foley}, {Campana}, {Mao}, {Jakobsson}, {Covino}, {Fynbo}, {Gorosabel},
  {Castro-Tirado}, {Amati}, \& {Nardini}}]{deUgartePostigo2011a}
{de Ugarte Postigo}, A., {Horv{\'a}th}, I., {Veres}, P., {et~al.} 2011, \aap,
  525, A109, \dodoi{10.1051/0004-6361/201015261}

\bibitem[{{D'Elia} {et~al.}(2015){D'Elia}, {Pian}, {Melandri}, {D'Avanzo},
  {Della Valle}, {Mazzali}, {Piranomonte}, {Tagliaferri}, {Antonelli},
  {Bufano}, {Covino}, {Fugazza}, {Malesani}, {M{\o}ller}, \&
  {Palazzi}}]{Delia2015AA}
{D'Elia}, V., {Pian}, E., {Melandri}, A., {et~al.} 2015, \aap, 577, A116,
  \dodoi{10.1051/0004-6361/201425381}

\bibitem[{{Della Valle} {et~al.}(2006){Della Valle}, {Malesani}, {Bloom},
  {Benetti}, {Chincarini}, {D'Avanzo}, {Foley}, {Covino}, {Melandri},
  {Piranomonte}, {Tagliaferri}, {Stella}, {Gilmozzi}, {Antonelli}, {Campana},
  {Chen}, {Filliatre}, {Fiore}, {Fugazza}, {Gehrels}, {Hurley}, {Mirabel},
  {Pellizza}, {Piro}, \& {Prochaska}}]{DellaValle2006a}
{Della Valle}, M., {Malesani}, D., {Bloom}, J.~S., {et~al.} 2006, \apjl, 642,
  L103, \dodoi{10.1086/504636}

\bibitem[{{Dichiara} {et~al.}(2020){Dichiara}, {Cenko}, {Troja}, {Gatkine},
  {Durbak}, {Kutyrev}, \& {Veilleux}}]{Dichiara2020GCNlowell}
{Dichiara}, S., {Cenko}, S.~B., {Troja}, E., {et~al.} 2020, GRB Coordinates
  Network, 28312, 1

\bibitem[{{Evans} {et~al.}(2010){Evans}, {Willingale}, {Osborne}, {O'Brien},
  {Page}, {Markwardt}, {Barthelmy}, {Beardmore}, {Burrows}, {Pagani},
  {Starling}, {Gehrels}, \& {Romano}}]{Evans2010a}
{Evans}, P.~A., {Willingale}, R., {Osborne}, J.~P., {et~al.} 2010, \aap, 519,
  A102, \dodoi{10.1051/0004-6361/201014819}

\bibitem[{{Ferrero} {et~al.}(2006){Ferrero}, {Kann}, {Zeh}, {Klose}, {Pian},
  {Palazzi}, {Masetti}, {Hartmann}, {Sollerman}, {Deng}, {Filippenko},
  {Greiner}, {Hughes}, {Mazzali}, {Li}, {Rol}, {Smith}, \&
  {Tanvir}}]{Ferrero2006AA}
{Ferrero}, P., {Kann}, D.~A., {Zeh}, A., {et~al.} 2006, \aap, 457, 857,
  \dodoi{10.1051/0004-6361:20065530}

\bibitem[{{Fong} {et~al.}(2021){Fong}, {Laskar}, {Rastinejad}, {Escorial},
  {Schroeder}, {Barnes}, {Kilpatrick}, {Paterson}, {Berger}, {Metzger}, {Dong},
  {Nugent}, {Strausbaugh}, {Blanchard}, {Goyal}, {Cucchiara}, {Terreran},
  {Alexander}, {Eftekhari}, {Fryer}, {Margalit}, {Margutti}, \&
  {Nicholl}}]{Fong2021a}
{Fong}, W., {Laskar}, T., {Rastinejad}, J., {et~al.} 2021, \apj, 906, 127,
  \dodoi{10.3847/1538-4357/abc74a}

\bibitem[{{Fontana} {et~al.}(2014){Fontana}, {Dunlop}, {Paris}, {Targett},
  {Boutsia}, {Castellano}, {Galametz}, {Grazian}, {McLure}, {Merlin},
  {Pentericci}, {Wuyts}, {Almaini}, {Caputi}, {Chary}, {Cirasuolo},
  {Conselice}, {Cooray}, {Daddi}, {Dickinson}, {Faber}, {Fazio}, {Ferguson},
  {Giallongo}, {Giavalisco}, {Grogin}, {Hathi}, {Koekemoer}, {Koo}, {Lucas},
  {Nonino}, {Rix}, {Renzini}, {Rosario}, {Santini}, {Scarlata}, {Sommariva},
  {Stark}, {van der Wel}, {Vanzella}, {Wild}, {Yan}, \&
  {Zibetti}}]{Fontana2014a}
{Fontana}, A., {Dunlop}, J.~S., {Paris}, D., {et~al.} 2014, \aap, 570, A11,
  \dodoi{10.1051/0004-6361/201423543}

\bibitem[{{Fynbo} {et~al.}(2006){Fynbo}, {Watson}, {Th{\"o}ne}, {Sollerman},
  {Bloom}, {Davis}, {Hjorth}, {Jakobsson}, {J{\o}rgensen}, {Graham},
  {Fruchter}, {Bersier}, {Kewley}, {Cassan}, {Castro Cer{\'o}n}, {Foley},
  {Gorosabel}, {Hinse}, {Horne}, {Jensen}, {Klose}, {Kocevski}, {Marquette},
  {Perley}, {Ramirez-Ruiz}, {Stritzinger}, {Vreeswijk}, {Wijers}, {Woller},
  {Xu}, \& {Zub}}]{Fynbo2006a}
{Fynbo}, J. P.~U., {Watson}, D., {Th{\"o}ne}, C.~C., {et~al.} 2006, \nat, 444,
  1047, \dodoi{10.1038/nature05375}

\bibitem[{{Gaia Collaboration} {et~al.}(2018){Gaia Collaboration}, {Brown},
  {Vallenari}, {Prusti}, {de Bruijne}, {Babusiaux}, {Bailer-Jones}, {Biermann},
  {Evans}, {Eyer}, \& et~al.}]{Gaia2018a}
{Gaia Collaboration}, {Brown}, A.~G.~A., {Vallenari}, A., {et~al.} 2018, \aap,
  616, A1, \dodoi{10.1051/0004-6361/201833051}

\bibitem[{{Gal-Yam} {et~al.}(2006){Gal-Yam}, {Fox}, {Price}, {Ofek}, {Davis},
  {Leonard}, {Soderberg}, {Schmidt}, {Lewis}, {Peterson}, {Kulkarni}, {Berger},
  {Cenko}, {Sari}, {Sharon}, {Frail}, {Moon}, {Brown}, {Cucchiara}, {Harrison},
  {Piran}, {Persson}, {McCarthy}, {Penprase}, {Chevalier}, \&
  {MacFadyen}}]{Gal-Yam2006a}
{Gal-Yam}, A., {Fox}, D.~B., {Price}, P.~A., {et~al.} 2006, \nat, 444, 1053,
  \dodoi{10.1038/nature05373}

\bibitem[{{Galama} {et~al.}(1998){Galama}, {Vreeswijk}, {van Paradijs},
  {Kouveliotou}, {Augusteijn}, {B{\"o}hnhardt}, {Brewer}, {Doublier},
  {Gonzalez}, {Leibundgut}, {Lidman}, {Hainaut}, {Patat}, {Heise}, {in't Zand},
  {Hurley}, {Groot}, {Strom}, {Mazzali}, {Iwamoto}, {Nomoto}, {Umeda},
  {Nakamura}, {Young}, {Suzuki}, {Shigeyama}, {Koshut}, {Kippen}, {Robinson},
  {de Wildt}, {Wijers}, {Tanvir}, {Greiner}, {Pian}, {Palazzi}, {Frontera},
  {Masetti}, {Nicastro}, {Feroci}, {Costa}, {Piro}, {Peterson}, {Tinney},
  {Boyle}, {Cannon}, {Stathakis}, {Sadler}, {Begam}, \& {Ianna}}]{Galama1998a}
{Galama}, T.~J., {Vreeswijk}, P.~M., {van Paradijs}, J., {et~al.} 1998, \nat,
  395, 670, \dodoi{10.1038/27150}

\bibitem[{{Gao} {et~al.}(2017){Gao}, {Zhang}, {L{\"u}}, \& {Li}}]{Gao2017a}
{Gao}, H., {Zhang}, B., {L{\"u}}, H.-J., \& {Li}, Y. 2017, \apj, 837, 50,
  \dodoi{10.3847/1538-4357/aa5be3}

\bibitem[{{Gehrels} {et~al.}(2008){Gehrels}, {Barthelmy}, {Burrows},
  {Cannizzo}, {Chincarini}, {Fenimore}, {Kouveliotou}, {O'Brien}, {Palmer},
  {Racusin}, {Roming}, {Sakamoto}, {Tueller}, {Wijers}, \&
  {Zhang}}]{Gehrels2008ApJ}
{Gehrels}, N., {Barthelmy}, S.~D., {Burrows}, D.~N., {et~al.} 2008, \apj, 689,
  1161, \dodoi{10.1086/592766}

\bibitem[{{Ghirlanda} {et~al.}(2021){Ghirlanda}, {Salvaterra}, {Toffano},
  {Ronchini}, {Guidorzi}, {Oganesyan}, {Ascenzi}, {Bernardini}, {Camisasca},
  {Mereghetti}, {Nava}, {Ravasio}, {Branchesi}, {Castro-Tirado}, {Amati},
  {Blain}, {Bozzo}, {O'Brien}, {G{\"o}tz}, {Le Floch}, {Osborne}, {Rosati},
  {Stratta}, {Tanvir}, {Bogomazov}, {D'Avanzo}, {Hafizi}, {Mandhai},
  {Melandri}, {Peer}, {Topinka}, {Vergani}, \& {Zane}}]{Ghirlanda21_WPTHESEUS}
{Ghirlanda}, G., {Salvaterra}, R., {Toffano}, M., {et~al.} 2021, Experimental
  Astronomy, \dodoi{10.1007/s10686-021-09763-3}

\bibitem[{{Ghisellini} {et~al.}(2020){Ghisellini}, {Ghirlanda}, {Oganesyan},
  {Ascenzi}, {Nava}, {Celotti}, {Salafia}, {Ravasio}, \&
  {Ronchi}}]{Ghisellini2020a}
{Ghisellini}, G., {Ghirlanda}, G., {Oganesyan}, G., {et~al.} 2020, \aap, 636,
  A82, \dodoi{10.1051/0004-6361/201937244}

\bibitem[{{Giallongo} {et~al.}(2008){Giallongo}, {Ragazzoni}, {Grazian},
  {Baruffolo}, {Beccari}, {de Santis}, {Diolaiti}, {di Paola}, {Farinato},
  {Fontana}, {Gallozzi}, {Gasparo}, {Gentile}, {Green}, {Hill}, {Kuhn},
  {Pasian}, {Pedichini}, {Radovich}, {Salinari}, {Smareglia}, {Speziali},
  {Testa}, {Thompson}, {Vernet}, \& {Wagner}}]{Giallongo2008a}
{Giallongo}, E., {Ragazzoni}, R., {Grazian}, A., {et~al.} 2008, \aap, 482, 349,
  \dodoi{10.1051/0004-6361:20078402}

\bibitem[{{Gompertz} {et~al.}(2018){Gompertz}, {Levan}, {Tanvir}, {Hjorth},
  {Covino}, {Evans}, {Fruchter}, {Gonz{\'a}lez-Fern{\'a}ndez}, {Jin}, {Lyman},
  {Oates}, {O'Brien}, \& {Wiersema}}]{Gompertz2018a}
{Gompertz}, B.~P., {Levan}, A.~J., {Tanvir}, N.~R., {et~al.} 2018, \apj, 860,
  62, \dodoi{10.3847/1538-4357/aac206}

\bibitem[{{Gonz{\'a}lez-Gait{\'a}n} {et~al.}(2015){Gonz{\'a}lez-Gait{\'a}n},
  {Tominaga}, {Molina}, {Galbany}, {Bufano}, {Anderson}, {Gutierrez},
  {F{\"o}rster}, {Pignata}, {Bersten}, {Howell}, {Sullivan}, {Carlberg}, {de
  Jaeger}, {Hamuy}, {Baklanov}, \& {Blinnikov}}]{Gonzalez-Gaitan2015a}
{Gonz{\'a}lez-Gait{\'a}n}, S., {Tominaga}, N., {Molina}, J., {et~al.} 2015,
  \mnras, 451, 2212, \dodoi{10.1093/mnras/stv1097}

\bibitem[{{Granot} \& {Sari}(2002)}]{GranotSari2002a}
{Granot}, J., \& {Sari}, R. 2002, \apj, 568, 820, \dodoi{10.1086/338966}

\bibitem[{{Greiner} {et~al.}(2009){Greiner}, {Kr{\"u}hler}, {Fynbo}, {Rossi},
  {Schwarz}, {Klose}, {Savaglio}, {Tanvir}, {McBreen}, {Totani}, {Zhang}, {Wu},
  {Watson}, {Barthelmy}, {Beardmore}, {Ferrero}, {Gehrels}, {Kann}, {Kawai},
  {Yolda{\c{s}}}, {M{\'e}sz{\'a}ros}, {Milvang-Jensen}, {Oates}, {Pierini},
  {Schady}, {Toma}, {Vreeswijk}, {Yolda{\c{s}}}, {Zhang}, {Afonso}, {Aoki},
  {Burrows}, {Clemens}, {Filgas}, {Haiman}, {Hartmann}, {Hasinger}, {Hjorth},
  {Jehin}, {Levan}, {Liang}, {Malesani}, {Pyo}, {Schulze}, {Szokoly}, {Terada},
  \& {Wiersema}}]{Greiner2009a}
{Greiner}, J., {Kr{\"u}hler}, T., {Fynbo}, J.~P.~U., {et~al.} 2009, \apj, 693,
  1610, \dodoi{10.1088/0004-637X/693/2/1610}

\bibitem[{{Greiner} {et~al.}(2015){Greiner}, {Mazzali}, {Kann}, {Kr{\"u}hler},
  {Pian}, {Prentice}, {Olivares E.}, {Rossi}, {Klose}, {Taubenberger}, {Knust},
  {Afonso}, {Ashall}, {Bolmer}, {Delvaux}, {Diehl}, {Elliott}, {Filgas},
  {Fynbo}, {Graham}, {Guelbenzu}, {Kobayashi}, {Leloudas}, {Savaglio},
  {Schady}, {Schmidl}, {Schweyer}, {Sudilovsky}, {Tanga}, {Updike}, {van
  Eerten}, \& {Varela}}]{Greiner2015Nat}
{Greiner}, J., {Mazzali}, P.~A., {Kann}, D.~A., {et~al.} 2015, \nat, 523, 189,
  \dodoi{10.1038/nature14579}

\bibitem[{{Gupta} {et~al.}(2020){Gupta}, {Sharma}, {Vibhute}, {Bhattacharya},
  {Bhalerao}, {Rao}, {Vadawale}, \& {AstroSat CZTI
  Collaboration}}]{Gupta2020GCNastrosat}
{Gupta}, S., {Sharma}, V., {Vibhute}, A., {et~al.} 2020, GRB Coordinates
  Network, 28288, 1

\bibitem[{{Hamburg} {et~al.}(2020){Hamburg}, {Malacaria}, {Meegan}, \& {Fermi
  GBM Team}}]{Hamburg2020a}
{Hamburg}, R., {Malacaria}, C., {Meegan}, C., \& {Fermi GBM Team}. 2020, GRB
  Coordinates Network, 29140, 1

\bibitem[{{Horv{\'a}th} \& {T{\'o}th}(2016)}]{Horvath2016a}
{Horv{\'a}th}, I., \& {T{\'o}th}, B.~G. 2016, \apss, 361, 155,
  \dodoi{10.1007/s10509-016-2748-6}

\bibitem[{{Hu} {et~al.}(2021){Hu}, {Castro-Tirado}, {Kumar}, {Gupta}, {Valeev},
  {Pandey}, {Kann}, {Castell{\'o}n}, {Agudo}, {Aryan}, {Caballero-Garc{\'\i}a},
  {Guziy}, {Martin-Carrillo}, {Oates}, {Pian}, {S{\'a}nchez-Ram{\'\i}rez},
  {Sokolov}, \& {Zhang}}]{Hu2021a}
{Hu}, Y.~D., {Castro-Tirado}, A.~J., {Kumar}, A., {et~al.} 2021, \aap, 646,
  A50, \dodoi{10.1051/0004-6361/202039349}

\bibitem[{{Huang} {et~al.}(2005){Huang}, {Urata}, {Filippenko}, {Hu}, {Ip},
  {Kuo}, {Li}, {Lin}, {Lin}, {Makishima}, {Onda}, {Qiu}, \&
  {Tamagawa}}]{Huang2005a}
{Huang}, K.~Y., {Urata}, Y., {Filippenko}, A.~V., {et~al.} 2005, \apjl, 628,
  L93, \dodoi{10.1086/432612}

\bibitem[{{Hullinger} {et~al.}(2006){Hullinger}, {Barbier}, {Barthelmy},
  {Cummings}, {Fenimore}, {Gehrels}, {Krimm}, {Koss}, {Markwardt}, {Palmer},
  {Parsons}, {Sakamoto}, {Sato}, {Stamatikos}, \&
  {Tueller}}]{HullingerGCN2006a}
{Hullinger}, D., {Barbier}, L., {Barthelmy}, S., {et~al.} 2006, GRB Coordinates
  Network, 5142, 1

\bibitem[{{Hunt} {et~al.}(2014){Hunt}, {Palazzi}, {Micha{\l}owski}, {Rossi},
  {Savaglio}, {Basa}, {Berta}, {Bianchi}, {Covino}, {D'Elia}, {Ferrero},
  {G{\"o}tz}, {Greiner}, {Klose}, {Le Borgne}, {Le Floc'h}, {Pian},
  {Piranomonte}, {Schady}, \& {Vergani}}]{Hunt2014a}
{Hunt}, L.~K., {Palazzi}, E., {Micha{\l}owski}, M.~J., {et~al.} 2014, \aap,
  565, A112, \dodoi{10.1051/0004-6361/201323340}

\bibitem[{{Hurley} {et~al.}(2020){Hurley}, {Mitrofanov}, {Golovin}, {Litvak},
  {Sanin}, {Kozlova}, {Golenetskii}, {Aptekar}, {Frederiks}, {Svinkin},
  {Cline}, {Goldstein}, {Briggs}, {Wilson-Hodge}, {von Kienlin}, {Zhang},
  {Rau}, {Savchenko}, {E. Bozzo}, {Ferrigno}, {Barthelmy}, {Cummings}, {Krimm},
  {Palmer}, {Boynton}, {Fellows}, {Harshman}, {Enos}, \&
  {Starr}}]{Hurley2020GCNipn}
{Hurley}, K., {Mitrofanov}, I.~G., {Golovin}, D., {et~al.} 2020, GRB
  Coordinates Network, 28291, 1

\bibitem[{{Ilbert} {et~al.}(2006){Ilbert}, {Arnouts}, {McCracken},
  {Bolzonella}, {Bertin}, {Le F{\`e}vre}, {Mellier}, {Zamorani}, {Pell{\`o}},
  {Iovino}, {Tresse}, {Le Brun}, {Bottini}, {Garilli}, {Maccagni}, {Picat},
  {Scaramella}, {Scodeggio}, {Vettolani}, {Zanichelli}, {Adami}, {Bardelli},
  {Cappi}, {Charlot}, {Ciliegi}, {Contini}, {Cucciati}, {Foucaud}, {Franzetti},
  {Gavignaud}, {Guzzo}, {Marano}, {Marinoni}, {Mazure}, {Meneux}, {Merighi},
  {Paltani}, {Pollo}, {Pozzetti}, {Radovich}, {Zucca}, {Bondi}, {Bongiorno},
  {Busarello}, {de La Torre}, {Gregorini}, {Lamareille}, {Mathez}, {Merluzzi},
  {Ripepi}, {Rizzo}, \& {Vergani}}]{Ilbert2006a}
{Ilbert}, O., {Arnouts}, S., {McCracken}, H.~J., {et~al.} 2006, \aap, 457, 841,
  \dodoi{10.1051/0004-6361:20065138}

\bibitem[{{Im} {et~al.}(2010){Im}, {Ko}, {Cho}, {Choi}, {Jeon}, {Lee}, \&
  {Ibrahimov}}]{Im2010}
{Im}, M., {Ko}, J., {Cho}, Y., {et~al.} 2010, Journal of Korean Astronomical
  Society, 43, 75, \dodoi{10.5303/JKAS.2010.43.3.075}

\bibitem[{{Izzo} {et~al.}(2019){Izzo}, {de Ugarte Postigo}, {Maeda},
  {Th{\"o}ne}, {Kann}, {Della Valle}, {Sagues Carracedo}, {Micha{\l}owski},
  {Schady}, {Schmidl}, {Selsing}, {Starling}, {Suzuki}, {Bensch}, {Bolmer},
  {Campana}, {Cano}, {Covino}, {Fynbo}, {Hartmann}, {Heintz}, {Hjorth},
  {Japelj}, {Kami{\'n}ski}, {Kaper}, {Kouveliotou}, {Kru{\.Z}y{\'n}ski},
  {Kwiatkowski}, {Leloudas}, {Levan}, {Malesani}, {Micha{\l}owski},
  {Piranomonte}, {Pugliese}, {Rossi}, {S{\'a}nchez-Ram{\'\i}rez}, {Schulze},
  {Steeghs}, {Tanvir}, {Ulaczyk}, {Vergani}, \& {Wiersema}}]{Izzo2019a}
{Izzo}, L., {de Ugarte Postigo}, A., {Maeda}, K., {et~al.} 2019, \nat, 565,
  324, \dodoi{10.1038/s41586-018-0826-3}

\bibitem[{{Jakobsson} {et~al.}(2004){Jakobsson}, {Hjorth}, {Fynbo}, {Watson},
  {Pedersen}, {Bj{\"o}rnsson}, \& {Gorosabel}}]{Jakobsson2004a}
{Jakobsson}, P., {Hjorth}, J., {Fynbo}, J.~P.~U., {et~al.} 2004, \apjl, 617,
  L21, \dodoi{10.1086/427089}

\bibitem[{{Japelj} {et~al.}(2016){Japelj}, {Vergani}, {Salvaterra}, {D'Avanzo},
  {Mannucci}, {Fernandez-Soto}, {Boissier}, {Hunt}, {Atek},
  {Rodr{\'\i}guez-Mu{\~n}oz}, {Scodeggio}, {Cristiani}, {Le Floc'h}, {Flores},
  {Gallego}, {Ghirlanda}, {Gomboc}, {Hammer}, {Perley}, {Pescalli},
  {Petitjean}, {Puech}, {Rafelski}, \& {Tagliaferri}}]{Japeli2016a}
{Japelj}, J., {Vergani}, S.~D., {Salvaterra}, R., {et~al.} 2016, \aap, 590,
  A129, \dodoi{10.1051/0004-6361/201628314}

\bibitem[{{Jin} {et~al.}(2015){Jin}, {Li}, {Cano}, {Covino}, {Fan}, \&
  {Wei}}]{Jin2015a}
{Jin}, Z.-P., {Li}, X., {Cano}, Z., {et~al.} 2015, \apjl, 811, L22,
  \dodoi{10.1088/2041-8205/811/2/L22}

\bibitem[{{Jin} {et~al.}(2021){Jin}, {Zhou}, {Covino}, {Liao}, {Li}, {Lei},
  {D'Avanzo}, {Fan}, \& {Wei}}]{Jin2021a}
{Jin}, Z.-P., {Zhou}, H., {Covino}, S., {et~al.} 2021, arXiv e-prints,
  arXiv:2109.07694.
\newblock \doarXiv{2109.07694}

\bibitem[{{Kann} {et~al.}(2006){Kann}, {Klose}, \& {Zeh}}]{Kann2006ApJ}
{Kann}, D.~A., {Klose}, S., \& {Zeh}, A. 2006, \apj, 641, 993,
  \dodoi{10.1086/500652}

\bibitem[{{Kann} {et~al.}(2021){Kann}, {Oates}, {Rossi}, {Klose}, {Blazek},
  {Ag{\"u}{\'\i} Fern{\'a}ndez}, {de Ugarte Postigo}, \&
  {Th{\"o}ne}}]{Kann2021a}
{Kann}, D.~A., {Oates}, S.~R., {Rossi}, A., {et~al.} 2021, arXiv e-prints,
  arXiv:2110.00110.
\newblock \doarXiv{2110.00110}

\bibitem[{{Kann} {et~al.}(2010){Kann}, {Klose}, {Zhang}, {Malesani}, {Nakar},
  {Pozanenko}, {Wilson}, {Butler}, {Jakobsson}, {Schulze}, {Andreev},
  {Antonelli}, {Bikmaev}, {Biryukov}, {B{\"o}ttcher}, {Burenin}, {Castro
  Cer{\'o}n}, {Castro-Tirado}, {Chincarini}, {Cobb}, {Covino}, {D'Avanzo},
  {D'Elia}, {Della Valle}, {de Ugarte Postigo}, {Efimov}, {Ferrero}, {Fugazza},
  {Fynbo}, {G{\aa}lfalk}, {Grundahl}, {Gorosabel}, {Gupta}, {Guziy}, {Hafizov},
  {Hjorth}, {Holhjem}, {Ibrahimov}, {Im}, {Israel}, {Je{\'l}inek}, {Jensen},
  {Karimov}, {Khamitov}, {Kizilo{\v g}lu}, {Klunko}, {Kub{\'a}nek}, {Kutyrev},
  {Laursen}, {Levan}, {Mannucci}, {Martin}, {Mescheryakov}, {Mirabal},
  {Norris}, {Ovaldsen}, {Paraficz}, {Pavlenko}, {Piranomonte}, {Rossi},
  {Rumyantsev}, {Salinas}, {Sergeev}, {Sharapov}, {Sollerman}, {Stecklum},
  {Stella}, {Tagliaferri}, {Tanvir}, {Telting}, {Testa}, {Updike}, {Volnova},
  {Watson}, {Wiersema}, \& {Xu}}]{Kann2010a}
{Kann}, D.~A., {Klose}, S., {Zhang}, B., {et~al.} 2010, \apj, 720, 1513,
  \dodoi{10.1088/0004-637X/720/2/1513}

\bibitem[{{Kann} {et~al.}(2011){Kann}, {Klose}, {Zhang}, {Covino}, {Butler},
  {Malesani}, {Nakar}, {Wilson}, {Antonelli}, {Chincarini}, {Cobb}, {D'Avanzo},
  {D'Elia}, {Della Valle}, {Ferrero}, {Fugazza}, {Gorosabel}, {Israel},
  {Mannucci}, {Piranomonte}, {Schulze}, {Stella}, {Tagliaferri}, \&
  {Wiersema}}]{Kann2011a}
---. 2011, \apj, 734, 96, \dodoi{10.1088/0004-637X/734/2/96}

\bibitem[{{Kann} {et~al.}(2019){Kann}, {Schady}, {Olivares E.}, {Klose},
  {Rossi}, {Perley}, {Kr{\"u}hler}, {Greiner}, {Nicuesa Guelbenzu}, {Elliott},
  {Knust}, {Filgas}, {Pian}, {Mazzali}, {Fynbo}, {Leloudas}, {Afonso},
  {Delvaux}, {Graham}, {Rau}, {Schmidl}, {Schulze}, {Tanga}, {Updike}, \&
  {Varela}}]{Kann2019AA}
{Kann}, D.~A., {Schady}, P., {Olivares E.}, F., {et~al.} 2019, \aap, 624, A143,
  \dodoi{10.1051/0004-6361/201629162}

\bibitem[{Kasliwal {et~al.}(2008)Kasliwal, Cenko, Kulkarni, Cameron, Nakar,
  Ofek, Rau, Soderberg, Campana, Bloom, Perley, Pollack, Barthelmy, Cummings,
  Gehrels, Krimm, Markwardt, Sato, Chandra, Frail, Fox, Price, Berger,
  Grebenev, Krivonos, \& Sunyaev}]{Kasliwal2008a}
Kasliwal, M.~M., Cenko, S.~B., Kulkarni, S.~R., {et~al.} 2008, The
  Astrophysical Journal, 678, 1127, \dodoi{10.1086/526407}

\bibitem[{{Kawaguchi} {et~al.}(2020){Kawaguchi}, {Shibata}, \&
  {Tanaka}}]{Kawaguchi2020}
{Kawaguchi}, K., {Shibata}, M., \& {Tanaka}, M. 2020, \apj, 889, 171,
  \dodoi{10.3847/1538-4357/ab61f6}

\bibitem[{{Kelly} {et~al.}(2014){Kelly}, {Filippenko}, {Modjaz}, \&
  {Kocevski}}]{Kelly2014a}
{Kelly}, P.~L., {Filippenko}, A.~V., {Modjaz}, M., \& {Kocevski}, D. 2014,
  \apj, 789, 23, \dodoi{10.1088/0004-637X/789/1/23}

\bibitem[{{Kennicutt}(1998)}]{Kennicutt1998a}
{Kennicutt}, Robert~C., J. 1998, \araa, 36, 189,
  \dodoi{10.1146/annurev.astro.36.1.189}

\bibitem[{{Klose} {et~al.}(2019){Klose}, {Schmidl}, {Kann}, {Nicuesa
  Guelbenzu}, {Schulze}, {Greiner}, {Olivares E.}, {Kr{\"u}hler}, {Schady},
  {Afonso}, {Filgas}, {Fynbo}, {Rau}, {Rossi}, {Takats}, {Tanga}, {Updike}, \&
  {Varela}}]{Klose2019a}
{Klose}, S., {Schmidl}, S., {Kann}, D.~A., {et~al.} 2019, \aap, 622, A138,
  \dodoi{10.1051/0004-6361/201832728}

\bibitem[{{Korobkin} {et~al.}(2021){Korobkin}, {Wollaeger}, {Fryer},
  {Hungerford}, {Rosswog}, {Fontes}, {Mumpower}, {Chase}, {Even}, {Miller},
  {Misch}, \& {Lippuner}}]{Korobkin:2021}
{Korobkin}, O., {Wollaeger}, R.~T., {Fryer}, C.~L., {et~al.} 2021, \apj, 910,
  116, \dodoi{10.3847/1538-4357/abe1b5}

\bibitem[{{Kouveliotou} {et~al.}(1993){Kouveliotou}, {Meegan}, {Fishman},
  {Bhat}, {Briggs}, {Koshut}, {Paciesas}, \& {Pendleton}}]{Kouveliotou1993a}
{Kouveliotou}, C., {Meegan}, C.~A., {Fishman}, G.~J., {et~al.} 1993, \apjl,
  413, L101, \dodoi{10.1086/186969}

\bibitem[{{Krimm} {et~al.}(2020){Krimm}, {Barthelmy}, {Laha}, {Lien},
  {Markwardt}, {Page}, {Palmer}, {Sakamoto}, {Stamatikos}, \&
  {Ukwatta}}]{Krimm2020a}
{Krimm}, H.~A., {Barthelmy}, S.~D., {Laha}, S., {et~al.} 2020, GRB Coordinates
  Network, 29139, 1

\bibitem[{{Labrie} {et~al.}(2019){Labrie}, {Anderson}, {C{\'a}rdenes},
  {Simpson}, \& {Turner}}]{dragons2019}
{Labrie}, K., {Anderson}, K., {C{\'a}rdenes}, R., {Simpson}, C., \& {Turner},
  J. E.~H. 2019, in Astronomical Society of the Pacific Conference Series, Vol.
  523, Astronomical Data Analysis Software and Systems XXVII, ed. P.~J.
  {Teuben}, M.~W. {Pound}, B.~A. {Thomas}, \& E.~M. {Warner}, 321

\bibitem[{{Lamareille}(2010)}]{Lamareille2010a}
{Lamareille}, F. 2010, \aap, 509, A53, \dodoi{10.1051/0004-6361/200913168}

\bibitem[{{Maiolino} \& {Mannucci}(2019)}]{MaiolinoMannucci2019a}
{Maiolino}, R., \& {Mannucci}, F. 2019, \aapr, 27, 3,
  \dodoi{10.1007/s00159-018-0112-2}

\bibitem[{{Maiorano} {et~al.}(2018){Maiorano}, {Amati}, {Rossi}, {Stratta},
  {Palazzi}, \& {Nicastro}}]{Maiorano2018a}
{Maiorano}, E., {Amati}, L., {Rossi}, A., {et~al.} 2018, \memsai, 89, 181.
\newblock \doarXiv{1802.01679}

\bibitem[{{Mangan} {et~al.}(2020){Mangan}, {Dunwoody}, {Meegan}, \& {Fermi GBM
  Team}}]{Mangan2020GCNgbm}
{Mangan}, J., {Dunwoody}, R., {Meegan}, C., \& {Fermi GBM Team}. 2020, GRB
  Coordinates Network, 28287, 1

\bibitem[{{Marocco} {et~al.}(2011){Marocco}, {Hache}, \&
  {Lamareille}}]{Marocco2011a}
{Marocco}, J., {Hache}, E., \& {Lamareille}, F. 2011, \aap, 531, A71,
  \dodoi{10.1051/0004-6361/201016143}

\bibitem[{{Mazets} {et~al.}(1981){Mazets}, {Golenetskii}, {Ilinskii}, {Panov},
  {Aptekar}, {Gurian}, {Proskura}, {Sokolov}, {Sokolova}, \&
  {Kharitonova}}]{Mazets1981a}
{Mazets}, E.~P., {Golenetskii}, S.~V., {Ilinskii}, V.~N., {et~al.} 1981, \apss,
  80, 3, \dodoi{10.1007/BF00649140}

\bibitem[{{Mazzali} {et~al.}(2021){Mazzali}, {Pian}, {Bufano}, \&
  {Ashall}}]{Mazzali2021a}
{Mazzali}, P.~A., {Pian}, E., {Bufano}, F., \& {Ashall}, C. 2021, \mnras, 505,
  4106, \dodoi{10.1093/mnras/stab1594}

\bibitem[{{McBreen} {et~al.}(2008){McBreen}, {Foley}, {Watson}, {Hanlon},
  {Malesani}, {Fynbo}, {Kann}, {Gehrels}, {McGlynn}, \&
  {Palmer}}]{McBreen2008a}
{McBreen}, S., {Foley}, S., {Watson}, D., {et~al.} 2008, \apjl, 677, L85,
  \dodoi{10.1086/588189}

\bibitem[{{Melandri} {et~al.}(2019){Melandri}, {Malesani}, {Izzo}, {Japelj},
  {Vergani}, {Schady}, {Sagu{\'e}s Carracedo}, {de Ugarte Postigo}, {Anderson},
  {Barbarino}, {Bolmer}, {Breeveld}, {Calissendorff}, {Campana}, {Cano},
  {Carini}, {Covino}, {D'Avanzo}, {D'Elia}, {della Valle}, {De Pasquale},
  {Fynbo}, {Gromadzki}, {Hammer}, {Hartmann}, {Heintz}, {Inserra}, {Jakobsson},
  {Kann}, {Kotilainen}, {Maguire}, {Masetti}, {Nicholl}, {Olivares E},
  {Pugliese}, {Rossi}, {Salvaterra}, {Sollerman}, {Stone}, {Tagliaferri},
  {Tomasella}, {Th{\"o}ne}, {Xu}, \& {Young}}]{Melandri2019a}
{Melandri}, A., {Malesani}, D.~B., {Izzo}, L., {et~al.} 2019, \mnras, 490,
  5366, \dodoi{10.1093/mnras/stz2900}

\bibitem[{{Melandri} {et~al.}(2021){Melandri}, {Izzo}, {Pian}, {Malesani},
  {Della Valle}, {Rossi}, {D'Avanzo}, {Guetta}, {Mazzali}, {Benetti},
  {Masetti}, {Palazzi}, {Savaglio}, {Amati}, {Antonelli}, {Ashall},
  {Bernardini}, {Campana}, {Carini}, {Covino}, {D'Elia}, {de Ugarte Postigo},
  {De Pasquale}, {Filippenko}, {Fruchter}, {Fynbo}, {Giunta}, {Hartmann},
  {Jakobsson}, {Japelj}, {Jonker}, {Kann}, {Lamb}, {Levan}, {Martin-Carrillo},
  {Moller}, {Piranomonte}, {Pugliese}, {Salvaterra}, {Schulze}, {Starling},
  {Stella}, {Tagliaferri}, {Tanvir}, \& {Watson}}]{Melandri2022AA}
{Melandri}, A., {Izzo}, L., {Pian}, E., {et~al.} 2021, arXiv e-prints,
  arXiv:2112.04759.
\newblock \doarXiv{2112.04759}

\bibitem[{{Metzger} \& {Piro}(2014)}]{MetzgerPiro2014a}
{Metzger}, B.~D., \& {Piro}, A.~L. 2014, \mnras, 439, 3916,
  \dodoi{10.1093/mnras/stu247}

\bibitem[{{Micha{\l}owski} {et~al.}(2018){Micha{\l}owski}, {Xu}, {Stevens},
  {Levan}, {Yang}, {Paragi}, {Kamble}, {Tsai}, {Dannerbauer}, {van der Horst},
  {Shao}, {Crosby}, {Gentile}, {Stanway}, {Wiersema}, {Fynbo}, {Tanvir},
  {Kamphuis}, {Garrett}, \& {Bartczak}}]{Michalowski2018a}
{Micha{\l}owski}, M.~J., {Xu}, D., {Stevens}, J., {et~al.} 2018, \aap, 616,
  A169, \dodoi{10.1051/0004-6361/201629942}

\bibitem[{{Minaev} \& {Pozanenko}(2020{\natexlab{a}})}]{MinaevPozanenko2020a}
{Minaev}, P.~Y., \& {Pozanenko}, A.~S. 2020{\natexlab{a}}, \mnras, 492, 1919,
  \dodoi{10.1093/mnras/stz3611}

\bibitem[{{Minaev} \& {Pozanenko}(2020{\natexlab{b}})}]{MinaevPozanenko2020b}
---. 2020{\natexlab{b}}, Astronomy Letters, 46, 573,
  \dodoi{10.1134/S1063773720090042}

\bibitem[{{Mink}(2019)}]{wcstools2019}
{Mink}, J. 2019, in Astronomical Society of the Pacific Conference Series, Vol.
  523, Astronomical Data Analysis Software and Systems XXVII, ed. P.~J.
  {Teuben}, M.~W. {Pound}, B.~A. {Thomas}, \& E.~M. {Warner}, 281

\bibitem[{{Modjaz} {et~al.}(2020){Modjaz}, {Bianco}, {Siwek}, {Huang},
  {Perley}, {Fierroz}, {Liu}, {Arcavi}, {Gal-Yam}, {Filippenko},
  {Blagorodnova}, {Cenko}, {Kasliwal}, {Kulkarni}, {Schulze}, {Taggart}, \&
  {Zheng}}]{Modjaz2020a}
{Modjaz}, M., {Bianco}, F.~B., {Siwek}, M., {et~al.} 2020, \apj, 892, 153,
  \dodoi{10.3847/1538-4357/ab4185}

\bibitem[{{Nicuesa Guelbenzu} {et~al.}(2011){Nicuesa Guelbenzu}, {Klose},
  {Rossi}, {Kann}, {Kr{\"u}hler}, {Greiner}, {Rau}, {Olivares E.}, {Afonso},
  {Filgas}, {K{\"u}pc{\"u} Yolda{\c s}}, {McBreen}, {Nardini}, {Schady},
  {Schmidl}, {Updike}, \& {Yolda{\c s}}}]{Nicuesa2011a}
{Nicuesa Guelbenzu}, A., {Klose}, S., {Rossi}, A., {et~al.} 2011, \aap, 531,
  L6, \dodoi{10.1051/0004-6361/201116657}

\bibitem[{{Nicuesa Guelbenzu} {et~al.}(2012){Nicuesa Guelbenzu}, {Klose},
  {Greiner}, {Kann}, {Kr{\"u}hler}, {Rossi}, {Schulze}, {Afonso}, {Elliott},
  {Filgas}, {Hartmann}, {K{\"u}pc{\"u} Yolda{\c s}}, {McBreen}, {Nardini},
  {Olivares E.}, {Rau}, {Schmidl}, {Schady}, {Sudilovsky}, {Updike}, \&
  {Yolda{\c s}}}]{Nicuesa2012a}
{Nicuesa Guelbenzu}, A., {Klose}, S., {Greiner}, J., {et~al.} 2012, \aap, 548,
  A101, \dodoi{10.1051/0004-6361/201219551}

\bibitem[{{Norris} {et~al.}(1996){Norris}, {Nemiroff}, {Bonnell}, {Scargle},
  {Kouveliotou}, {Paciesas}, {Meegan}, \& {Fishman}}]{Norris1996a}
{Norris}, J.~P., {Nemiroff}, R.~J., {Bonnell}, J.~T., {et~al.} 1996, \apj, 459,
  393, \dodoi{10.1086/176902}

\bibitem[{{Nysewander} {et~al.}(2009){Nysewander}, {Fruchter}, \&
  {Pe'er}}]{Nysewander2009ApJ}
{Nysewander}, M., {Fruchter}, A.~S., \& {Pe'er}, A. 2009, \apj, 701, 824,
  \dodoi{10.1088/0004-637X/701/1/824}

\bibitem[{{O'Connor} {et~al.}(2021){O'Connor}, {Troja}, {Dichiara}, {Chase},
  {Ryan}, {Cenko}, {Fryer}, {Ricci}, {Marshall}, {Kouveliotou}, {Wollaeger},
  {Fontes}, {Korobkin}, {Gatkine}, {Kutyrev}, {Veilleux}, {Kawai}, \&
  {Sakamoto}}]{Oconnor2021a}
{O'Connor}, B., {Troja}, E., {Dichiara}, S., {et~al.} 2021, \mnras, 502, 1279,
  \dodoi{10.1093/mnras/stab132}

\bibitem[{{Ofek} {et~al.}(2007){Ofek}, {Cenko}, {Gal-Yam}, {Fox}, {Nakar},
  {Rau}, {Frail}, {Kulkarni}, {Price}, {Schmidt}, {Soderberg}, {Peterson},
  {Berger}, {Sharon}, {Shemmer}, {Penprase}, {Chevalier}, {Brown}, {Burrows},
  {Gehrels}, {Harrison}, {Holland}, {Mangano}, {McCarthy}, {Moon}, {Nousek},
  {Persson}, {Piran}, \& {Sari}}]{Ofek2007a}
{Ofek}, E.~O., {Cenko}, S.~B., {Gal-Yam}, A., {et~al.} 2007, \apj, 662, 1129,
  \dodoi{10.1086/518082}

\bibitem[{{Olivares E.} {et~al.}(2012){Olivares E.}, {Greiner}, {Schady},
  {Rau}, {Klose}, {Kr{\"u}hler}, {Afonso}, {Updike}, {Nardini}, {Filgas},
  {Nicuesa Guelbenzu}, {Clemens}, {Elliott}, {Kann}, {Rossi}, \&
  {Sudilovsky}}]{Olivares2012a}
{Olivares E.}, F., {Greiner}, J., {Schady}, P., {et~al.} 2012, \aap, 539, A76,
  \dodoi{10.1051/0004-6361/201117929}

\bibitem[{{Osterbrock}(1989)}]{Osterbrock1989}
{Osterbrock}, D.~E. 1989, {Astrophysics of gaseous nebulae and active galactic
  nuclei} (University Science Books)

\bibitem[{{Paul} {et~al.}(2011){Paul}, {Wei}, {Basa}, \& {Zhang}}]{svom2011a}
{Paul}, J., {Wei}, J., {Basa}, S., \& {Zhang}, S.-N. 2011, Comptes Rendus
  Physique, 12, 298, \dodoi{10.1016/j.crhy.2011.01.009}

\bibitem[{{Pei}(1992)}]{Pei1992a}
{Pei}, Y.~C. 1992, \apj, 395, 130, \dodoi{10.1086/171637}

\bibitem[{{Peng} {et~al.}(2021){Peng}, {Liu}, \& {Zhang}}]{Peng2021a}
{Peng}, Z.-K., {Liu}, Z.-K., \& {Zhang}, B.-B. 2021, arXiv e-prints,
  arXiv:2109.06041.
\newblock \doarXiv{2109.06041}

\bibitem[{{Perego} {et~al.}(2017){Perego}, {Radice}, \&
  {Bernuzzi}}]{Perego2017a}
{Perego}, A., {Radice}, D., \& {Bernuzzi}, S. 2017, \apjl, 850, L37,
  \dodoi{10.3847/2041-8213/aa9ab9}

\bibitem[{{Perley} {et~al.}(2008){Perley}, {Li}, {Chornock}, {Prochaska},
  {Butler}, {Chandra}, {Pollack}, {Bloom}, {Filippenko}, {Swan}, {Yuan},
  {Akerlof}, {Auger}, {Cenko}, {Chen}, {Fassnacht}, {Fox}, {Frail},
  {Johansson}, {McKay}, {Le Mignant}, {Modjaz}, {Rujopakarn}, {Russel},
  {Skinner}, {Smith}, {Smith}, {van Dam}, \& {Yost}}]{Perley2008a}
{Perley}, D.~A., {Li}, W., {Chornock}, R., {et~al.} 2008, \apj, 688, 470,
  \dodoi{10.1086/591961}

\bibitem[{{Perley} {et~al.}(2013){Perley}, {Levan}, {Tanvir}, {Cenko}, {Bloom},
  {Hjorth}, {Kr{\"u}hler}, {Filippenko}, {Fruchter}, {Fynbo}, {Jakobsson},
  {Kalirai}, {Milvang-Jensen}, {Morgan}, {Prochaska}, \&
  {Silverman}}]{Perley2013a}
{Perley}, D.~A., {Levan}, A.~J., {Tanvir}, N.~R., {et~al.} 2013, \apj, 778,
  128, \dodoi{10.1088/0004-637X/778/2/128}

\bibitem[{{Pian} {et~al.}(2017){Pian}, {D'Avanzo}, {Benetti}, {Branchesi},
  {Brocato}, {Campana}, {Cappellaro}, {Covino}, {D'Elia}, {Fynbo}, {Getman},
  {Ghirlanda}, {Ghisellini}, {Grado}, {Greco}, {Hjorth}, {Kouveliotou},
  {Levan}, {Limatola}, {Malesani}, {Mazzali}, {Melandri}, {M{\o}ller},
  {Nicastro}, {Palazzi}, {Piranomonte}, {Rossi}, {Salafia}, {Selsing},
  {Stratta}, {Tanaka}, {Tanvir}, {Tomasella}, {Watson}, {Yang}, {Amati},
  {Antonelli}, {Ascenzi}, {Bernardini}, {Bo{\"e}r}, {Bufano}, {Bulgarelli},
  {Capaccioli}, {Casella}, {Castro-Tirado}, {Chassande-Mottin}, {Ciolfi},
  {Copperwheat}, {Dadina}, {De Cesare}, {di Paola}, {Fan}, {Gendre},
  {Giuffrida}, {Giunta}, {Hunt}, {Israel}, {Jin}, {Kasliwal}, {Klose}, {Lisi},
  {Longo}, {Maiorano}, {Mapelli}, {Masetti}, {Nava}, {Patricelli}, {Perley},
  {Pescalli}, {Piran}, {Possenti}, {Pulone}, {Razzano}, {Salvaterra},
  {Schipani}, {Spera}, {Stamerra}, {Stella}, {Tagliaferri}, {Testa}, {Troja},
  {Turatto}, {Vergani}, \& {Vergani}}]{Pian2017a}
{Pian}, E., {D'Avanzo}, P., {Benetti}, S., {et~al.} 2017, \nat, 551, 67,
  \dodoi{10.1038/nature24298}

\bibitem[{{Pinna} {et~al.}(2021){Pinna}, {Rossi}, {Puglisi}, {Agapito},
  {Bonaglia}, {Plantet}, {Mazzoni}, {Briguglio}, {Carbonaro}, {Xompero},
  {Grani}, {Riccardi}, {Esposito}, {Hinz}, {Vaz}, {Ertel}, {Montoya}, {Durney},
  {Christou}, {Miller}, {Taylor}, {Cavallaro}, \& {Lefebvre}}]{Pinna2021a}
{Pinna}, E., {Rossi}, F., {Puglisi}, A., {et~al.} 2021, arXiv e-prints,
  arXiv:2101.07091.
\newblock \doarXiv{2101.07091}

\bibitem[{{Pittori} {et~al.}(2020){Pittori}, {Verrecchia}, {Ursi}, {Tavani},
  {Longo}, {Parmiggiani}, \& {Lucarelli}}]{Pittori2020GCNagile}
{Pittori}, C., {Verrecchia}, F., {Ursi}, A., {et~al.} 2020, GRB Coordinates
  Network, 28289, 1

\bibitem[{{Planck Collaboration} {et~al.}(2016){Planck Collaboration}, {Ade},
  {Aghanim}, {Arnaud}, {Ashdown}, {Aumont}, {Baccigalupi}, {Banday},
  {Barreiro}, {Bartlett}, {Bartolo}, {Battaner}, {Battye}, {Benabed},
  {Beno{\^\i}t}, {Benoit-L{\'e}vy}, {Bernard}, {Bersanelli}, {Bielewicz},
  {Bock}, {Bonaldi}, {Bonavera}, {Bond}, {Borrill}, {Bouchet}, {Boulanger},
  {Bucher}, {Burigana}, {Butler}, {Calabrese}, {Cardoso}, {Catalano},
  {Challinor}, {Chamballu}, {Chary}, {Chiang}, {Chluba}, {Christensen},
  {Church}, {Clements}, {Colombi}, {Colombo}, {Combet}, {Coulais}, {Crill},
  {Curto}, {Cuttaia}, {Danese}, {Davies}, {Davis}, {de Bernardis}, {de Rosa},
  {de Zotti}, {Delabrouille}, {D{\'e}sert}, {Di Valentino}, {Dickinson},
  {Diego}, {Dolag}, {Dole}, {Donzelli}, {Dor{\'e}}, {Douspis}, {Ducout},
  {Dunkley}, {Dupac}, {Efstathiou}, {Elsner}, {En{\ss}lin}, {Eriksen},
  {Farhang}, {Fergusson}, {Finelli}, {Forni}, {Frailis}, {Fraisse},
  {Franceschi}, {Frejsel}, {Galeotta}, {Galli}, {Ganga}, {Gauthier}, {Gerbino},
  {Ghosh}, {Giard}, {Giraud-H{\'e}raud}, {Giusarma}, {Gjerl{\o}w},
  {Gonz{\'a}lez-Nuevo}, {G{\'o}rski}, {Gratton}, {Gregorio}, {Gruppuso},
  {Gudmundsson}, {Hamann}, {Hansen}, {Hanson}, {Harrison}, {Helou},
  {Henrot-Versill{\'e}}, {Hern{\'a}ndez-Monteagudo}, {Herranz}, {Hildebrand t},
  {Hivon}, {Hobson}, {Holmes}, {Hornstrup}, {Hovest}, {Huang}, {Huffenberger},
  {Hurier}, {Jaffe}, {Jaffe}, {Jones}, {Juvela}, {Keih{\"a}nen}, {Keskitalo},
  {Kisner}, {Kneissl}, {Knoche}, {Knox}, {Kunz}, {Kurki-Suonio}, {Lagache},
  {L{\"a}hteenm{\"a}ki}, {Lamarre}, {Lasenby}, {Lattanzi}, {Lawrence}, {Leahy},
  {Leonardi}, {Lesgourgues}, {Levrier}, {Lewis}, {Liguori}, {Lilje},
  {Linden-V{\o}rnle}, {L{\'o}pez-Caniego}, {Lubin}, {Mac{\'\i}as-P{\'e}rez},
  {Maggio}, {Maino}, {Mandolesi}, {Mangilli}, {Marchini}, {Maris}, {Martin},
  {Martinelli}, {Mart{\'\i}nez-Gonz{\'a}lez}, {Masi}, {Matarrese}, {McGehee},
  {Meinhold}, {Melchiorri}, {Melin}, {Mendes}, {Mennella}, {Migliaccio},
  {Millea}, {Mitra}, {Miville-Desch{\^e}nes}, {Moneti}, {Montier}, {Morgante},
  {Mortlock}, {Moss}, {Munshi}, {Murphy}, {Naselsky}, {Nati}, {Natoli},
  {Netterfield}, {N{\o}rgaard-Nielsen}, {Noviello}, {Novikov}, {Novikov},
  {Oxborrow}, {Paci}, {Pagano}, {Pajot}, {Paladini}, {Paoletti}, {Partridge},
  {Pasian}, {Patanchon}, {Pearson}, {Perdereau}, {Perotto}, {Perrotta},
  {Pettorino}, {Piacentini}, {Piat}, {Pierpaoli}, {Pietrobon}, {Plaszczynski},
  {Pointecouteau}, {Polenta}, {Popa}, {Pratt}, {Pr{\'e}zeau}, {Prunet},
  {Puget}, {Rachen}, {Reach}, {Rebolo}, {Reinecke}, {Remazeilles}, {Renault},
  {Renzi}, {Ristorcelli}, {Rocha}, {Rosset}, {Rossetti}, {Roudier},
  {Rouill{\'e} d'Orfeuil}, {Rowan-Robinson}, {Rubi{\~n}o-Mart{\'\i}n},
  {Rusholme}, {Said}, {Salvatelli}, {Salvati}, {Sandri}, {Santos},
  {Savelainen}, {Savini}, {Scott}, {Seiffert}, {Serra}, {Shellard}, {Spencer},
  {Spinelli}, {Stolyarov}, {Stompor}, {Sudiwala}, {Sunyaev}, {Sutton},
  {Suur-Uski}, {Sygnet}, {Tauber}, {Terenzi}, {Toffolatti}, {Tomasi},
  {Tristram}, {Trombetti}, {Tucci}, {Tuovinen}, {T{\"u}rler}, {Umana},
  {Valenziano}, {Valiviita}, {Van Tent}, {Vielva}, {Villa}, {Wade}, {Wandelt},
  {Wehus}, {White}, {White}, {Wilkinson}, {Yvon}, {Zacchei}, \&
  {Zonca}}]{Planck2016a}
{Planck Collaboration}, {Ade}, P.~A.~R., {Aghanim}, N., {et~al.} 2016, \aap,
  594, A13, \dodoi{10.1051/0004-6361/201525830}

\bibitem[{{Pogge} {et~al.}(2010){Pogge}, {Atwood}, {Brewer}, {Byard},
  {Derwent}, {Gonzalez}, {Martini}, {Mason}, {O'Brien}, {Osmer}, {Pappalardo},
  {Steinbrecher}, {Teiga}, \& {Zhelem}}]{Pogge2010a}
{Pogge}, R.~W., {Atwood}, B., {Brewer}, D.~F., {et~al.} 2010, in Society of
  Photo-Optical Instrumentation Engineers (SPIE) Conference Series, Vol. 7735,
  Ground-based and Airborne Instrumentation for Astronomy III, ed. I.~S.
  {McLean}, S.~K. {Ramsay}, \& H.~{Takami}, 77350A, \dodoi{10.1117/12.857215}

\bibitem[{{Politsch} {et~al.}(2020){Politsch}, {Cisewski-Kehe}, {Croft}, \&
  {Wasserman}}]{Politsch2020a}
{Politsch}, C.~A., {Cisewski-Kehe}, J., {Croft}, R. A.~C., \& {Wasserman}, L.
  2020, \mnras, 492, 4005, \dodoi{10.1093/mnras/staa106}

\bibitem[{{Pollack} {et~al.}(2009){Pollack}, {Chen}, {Prochaska}, \&
  {Bloom}}]{Pollack2009a}
{Pollack}, L.~K., {Chen}, H.~W., {Prochaska}, J.~X., \& {Bloom}, J.~S. 2009,
  \apj, 701, 1605, \dodoi{10.1088/0004-637X/701/2/1605}

\bibitem[{{Radice} {et~al.}(2018{\natexlab{a}}){Radice}, {Perego}, {Bernuzzi},
  \& {Zhang}}]{Radice2018b}
{Radice}, D., {Perego}, A., {Bernuzzi}, S., \& {Zhang}, B. 2018{\natexlab{a}},
  \mnras, 481, 3670, \dodoi{10.1093/mnras/sty2531}

\bibitem[{{Radice} {et~al.}(2018{\natexlab{b}}){Radice}, {Perego},
  {Hotokezaka}, {Fromm}, {Bernuzzi}, \& {Roberts}}]{Radice2018a}
{Radice}, D., {Perego}, A., {Hotokezaka}, K., {et~al.} 2018{\natexlab{b}},
  \apj, 869, 130, \dodoi{10.3847/1538-4357/aaf054}

\bibitem[{{Rastinejad} {et~al.}(2021){Rastinejad}, {Fong}, {Kilpatrick},
  {Paterson}, {Tanvir}, {Levan}, {Metzger}, {Berger}, {Chornock}, {Cobb},
  {Laskar}, {Milne}, {Nugent}, \& {Smith}}]{Rastinejad2021a}
{Rastinejad}, J.~C., {Fong}, W., {Kilpatrick}, C.~D., {et~al.} 2021, \apj, 916,
  89, \dodoi{10.3847/1538-4357/ac04b4}

\bibitem[{{Rhodes} {et~al.}(2021){Rhodes}, {Fender}, {Williams}, \&
  {Mooley}}]{Rhodes2021a}
{Rhodes}, L., {Fender}, R., {Williams}, D. R.~A., \& {Mooley}, K. 2021, \mnras,
  503, 2966, \dodoi{10.1093/mnras/stab640}

\bibitem[{{Ridnaia} {et~al.}(2020){Ridnaia}, {Golenetskii}, {Aptekar},
  {Frederiks}, {Ulanov}, {Svinkin}, {Tsvetkova}, {Lysenko}, {Cline}, \&
  {Konus-Wind Team}}]{Ridnaia2020GCNkw}
{Ridnaia}, A., {Golenetskii}, S., {Aptekar}, R., {et~al.} 2020, GRB Coordinates
  Network, 28294, 1

\bibitem[{{Rossi} {et~al.}(2018){Rossi}, {Stratta}, {Maiorano}, {Amati},
  {Nicastro}, \& {Palazzi}}]{Rossi2018a}
{Rossi}, A., {Stratta}, G., {Maiorano}, E., {et~al.} 2018, \memsai, 89, 254.
\newblock \doarXiv{1802.01688}

\bibitem[{{Rossi} {et~al.}(2012){Rossi}, {Klose}, {Ferrero}, {Greiner},
  {Arnold}, {Gonsalves}, {Hartmann}, {Updike}, {Kann}, {Kr{\"u}hler},
  {Palazzi}, {Savaglio}, {Schulze}, {Afonso}, {Amati}, {Castro-Tirado},
  {Clemens}, {Filgas}, {Gorosabel}, {Hunt}, {K{\"u}pc{\"u} Yolda{\c{s}}},
  {Masetti}, {Nardini}, {Nicuesa Guelbenzu}, {Olivares}, {Pian}, {Rau},
  {Schady}, {Schmidl}, {Yolda{\c{s}}}, \& {de Ugarte Postigo}}]{Rossi2012a}
{Rossi}, A., {Klose}, S., {Ferrero}, P., {et~al.} 2012, \aap, 545, A77,
  \dodoi{10.1051/0004-6361/201117201}

\bibitem[{{Rossi} {et~al.}(2020{\natexlab{a}}){Rossi}, {D'Avanzo}, {D'Elia},
  {Melandri}, {Palazzi}, {CIBO Collaboration}, {Rothberg}, {Kuhn}, \&
  {Veillet}}]{Rossi2020GCN}
{Rossi}, A., {D'Avanzo}, P., {D'Elia}, V., {et~al.} 2020{\natexlab{a}}, GRB
  Coordinates Network, 28949, 1

\bibitem[{{Rossi} {et~al.}(2020{\natexlab{b}}){Rossi}, {Stratta}, {Maiorano},
  {Spighi}, {Masetti}, {Palazzi}, {Gardini}, {Melandri}, {Nicastro}, {Pian},
  {Branchesi}, {Dadina}, {Testa}, {Brocato}, {Benetti}, {Ciolfi}, {Covino},
  {D'Elia}, {Grado}, {Izzo}, {Perego}, {Piranomonte}, {Salvaterra}, {Selsing},
  {Tomasella}, {Yang}, {Vergani}, {Amati}, \& {Stephen}}]{Rossi2020a}
{Rossi}, A., {Stratta}, G., {Maiorano}, E., {et~al.} 2020{\natexlab{b}},
  \mnras, 493, 3379, \dodoi{10.1093/mnras/staa479}

\bibitem[{{Rothberg} {et~al.}(2020{\natexlab{a}}){Rothberg}, {Kuhn}, {Veillet},
  \& {Allanson}}]{Rothberg2020GCNlbt}
{Rothberg}, B., {Kuhn}, O., {Veillet}, C., \& {Allanson}, S.
  2020{\natexlab{a}}, GRB Coordinates Network, 28319, 1

\bibitem[{{Rothberg} {et~al.}(2020{\natexlab{b}}){Rothberg}, {Power}, {Kuhn},
  {Thompson}, {Hill}, {Wagner}, \& {Veillet}}]{2020SPIE11447E..06R}
{Rothberg}, B., {Power}, J., {Kuhn}, O., {et~al.} 2020{\natexlab{b}}, in
  Society of Photo-Optical Instrumentation Engineers (SPIE) Conference Series,
  Vol. 11447, Society of Photo-Optical Instrumentation Engineers (SPIE)
  Conference Series, 1144706, \dodoi{10.1117/12.2563352}

\bibitem[{{Sagues Carracedo} {et~al.}(2020){Sagues Carracedo}, {Kumar},
  {Ahumada}, {Andreoni}, {Anand}, {Stein}, {Coughlin}, \&
  {Singer}}]{Carracedo2020GCN}
{Sagues Carracedo}, A., {Kumar}, H., {Ahumada}, T., {et~al.} 2020, GRB
  Coordinates Network, 28293, 1

\bibitem[{{Sari} {et~al.}(1998){Sari}, {Piran}, \& {Narayan}}]{Sari1998a}
{Sari}, R., {Piran}, T., \& {Narayan}, R. 1998, \apjl, 497, L17,
  \dodoi{10.1086/311269}

\bibitem[{{Savaglio} {et~al.}(2009){Savaglio}, {Glazebrook}, \& {Le
  Borgne}}]{Savaglio2009a}
{Savaglio}, S., {Glazebrook}, K., \& {Le Borgne}, D. 2009, \apj, 691, 182,
  \dodoi{10.1088/0004-637X/691/1/182}

\bibitem[{{Schlafly} \& {Finkbeiner}(2011)}]{SchlaflyFinkbeiner2011a}
{Schlafly}, E.~F., \& {Finkbeiner}, D.~P. 2011, \apj, 737, 103,
  \dodoi{10.1088/0004-637X/737/2/103}

\bibitem[{{Schreiber} {et~al.}(2018){Schreiber}, {Elbaz}, {Pannella}, {Ciesla},
  {Wang}, \& {Franco}}]{Schreiber2018a}
{Schreiber}, C., {Elbaz}, D., {Pannella}, M., {et~al.} 2018, \aap, 609, A30,
  \dodoi{10.1051/0004-6361/201731506}

\bibitem[{{Schulze} {et~al.}(2011){Schulze}, {Klose}, {Bj{\"o}rnsson},
  {Jakobsson}, {Kann}, {Rossi}, {Kr{\"u}hler}, {Greiner}, \&
  {Ferrero}}]{Schulze2011a}
{Schulze}, S., {Klose}, S., {Bj{\"o}rnsson}, G., {et~al.} 2011, \aap, 526, A23,
  \dodoi{10.1051/0004-6361/201015581}

\bibitem[{{Schulze} {et~al.}(2014){Schulze}, {Malesani}, {Cucchiara}, {Tanvir},
  {Kr{\"u}hler}, {de Ugarte Postigo}, {Leloudas}, {Lyman}, {Bersier},
  {Wiersema}, {Perley}, {Schady}, {Gorosabel}, {Anderson}, {Castro-Tirado},
  {Cenko}, {De Cia}, {Ellerbroek}, {Fynbo}, {Greiner}, {Hjorth}, {Kann},
  {Kaper}, {Klose}, {Levan}, {Mart{\'\i}n}, {O'Brien}, {Page}, {Pignata},
  {Rapaport}, {S{\'a}nchez-Ram{\'\i}rez}, {Sollerman}, {Smith}, {Sparre},
  {Th{\"o}ne}, {Watson}, {Xu}, {Bauer}, {Bayliss}, {Bj{\"o}rnsson}, {Bremer},
  {Cano}, {Covino}, {D'Elia}, {Frail}, {Geier}, {Goldoni}, {Hartoog},
  {Jakobsson}, {Korhonen}, {Lee}, {Milvang-Jensen}, {Nardini}, {Nicuesa
  Guelbenzu}, {Oguri}, {Pandey}, {Petitpas}, {Rossi}, {Sandberg}, {Schmidl},
  {Tagliaferri}, {Tilanus}, {Winters}, {Wright}, \& {Wuyts}}]{Schulze2014a}
{Schulze}, S., {Malesani}, D., {Cucchiara}, A., {et~al.} 2014, \aap, 566, A102,
  \dodoi{10.1051/0004-6361/201423387}

\bibitem[{{Schulze} {et~al.}(2018){Schulze}, {Kr{\"u}hler}, {Leloudas},
  {Gorosabel}, {Mehner}, {Buchner}, {Kim}, {Ibar}, {Amor{\'\i}n},
  {Herrero-Illana}, {Anderson}, {Bauer}, {Christensen}, {de Pasquale}, {de
  Ugarte Postigo}, {Gallazzi}, {Hjorth}, {Morrell}, {Malesani}, {Sparre},
  {Stalder}, {Stark}, {Th{\"o}ne}, \& {Wheeler}}]{Schulze2018a}
{Schulze}, S., {Kr{\"u}hler}, T., {Leloudas}, G., {et~al.} 2018, \mnras, 473,
  1258, \dodoi{10.1093/mnras/stx2352}

\bibitem[{{Science Software Branch at STScI}(2012)}]{pyraf2012}
{Science Software Branch at STScI}. 2012, {PyRAF: Python alternative for IRAF}.
\newblock \doeprint{1207.011}

\bibitem[{{Seifert} {et~al.}(2003){Seifert}, {Appenzeller}, {Baumeister},
  {Bizenberger}, {Bomans}, {Dettmar}, {Grimm}, {Herbst}, {Hofmann}, {Juette},
  {Laun}, {Lehmitz}, {Lemke}, {Lenzen}, {Mandel}, {Polsterer}, {Rohloff},
  {Schuetze}, {Seltmann}, {Thatte}, {Weiser}, \& {Xu}}]{Seifert2003a}
{Seifert}, W., {Appenzeller}, I., {Baumeister}, H., {et~al.} 2003, in Society
  of Photo-Optical Instrumentation Engineers (SPIE) Conference Series, Vol.
  4841, Instrument Design and Performance for Optical/Infrared Ground-based
  Telescopes, ed. M.~{Iye} \& A.~F.~M. {Moorwood}, 962--973,
  \dodoi{10.1117/12.459494}

\bibitem[{{Siegel} \& {Ciolfi}(2016)}]{Siegel2016a}
{Siegel}, D.~M., \& {Ciolfi}, R. 2016, \apj, 819, 14,
  \dodoi{10.3847/0004-637X/819/1/14}

\bibitem[{{Smartt} {et~al.}(2017){Smartt}, {Chen}, {Jerkstrand}, {Coughlin},
  {Kankare}, {Sim}, {Fraser}, {Inserra}, {Maguire}, {Chambers}, {Huber},
  {Kr{\"u}hler}, {Leloudas}, {Magee}, {Shingles}, {Smith}, {Young}, {Tonry},
  {Kotak}, {Gal-Yam}, {Lyman}, {Homan}, {Agliozzo}, {Anderson}, {Angus},
  {Ashall}, {Barbarino}, {Bauer}, {Berton}, {Botticella}, {Bulla}, {Bulger},
  {Cannizzaro}, {Cano}, {Cartier}, {Cikota}, {Clark}, {De Cia}, {Della Valle},
  {Denneau}, {Dennefeld}, {Dessart}, {Dimitriadis}, {Elias-Rosa}, {Firth},
  {Flewelling}, {Fl{\"o}rs}, {Franckowiak}, {Frohmaier}, {Galbany},
  {Gonz{\'a}lez-Gait{\'a}n}, {Greiner}, {Gromadzki}, {Guelbenzu},
  {Guti{\'e}rrez}, {Hamanowicz}, {Hanlon}, {Harmanen}, {Heintz}, {Heinze},
  {Hernandez}, {Hodgkin}, {Hook}, {Izzo}, {James}, {Jonker}, {Kerzendorf},
  {Klose}, {Kostrzewa-Rutkowska}, {Kowalski}, {Kromer}, {Kuncarayakti},
  {Lawrence}, {Lowe}, {Magnier}, {Manulis}, {Martin-Carrillo}, {Mattila},
  {McBrien}, {M{\"u}ller}, {Nordin}, {O'Neill}, {Onori}, {Palmerio},
  {Pastorello}, {Patat}, {Pignata}, {Podsiadlowski}, {Pumo}, {Prentice}, {Rau},
  {Razza}, {Rest}, {Reynolds}, {Roy}, {Ruiter}, {Rybicki}, {Salmon}, {Schady},
  {Schultz}, {Schweyer}, {Seitenzahl}, {Smith}, {Sollerman}, {Stalder},
  {Stubbs}, {Sullivan}, {Szegedi}, {Taddia}, {Taubenberger}, {Terreran}, {van
  Soelen}, {Vos}, {Wainscoat}, {Walton}, {Waters}, {Weiland}, {Willman},
  {Wiseman}, {Wright}, {Wyrzykowski}, \& {Yaron}}]{Smartt2017a}
{Smartt}, S.~J., {Chen}, T.-W., {Jerkstrand}, A., {et~al.} 2017, \nat, 551, 75,
  \dodoi{10.1038/nature24303}

\bibitem[{{Soderberg} {et~al.}(2006){Soderberg}, {Kulkarni}, {Price}, {Fox},
  {Berger}, {Moon}, {Cenko}, {Gal-Yam}, {Frail}, {Chevalier}, {Cowie}, {Da
  Costa}, {MacFadyen}, {McCarthy}, {Noel}, {Park}, {Peterson}, {Phillips},
  {Rauch}, {Rest}, {Rich}, {Roth}, {Roth}, {Schmidt}, {Smith}, \&
  {Wood}}]{Soderberg2006ApJ}
{Soderberg}, A.~M., {Kulkarni}, S.~R., {Price}, P.~A., {et~al.} 2006, \apj,
  636, 391, \dodoi{10.1086/498009}

\bibitem[{{Svinkin} {et~al.}(2020){Svinkin}, {Frederiks}, {Ridnaia},
  {Tsvetkova}, \& {Konus-Wind Team}}]{Svinkin2020GCNenergy}
{Svinkin}, D., {Frederiks}, D., {Ridnaia}, A., {Tsvetkova}, A., \& {Konus-Wind
  Team}. 2020, GRB Coordinates Network, 28301, 1

\bibitem[{{Tanga} {et~al.}(2018){Tanga}, {Kr{\"u}hler}, {Schady}, {Klose},
  {Graham}, {Greiner}, {Kann}, \& {Nardini}}]{Tanga2018a}
{Tanga}, M., {Kr{\"u}hler}, T., {Schady}, P., {et~al.} 2018, \aap, 615, A136,
  \dodoi{10.1051/0004-6361/201731799}

\bibitem[{{Th{\"o}ne} {et~al.}(2011){Th{\"o}ne}, {Campana}, {Lazzati}, {de
  Ugarte Postigo}, {Fynbo}, {Christensen}, {Levan}, {Aloy}, {Hjorth},
  {Jakobsson}, {Levesque}, {Malesani}, {Milvang-Jensen}, {Roming}, {Tanvir},
  {Wiersema}, {Gladders}, {Wuyts}, \& {Dahle}}]{Thone2011MNRAS}
{Th{\"o}ne}, C.~C., {Campana}, S., {Lazzati}, D., {et~al.} 2011, \mnras, 414,
  479, \dodoi{10.1111/j.1365-2966.2011.18408.x}

\bibitem[{{Tody}(1993)}]{Tody1993}
{Tody}, D. 1993, in Astronomical Society of the Pacific Conference Series,
  Vol.~52, Astronomical Data Analysis Software and Systems II, ed.
  {R.~J.~Hanisch, R.~J.~V.~Brissenden, \& J.~Barnes}, 173

\bibitem[{{Toy} {et~al.}(2016){Toy}, {Cenko}, {Silverman}, {Butler},
  {Cucchiara}, {Watson}, {Bersier}, {Perley}, {Margutti}, {Bellm}, {Bloom},
  {Cao}, {Capone}, {Clubb}, {Corsi}, {De Cia}, {de Diego}, {Filippenko}, {Fox},
  {Gal-Yam}, {Gehrels}, {Georgiev}, {Gonz{\'a}lez}, {Kasliwal}, {Kelly},
  {Kulkarni}, {Kutyrev}, {Lee}, {Prochaska}, {Ramirez-Ruiz}, {Richer},
  {Rom{\'a}n-Z{\'u}{\~n}iga}, {Singer}, {Stern}, {Troja}, \&
  {Veilleux}}]{Toy2016ApJ}
{Toy}, V.~L., {Cenko}, S.~B., {Silverman}, J.~M., {et~al.} 2016, \apj, 818, 79,
  \dodoi{10.3847/0004-637X/818/1/79}

\bibitem[{{Trouille} {et~al.}(2011){Trouille}, {Barger}, \&
  {Tremonti}}]{Trouille2011a}
{Trouille}, L., {Barger}, A.~J., \& {Tremonti}, C. 2011, \apj, 742, 46,
  \dodoi{10.1088/0004-637X/742/1/46}

\bibitem[{{Vergani} {et~al.}(2009){Vergani}, {Petitjean}, {Ledoux},
  {Vreeswijk}, {Smette}, \& {Meurs}}]{Vergani2009a}
{Vergani}, S.~D., {Petitjean}, P., {Ledoux}, C., {et~al.} 2009, \aap, 503, 771,
  \dodoi{10.1051/0004-6361/200911747}

\bibitem[{{Vergani} {et~al.}(2015){Vergani}, {Salvaterra}, {Japelj}, {Le
  Floc'h}, {D'Avanzo}, {Fernandez-Soto}, {Kr{\"u}hler}, {Melandri}, {Boissier},
  {Covino}, {Puech}, {Greiner}, {Hunt}, {Perley}, {Petitjean}, {Vinci},
  {Hammer}, {Levan}, {Mannucci}, {Campana}, {Flores}, {Gomboc}, \&
  {Tagliaferri}}]{Vergani2015a}
{Vergani}, S.~D., {Salvaterra}, R., {Japelj}, J., {et~al.} 2015, \aap, 581,
  A102, \dodoi{10.1051/0004-6361/201425013}

\bibitem[{{Volnova} {et~al.}(2017){Volnova}, {Pruzhinskaya}, {Pozanenko},
  {Blinnikov}, {Minaev}, {Burkhonov}, {Chernenko}, {Ehgamberdiev},
  {Inasaridze}, {Jelinek}, {Khorunzhev}, {Klunko}, {Krugly}, {Mazaeva},
  {Rumyantsev}, \& {Volvach}}]{Volnova2017MNRAS}
{Volnova}, A.~A., {Pruzhinskaya}, M.~V., {Pozanenko}, A.~S., {et~al.} 2017,
  \mnras, 467, 3500, \dodoi{10.1093/mnras/stw3297}

\bibitem[{{White} {et~al.}(2021){White}, {Bauer}, {Baumgartner}, {Bautz},
  {Berger}, {Cenko}, {Chang}, {Falcone}, {Fausey}, {Feldman}, {Fox}, {Fox},
  {Fruchter}, {Fryer}, {Ghirlanda}, {Gorski}, {Grant}, {Guiriec}, {Hart},
  {Hartmann}, {Hennawi}, {Kann}, {Kaplan}, {Kennea}, {Kocevski}, {Kouveliotou},
  {Lawrence}, {Levan}, {Lidz}, {Lien}, {Littenberg}, {Mas-Ribas}, {Moss},
  {O'Brien}, {O'Meara}, {Palmer}, {Pasham}, {Racusin}, {Remillard}, {Roberts},
  {Roming}, {Rud}, {Salvaterra}, {Sambruna}, {Seiffert}, {Sun}, {Tanvir},
  {Terrile}, {Thomas}, {van der Horst}, {Verstrand}, {Willems}, {Wilson-Hodge},
  {Young}, {Amati}, {Bozzo}, {Karczewski}, {Hernandez-Monteagudo}, {Rebolo
  Lopez}, {Genova-Santos}, {Martin}, {Granot}, {Bemiamini}, {Gil}, \&
  {Burns}}]{White2021SPIE}
{White}, N.~E., {Bauer}, F.~E., {Baumgartner}, W., {et~al.} 2021, in Society of
  Photo-Optical Instrumentation Engineers (SPIE) Conference Series, Vol. 11821,
  Society of Photo-Optical Instrumentation Engineers (SPIE) Conference Series,
  1182109, \dodoi{10.1117/12.2599293}

\bibitem[{{Wiersema} {et~al.}(2008){Wiersema}, {van der Horst}, {Kann}, {Rol},
  {Starling}, {Curran}, {Gorosabel}, {Levan}, {Fynbo}, {de Ugarte Postigo},
  {Wijers}, {Castro-Tirado}, {Guziy}, {Hornstrup}, {Hjorth}, {Jel{\'\i}nek},
  {Jensen}, {Kidger}, {Mart{\'\i}n-Luis}, {Tanvir}, {Tristram}, \&
  {Vreeswijk}}]{Wiersema2008a}
{Wiersema}, K., {van der Horst}, A.~J., {Kann}, D.~A., {et~al.} 2008, \aap,
  481, 319, \dodoi{10.1051/0004-6361:20078050}

\bibitem[{{Willingale} {et~al.}(2013){Willingale}, {Starling}, {Beardmore},
  {Tanvir}, \& {O'Brien}}]{Willingale2013a}
{Willingale}, R., {Starling}, R.~L.~C., {Beardmore}, A.~P., {Tanvir}, N.~R., \&
  {O'Brien}, P.~T. 2013, \mnras, 431, 394, \dodoi{10.1093/mnras/stt175}

\bibitem[{{Woosley} \& {Bloom}(2006)}]{WoosleyBloom2006a}
{Woosley}, S.~E., \& {Bloom}, J.~S. 2006, \araa, 44, 507,
  \dodoi{10.1146/annurev.astro.43.072103.150558}

\bibitem[{{Xu} {et~al.}(2009){Xu}, {Starling}, {Fynbo}, {Sollerman}, {Yost},
  {Watson}, {Foley}, {O'Brien}, \& {Hjorth}}]{Xu2009a}
{Xu}, D., {Starling}, R.~L.~C., {Fynbo}, J.~P.~U., {et~al.} 2009, \apj, 696,
  971, \dodoi{10.1088/0004-637X/696/1/971}

\bibitem[{{Yamazaki} {et~al.}(2004){Yamazaki}, {Ioka}, \&
  {Nakamura}}]{Yamazaki2004a}
{Yamazaki}, R., {Ioka}, K., \& {Nakamura}, T. 2004, \apjl, 607, L103,
  \dodoi{10.1086/421872}

\bibitem[{{Yang} {et~al.}(2015){Yang}, {Jin}, {Li}, {Covino}, {Zheng},
  {Hotokezaka}, {Fan}, {Piran}, \& {Wei}}]{Yang2015a}
{Yang}, B., {Jin}, Z.-P., {Li}, X., {et~al.} 2015, Nature Communications, 6,
  7323, \dodoi{10.1038/ncomms8323}

\bibitem[{{Yu} {et~al.}(2013){Yu}, {Zhang}, \& {Gao}}]{Yu2013a}
{Yu}, Y.-W., {Zhang}, B., \& {Gao}, H. 2013, \apjl, 776, L40,
  \dodoi{10.1088/2041-8205/776/2/L40}

\bibitem[{{Zeh} {et~al.}(2004){Zeh}, {Klose}, \& {Hartmann}}]{Zeh2004ApJ}
{Zeh}, A., {Klose}, S., \& {Hartmann}, D.~H. 2004, \apj, 609, 952,
  \dodoi{10.1086/421100}

\bibitem[{{Zeh} {et~al.}(2006){Zeh}, {Klose}, \& {Kann}}]{Zeh2006a}
{Zeh}, A., {Klose}, S., \& {Kann}, D.~A. 2006, \apj, 637, 889,
  \dodoi{10.1086/498442}

\bibitem[{{Zhang} {et~al.}(2009){Zhang}, {Zhang}, {Virgili}, {Liang}, {Kann},
  {Wu}, {Proga}, {Lv}, {Toma}, {M{\'e}sz{\'a}ros}, {Burrows}, {Roming}, \&
  {Gehrels}}]{Zhang2009ApJ}
{Zhang}, B., {Zhang}, B.-B., {Virgili}, F.~J., {et~al.} 2009, \apj, 703, 1696,
  \dodoi{10.1088/0004-637X/703/2/1696}

\bibitem[{{Zhang} {et~al.}(2021){Zhang}, {Liu}, {Peng}, {Li}, {L{\"u}}, {Yang},
  {Yang}, {Yang}, {Meng}, {Zou}, {Ye}, {Wang}, {Mao}, {Zhao}, {Bai},
  {Castro-Tirado}, {Hu}, {Dai}, {Liang}, \& {Zhang}}]{Zhang2021a}
{Zhang}, B.~B., {Liu}, Z.~K., {Peng}, Z.~K., {et~al.} 2021, Nature Astronomy,
  5, 911, \dodoi{10.1038/s41550-021-01395-z}

\bibitem[{{Zhu} {et~al.}(2020){Zhu}, {Yang}, {Liu}, {Huang}, {Zhang}, {Li},
  {Yu}, \& {Gao}}]{Zhu2020a}
{Zhu}, J.-P., {Yang}, Y.-P., {Liu}, L.-D., {et~al.} 2020, \apj, 897, 20,
  \dodoi{10.3847/1538-4357/ab93bf}

\end{thebibliography}

\end{document}